\documentclass[%
 reprint,
superscriptaddress,
preprintnumbers,
nofootinbib,
 amsmath,amssymb,
 aps,
prd,
]{revtex4-2}

\usepackage[margin=5pt,caption=false]{subfig}
\usepackage{slantsc}
\usepackage{comment}
\usepackage{threeparttable}
\usepackage{booktabs}
\usepackage[figuresright]{rotating}
\usepackage{textcase}
\usepackage{xspace} % for \redmapper
\usepackage[dvipsnames]{xcolor}
\usepackage{multirow}

\newcommand{\hiMsun}{h^{-1}M_{\odot}}
\newcommand{\hiMpc}{h^{-1}\mathrm{Mpc}}

\newcommand{\redmapper}{\texttt{redMaPPer}\xspace} % avoid typoes!
\newcommand{\XMM}{{\em XMM}\xspace}
\newcommand{\Chandra}{{\em Chandra}\xspace}

\newcommand{\pmem}{p_\mathrm{mem}}

\newcommand{\lambH}{\lambda_{\text{h}}}
\newcommand{\fhtop}{f_\mathrm{1h}} 
\newcommand{\fhN}{f_{N\mathrm{h}}}
\newcommand{\lambTop}{\lambda_\mathrm{1h}}

\newcommand{\lambOb}{\lambda_\mathrm{ob}}
\newcommand{\lambTr}{\lambda_\mathrm{tr}}

\newcommand{\fcent}{f_\mathrm{cen}}
\newcommand{\taumis}{\tau_\mathrm{mis}}
\newcommand{\sigcen}{\sigma_\mathrm{cen}}

\newcommand{\hunit}{$\rm{km\,s^{-1}Mpc^{-1}}$}
\newcommand{\lcdm}{$\Lambda$CDM}
\newcommand{\om}{$\Omega_{m}$}
\newcommand{\Om}{\Omega_{m}}
\newcommand{\ob}{$\Omega_{b}$}
\newcommand{\Ob}{\Omega_{b}}
\newcommand{\neff}{N_{\rm eff}}

\usepackage{array}
\usepackage[T1]{fontenc}
\usepackage{scalerel}
\usepackage{tikz}
\usetikzlibrary{svg.path}

\definecolor{orcidlogocol}{HTML}{A6CE39}
\tikzset{
  orcidlogo/.pic={
    \fill[orcidlogocol] svg{M256,128c0,70.7-57.3,128-128,128C57.3,256,0,198.7,0,128C0,57.3,57.3,0,128,0C198.7,0,256,57.3,256,128z};
    \fill[white] svg{M86.3,186.2H70.9V79.1h15.4v48.4V186.2z}
                 svg{M108.9,79.1h41.6c39.6,0,57,28.3,57,53.6c0,27.5-21.5,53.6-56.8,53.6h-41.8V79.1z M124.3,172.4h24.5c34.9,0,42.9-26.5,42.9-39.7c0-21.5-13.7-39.7-43.7-39.7h-23.7V172.4z}
                 svg{M88.7,56.8c0,5.5-4.5,10.1-10.1,10.1c-5.6,0-10.1-4.6-10.1-10.1c0-5.6,4.5-10.1,10.1-10.1C84.2,46.7,88.7,51.3,88.7,56.8z};
  }
}

\newcommand\orcidicon[1]{\href{https://orcid.org/#1}{\mbox{\scalerel*{
\begin{tikzpicture}[yscale=-1,transform shape]
\pic{orcidlogo};
\end{tikzpicture}
}{|}}}}

\usepackage{amsmath,bm}	% Advanced maths commands
\usepackage{amssymb}	% Extra maths symbols
\usepackage{fnpct}

\usepackage[colorlinks=true,allcolors=blue]{hyperref}
\begin{document}

%\preprint{APS/123-QED}

\title{Association between optically identified galaxy clusters and the underlying dark matter halos}
\preprint{DES-2025-0902}
\preprint{FERMILAB-PUB-25-0383-PPD}

\author{Shulei Cao$^{\orcidicon{0000-0003-2421-7071}}$}
 \email[Email: ]{shuleic@mail.smu.edu}
\affiliation{Department of Physics, Southern Methodist University, Dallas, TX 75205, USA}

\author{Hao-Yi Wu$^{\orcidicon{0000-0002-7904-1707}}$}
\email[Email: ]{hywu@mail.smu.edu}
\affiliation{Department of Physics, Southern Methodist University, Dallas, TX 75205, USA}

\author{Matteo Costanzi$^{\orcidicon{0000-0001-8158-1449}}$}
\affiliation{Dipartimento di Fisica---Sezione di Astronomia, Universit\`a di Trieste, 34131 Trieste, Italy, \\
INAF-Osservatorio Astronomico di Trieste, 34143 Trieste, Italy, and \\
IFPU---Institute for Fundamental Physics of the Universe, 34014 Trieste, Italy}

\author{Arya Farahi
$^{\orcidicon{0000-0003-0777-4618}}$}
\affiliation{Departments of Statistics and Data Sciences, University of Texas at Austin, Austin, TX 78712, USA}
\affiliation{The NSF-Simons AI Institute for Cosmic Origins, University of Texas at Austin, Austin, TX 78712, USA}

\author{Sebastian Grandis
$^{\orcidicon{0000-0002-4577-8217}}$}
\affiliation{Universit\"at Innsbruck, Institut f\"ur Astro- und Teilchenphysik, Technikerstr.~25/8, 6020 Innsbruck, Austria}

\author{David H. Weinberg$^{\orcidicon{0000-0001-7775-7261}}$}
\affiliation{Department of Astronomy and Center for Cosmology and AstroParticle Physics (CCAPP), The Ohio State University, Columbus, OH 43210, USA}

\author{August E. Evrard$^{\orcidicon{0000-0002-4876-956X}}$}
\affiliation{Departments of Physics and Astronomy, Leinweber Center for Theoretical Physics, University of Michigan, Ann Arbor, MI 48109, USA}

\author{Eduardo Rozo$^{\orcidicon{0000-0002-1666-6275}}$}
\affiliation{Department of Physics, University of Arizona, Tucson, AZ 85721, USA}

\author{Andr\'es N. Salcedo$^{\orcidicon{0000-0003-1420-527X}}$}
\affiliation{Department of Astronomy/Steward Observatory, University of Arizona, Tucson, AZ 85721, USA}
\affiliation{Department of Physics, University of Arizona, Tucson, AZ 85721, USA}

\author{Chun-Hao To$^{\orcidicon{0000-0001-7836-2261}}$}
\affiliation{Department of Astronomy and Astrophysics, University of Chicago, Chicago, IL 60637, USA}

\author{Lei Yang$^{\orcidicon{0000-0001-8297-0868}}$}
\affiliation{Department of Physics, Southern Methodist University, Dallas, TX 75205, USA}

\author{Conghao Zhou$^{\orcidicon{0000-0002-2897-6326}}$}
\affiliation{Department of Physics and Santa Cruz Institute for Particle Physics, University of California, Santa Cruz, CA 95064, USA}

\collaboration{DES Collaboration}%\noaffiliation

\date{\today}% It is always \today, today,
             %  but any date may be explicitly specified

\begin{abstract}
Clusters of galaxies trace massive dark matter halos in the Universe, but they can include multiple halos projected along lines of sight.  
As a case study, we quantify the properties of halos contributing to clusters identified by the \redmapper algorithm using the Cardinal simulation, which mimics the Dark Energy Survey data.
For each cluster, we identify the halos hosting its member galaxies, and we define the main halo as the one contributing the most to the cluster's richness ($\lambda$, the estimated number of member galaxies). At $z=0.3$, for clusters with $\lambda > 60$, the main halo typically contributes to $92\%$ of the richness, and this fraction drops to $67\%$ for $\lambda \approx 20$. Defining ``clean'' clusters as those with $\geq50\%$ of the richness contributed by the main halo, we find that $100\%$ of the $\lambda > 60$ clusters are clean, while $73\%$ of the $\lambda \approx 20$ clusters are clean. Three halos can usually account for more than $80\%$ of the richness of a cluster. The main halos associated with \redmapper clusters have a completeness ranging from 98\% at virial mass $10^{14.6}~\hiMsun$ to 64\% at $10^{14}~\hiMsun$. In addition, we compare the inferred cluster centers with true halo centers, finding that 30\% of the clusters are miscentered with a mean offset $40\%$ of the cluster radii, in agreement with recent X-ray studies. These systematics worsen as redshift increases, but we expect that upcoming surveys extending to longer wavelengths will improve the cluster finding at high redshifts. Our results affirm the robustness of the \redmapper algorithm and provide a framework for benchmarking other cluster-finding strategies.
\end{abstract}
%\keywords{Suggested keywords}%Use showkeys class option if keyword
                              %display desired
\maketitle

\section{Introduction}

Galaxy clusters, as the name suggests, are most easily identified as high-density regions of galaxies in the sky. Due to galaxies' distance uncertainties, a cluster inevitably includes galaxies in the foreground or background and, therefore, multiple dark matter halos. This effect can lead to systematics in cluster cosmology analyses. In this paper, we attempt to answer the question: What are we really finding with optical cluster finders?

The number of dark matter halos as a function of mass and redshift is a powerful probe of cosmology. In particular, the massive end of the halo mass function, probed by galaxy clusters, is sensitive to the density fluctuation parameter $\sigma_8$ \citep{Henryetal2009,Vikhlininetal2009,MantzRapettiEbeling2010,Rozoetal2010,Weinbergetal2013,DES2021,Wuetal2021,DES2025}. Wide-field cluster surveys have been conducted in optical \citep{GladdersMichaelYee2000,Rykoffetal2014,Gonzalezetal2019,CFC,KiDS2019,DES2025}, 
X-ray \citep{Bohringeretal2001, Melnyketal2018, Kleinebreiletal2024, Bulbuletal2024}, and millimeter waves \citep{Vanderlindeetal2010, Bleemetal2015, Planck2020, Planck2016, Hiltonetal2021, SPT2024, ACT2024}. Among these survey techniques, optical imaging can detect a wide range of cluster masses, estimate richness (the number of galaxies in a cluster) as a cluster mass proxy, and provide weak lensing signals to calibrate the richness-mass relation \cite{Simet17, Melchior17, DES2019a}.

In this paper, we treat galaxy clusters as observational objects with sky coordinates and photometric redshifts, and we consider dark matter halos as 3D spheres with Cartesian coordinates found in $N$-body simulations. The association between optically identified clusters and the underlying dark matter halos is complicated by several factors. First, the member galaxies of a cluster can belong to multiple halos along the line of sight (LOS) due to the distance uncertainties, a systematic uncertainty referred to as {\em projection effects} \citep{Cohnetal2007, Rozoetal2010, Farahietal2016, BuschWhite2017, Zuetal2017, Costanzietal2019, Sunayamaetal2020, Wuetal2022, Sunayamaetal2024, EuclidProj2025}. Second, a cluster sample selected above a richness threshold is not a {\em complete} mass-selected sample due to the scatter between richness and mass \cite{Rozoetal2015, Evrard14}. Third, the inferred cluster center may not be at the minimum of the gravitational potential of the dark matter halos, a systematic effect referred to as {\em miscentering} \cite{BeckerKravtsov2011, Dietrichetal2019, Grandisetal2021a, Schrabbacketal2021, Sommeretal2022}.

We quantify these systematics using the Cardinal simulation \cite{Toetal2024}, which mimics the Dark Energy Survey data.  We use the clusters found by the \redmapper algorithm \citep{Rykoffetal2014, Rozoetal2015, DES2016}, which uses the red sequence---the tight color-magnitude relation for galaxies in clusters---to detect clusters, identify cluster members, and estimate photometric redshifts. For each cluster, we identify the halos contributing to its member galaxies, and we quantify the contribution of main and projected halos.

Our study has been inspired by \citet{Cohnetal2007}, a pioneering study of the cluster-halo association of red-sequence cluster finders. They have used the semi-analytic galaxy catalog from the Millennium Simulation \cite{Crotonetal2006}, added passive color and magnitude evolution, and performed the mock cluster finding based on projected spatial overdensities of red-sequence galaxies. They have found that 90\% of the clusters in their simulations are ``clean,'' and they have quantified the completeness and mass distribution of halos under the richness selection. Although it is difficult to compare our results with theirs due to the different galaxy models and richness definitions, we have adopted several key concepts from their paper, including the definition of ``top-ranked'' halos and ``clean'' clusters.

More recently, Ref.~\cite{Farahietal2016} presents the cluster-halo association for the \redmapper catalog derived from the Aardvark simulation, a precursor of Buzzard \cite{DeRoseetal2019} and Cardinal \cite{Toetal2024}. Ref.~\cite{Costanzietal2019} presents a projection effect model that splits the cluster richness into three components: the main halo, the uncorrelated galaxy background, and the correlated large-scale structure. In a similar spirit, we quantify the contribution from main and projected halos, focusing on the modeling and validation of DES cluster cosmology analyses.

The paper is organized as follows. 
Section~\ref{data} describes the Cardinal simulation and its \redmapper cluster catalog. Section~\ref{sec:matching} presents our cluster-halo matching and quantifies the contribution of main and projected halos to cluster richness. 
Section~\ref{sec:completeness} presents the completeness of halos found by \redmapper as a function of mass.
Section~\ref{sec:miscenter} quantifies the centering offset between clusters and their main halos. 
We compare our results with those from previous studies in Sec.~\ref{sec:previous}
and summarize in Sec.~\ref{sec:summary}.

%%%%%%%%%%%%%%%%%%%%%%%%%%%%%%
%%%%%%%%%%%%%%%%%%%%%%%%%%%%%%
\section{\label{data} Data}

%%%%%%%%%%%%%%%%%%%%%%%%%%%%%%
\begin{figure*}[htbp!]
    \centering
    \includegraphics[width=1\linewidth]{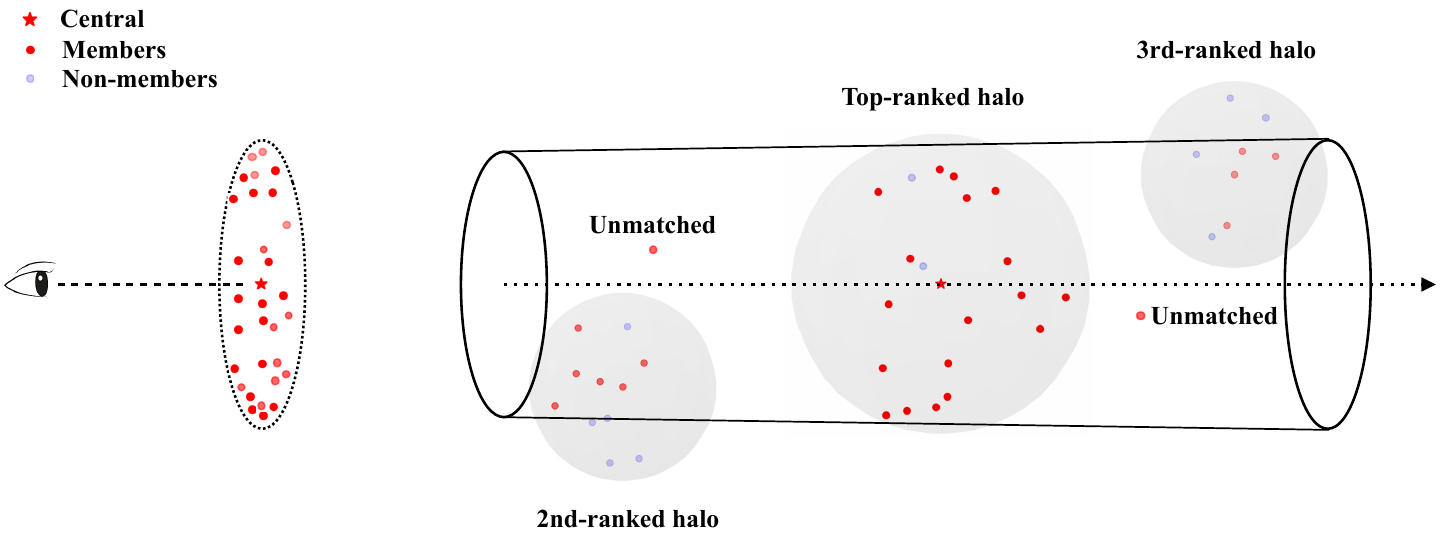}
    \caption{Cartoon picture for our cluster-halo matching process. The ellipse on the left-hand side shows a cluster observed in the sky.  The red star and points correspond to the central and member galaxies. These members are matched to three halos (gray spheres) or are sometimes unmatched due to the limited simulation resolution. We rank halos by their contribution to the richness $\lambH$. 
    For example, the top-ranked halo has $\lambH=19$ (19 galaxies, each with $\pmem=1$), the second-ranked halo has $\lambH=4.8$ (6 galaxies, each with $\pmem=0.8$), and the third-ranked halo has $\lambH=2.4$ (4 galaxies, each with $\pmem=0.6$). Unmatched members contribute to richness $0.8$ (2 galaxies, each with $\pmem=0.4$).
    The total cluster richness is $\lambda=27$, and the top-ranked richness fraction is $\fhtop=70.4\%$. Note that $\pmem$ values do not necessarily decrease with the halo rank.  The blue points are galaxies deemed non-members by the cluster finder.   }
    \label{fig:cartoon}
\end{figure*}
%%%%%%%%%%%%%%%%%%%%%%%%%%%%%%

In this section, we introduce the Cardinal simulation and the \redmapper cluster catalog used in our analyses.

%%%%%%%%%%%%%%%%%%%%%%%%%%%%%%
%%%%%%%%%%%%%%%%%%%%%%%%%%%%%%
\subsection{The Cardinal simulation}

Cardinal \citep{Toetal2024} is a lightcone mock catalog of 10,313 square degrees based on DES Year 6 (Y6) photometric noises and masks. It uses the \texttt{ADDGALS} algorithm \cite{Wechsleretal2022} to populate galaxies in large-volume $N$-body simulations that cannot resolve halos hosting faint galaxies. Compared with its precursor Buzzard \citep{DeRoseetal2019}, Cardinal improves the modeling of cluster galaxies by accounting for the tidal disruption of subhalos. As a result, it has more realistic cluster richness and cluster-galaxy correlation functions and is uniquely suitable for our analysis. One caveat is that some galaxies in Cardinal are associated with density peaks rather than resolved halos. As we will see, on average, 19\% of the richness is not associated with resolved halos at $z=0.3$.

Cardinal has a dark matter particle mass resolution $3.3\times 10^{10}~\hiMsun$ for $z<0.315$ and $1.6\times 10^{11}~\hiMsun$ for $0.315<z<0.955$. Because of this resolution difference, the number of galaxies associated with unresolved halos increases with redshift. We use the isolated halos defined by the \textsc{Rockstar} halo finder \cite{BehrooziWechslerWu2013} and do not consider subhalos in this work. We use the virial halo mass definition based on the spherical overdensity of \cite{BryanNorman1998}, and $M_{\rm h}\equiv{M_\mathrm{vir}}$ throughout this paper. We only consider halos with mass above $6\times10^{12}~\hiMsun$; halos below this mass are incomplete. 
For $z<0.5$, richness is contributed by halos well above this mass. However, for $z\geq0.5$ and $\lambda<60$, the projection effects worsen, and galaxies associated with unresolved halos can contribute to $\approx$40\% of the richness.  As a result, the projection effects for $z\geq0.5$ from Cardinal should be considered as a lower limit.

Cardinal adopts a flat \lcdm\ cosmology with $\Om=0.286$, $\Ob=0.047$, $\sigma_8=0.82$, $n_s=0.96$, $h=0.7$, and $\neff=3.046$. Here \om\ and \ob\ are the current values of the nonrelativistic matter and baryon density parameters, respectively; $\sigma_8$ is the root mean square of the amplitude of matter perturbations on scales of $8~\hiMpc$; $n_s$ is the spectral index of the scalar power spectrum; $h$ is the Hubble constant in units of 100 \hunit; $\neff$ is the effective number of three massless neutrino species.

%%%%%%%%%%%%%%%%%%%%%%%%%%%%%%
\subsection{The \redmapper cluster catalog from Cardinal}

The \redmapper (red-sequence Matched-filter Probabilistic Percolation) algorithm identifies clusters from photometric data by looking for spatial overdensities of red-sequence galaxies \citep{Rykoffetal2014, Rozoetal2015, DES2016}. It iteratively performs red-sequence calibration and cluster finding. For the red-sequence calibration, \redmapper fits the color-magnitude relation using previously identified clusters to build a red-sequence template. For cluster finding, \redmapper considers all galaxies as potential cluster centers and searches member galaxies around them, assigning each member galaxy a membership probability. The cluster redshift ($z_\lambda$, referred to as $z$ hereafter) is determined by comparing the colors of the member galaxies with the red-sequence template.

The \redmapper algorithm defines richness $\lambda$ as 
\begin{equation}
\lambda= S \sum_i p_{\mathrm{mem},i}\theta^{L}_i\theta^{R}_i.
\end{equation}
This summation includes all galaxies within $R_\lambda=(\lambda/100)^{0.2} h^{-1}{\rm Mpc}$ (physical), which is iteratively determined together with $\lambda$.
Here $\pmem\equiv pp_{\rm free}$ is the membership probability; $p$ and $p_{\rm free}$ represent the photometric membership probability and the probability of the galaxy not belonging to a previously-identified cluster, respectively. The weights $\theta^{L}$ and $\theta^{R}$ correspond to the softening near the luminosity and radius threshold \cite{Rozoetal2015}.  The extra $S$ factor corrects for the missing members due to masking.

In this work, we only consider each member's $\pmem$ and omit the extra factors $\theta^{L}$, $\theta^{R}$, and $S$. On average, $\sum_i p_{\mathrm{mem},i}$ is higher than $\lambda$ by $9.4\%$ in our data.  Since we focus on the fractional contribution to richness, these factors do not impact our results.

We use the richness and redshift bins adopted in the DES-Y1 analysis \cite{Abbottetal2020}: $\lambda\in[20, 30)$, $[30, 45)$, $[45, 60)$, and $[60, \infty)$, and $z\in[0.2,0.35)$, $[0.35,0.5)$, and $[0.5,0.65)$. There are 713, 3471, and 3878 clusters in $z\in[0.2,0.35)$, $[0.35,0.5)$, and $[0.5,0.65)$, respectively. Clusters with $\lambda < 20$ are usually not used in cosmological analyses to avoid large fractional uncertainties in richness due to background galaxy density fluctuations.

\section{Halos contributing to a cluster} 
\label{sec:matching}

In this section, we quantify the projection effect present in \redmapper clusters in Cardinal.

\subsection{Cluster-halo matching procedure and visualization}

%%%%%%%%%%%%%%%%%%%%%%%%%%%%%%
\begin{figure*}[htbp!]
    \centering
    \includegraphics[width=0.49\textwidth,height=0.38\textwidth]{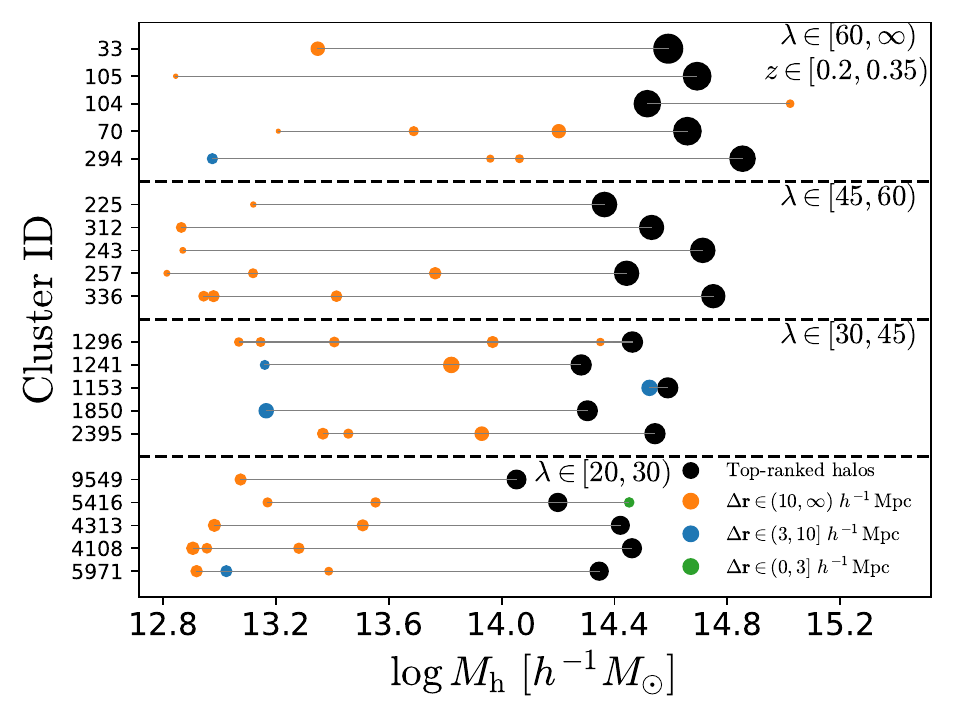}
    \hfill % Horizontal space between images
    \includegraphics[width=0.49\textwidth,height=0.38\textwidth]{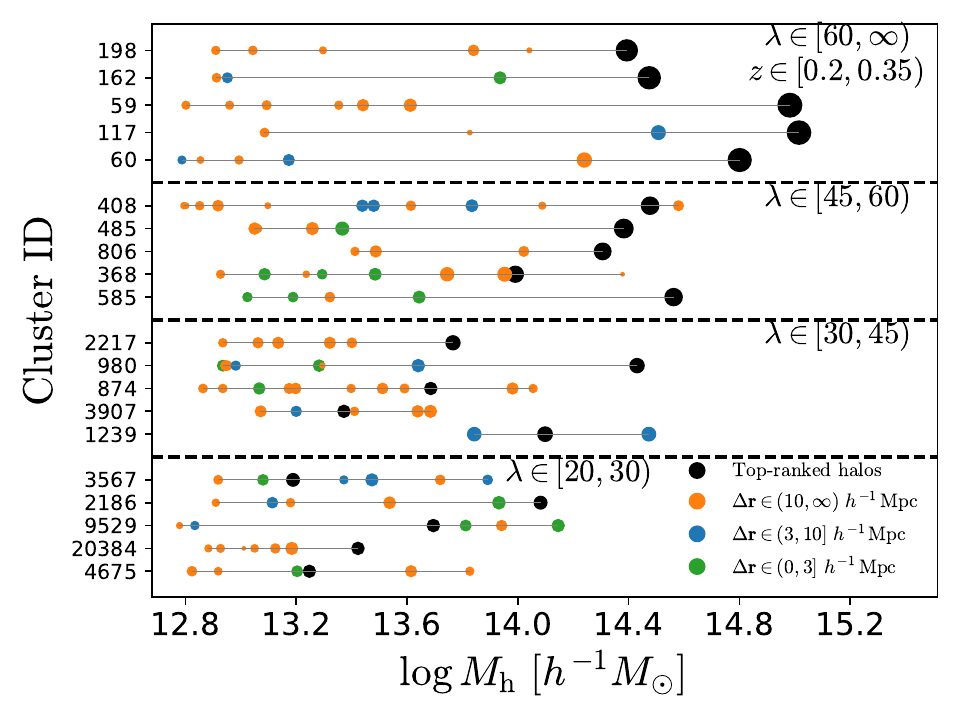}
    \caption{Examples of the halos matched to a cluster. Each row corresponds to one cluster, and the $x$-axis corresponds to the mass of matched halos. The left-hand panel showcases typical clusters (median $\fhtop$, fractional richness contributed by the main halo), while the right-hand panel showcases heavily projected clusters ($\fhtop$ in the bottom two percentile). The black circles correspond to the main halo, while the colored circles correspond to projected halos. The areas of the circles are proportional to the richness contributed by each halo, and the color indicates their 3D distances to the main halo.
    }
    \label{fig:halo_masses}
\end{figure*}
%%%%%%%%%%%%%%%%%%%%%%%%%%%%%%

For a given cluster, we perform membership matching to establish the cluster-halo association: For each member galaxy, we find its host halo, defined as the isolated halo that contains the galaxy within the virial radius. We thus obtain a list of halos contributing to this cluster. Galaxies with unresolved host halos are labeled as unmatched.

Figure~\ref{fig:cartoon} shows a cartoon picture illustrating the cluster-halo matching procedure. The LOS is indicated by a horizontal dashed line (with most of the distance omitted) followed by a dotted line. The dotted ellipse on the left-hand side shows the projected view of a cluster. The red star marks the cluster center, and the red points represent cluster members, with darker shades indicating higher $\pmem$. The gray spheres represent three halos hosting cluster members, ranked by their contribution to richness. The blue points correspond to galaxies considered non-members by the cluster finder because they do not meet the color or magnitude criteria. Two galaxies are not associated with resolved halos due to the limited simulation resolution.

For a given cluster, we rank the matched halos by their contribution to richness $\lambH=\sum_{i} p_{\mathrm{mem},i}$, summing over the galaxies in a given halo. We refer to $\lambH$ as {\em halo richness}; it is analogous to halo occupation but has extra $p_\mathrm{mem}$ weighting.
The halo with the highest $\lambH$ is defined as the main or top-ranked halo\footnote{Throughout the paper, we use the terms ``main halo'' and ``top-ranked halo'' interchangeably, and we use ``projected halos'' to refer to non-top-ranked halos.}. The $\lambH$ of the top-ranked halo is equivalent to the {\em true richness} in \cite{Costanzietal2019}; see Appendix~\ref{app:model_proj}.

Figure~\ref{fig:cartoon} only shows the top 3 halos of a cluster as an example. We use $\fhtop$ to denote the richness fraction contributed by the top-ranked halo; a lower $\fhtop$ indicates a strong projection effect. We consider a cluster {\em clean} if its main halo contributes to more than 50\% of the richness ($\fhtop\geq50\%$).

We have also considered ranking matched halos by their mass (see Appendix~\ref{app:rank_by_mass}). The two ranking methods give similar results; however, in rare cases, the most massive halo contributes to only a small fraction of members and is not a sensible choice for the main halo (see examples in Fig.~\ref{fig:halo_masses}). Therefore, we consider it more robust to rank halos by their contribution to richness. Although two clusters could have the same top-ranked halo, this duplicate only happens once out of the 8000 clusters in our sample and has negligible impact.

Figure~\ref{fig:halo_masses} shows several examples of the matched halos of a cluster. The left-hand panel shows examples of typical clusters (with $\fhtop$ near the median), while the right-hand panel shows examples of heavily projected clusters (the bottom two percentile of $\fhtop$), at $z\in[0.2,0.35)$, in four richness bins. Other redshift bins show similar behaviors. The circle area is proportional to the halo richness $\lambH$, while the color represents the 3D comoving distance from the main halo (shown in the legend). For typical clusters (left-hand panel), the contributions from projected halos are minimal. An interesting case is Cluster 104 (the third one from the top): its main halo has a mass of $10^{14.5}~\hiMsun$, while a projected halo $93~\hiMpc$ away has a higher mass of $10^{15}~\hiMsun$.  The right-hand panel shows that heavily projected clusters tend to be contaminated by several small halos instead of by a single large halo.

%%%%%%%%%%%%%%%%%%%%%%%%%%%%%%
%%%%%%%%%%%%%%%%%%%%%%%%%%%%%%
\subsection{Quantifying projection effects}
\subsubsection{\textbf{Contribution by the main halos}}
\begin{figure}[htbp!]
\centering
\includegraphics[width=\columnwidth]{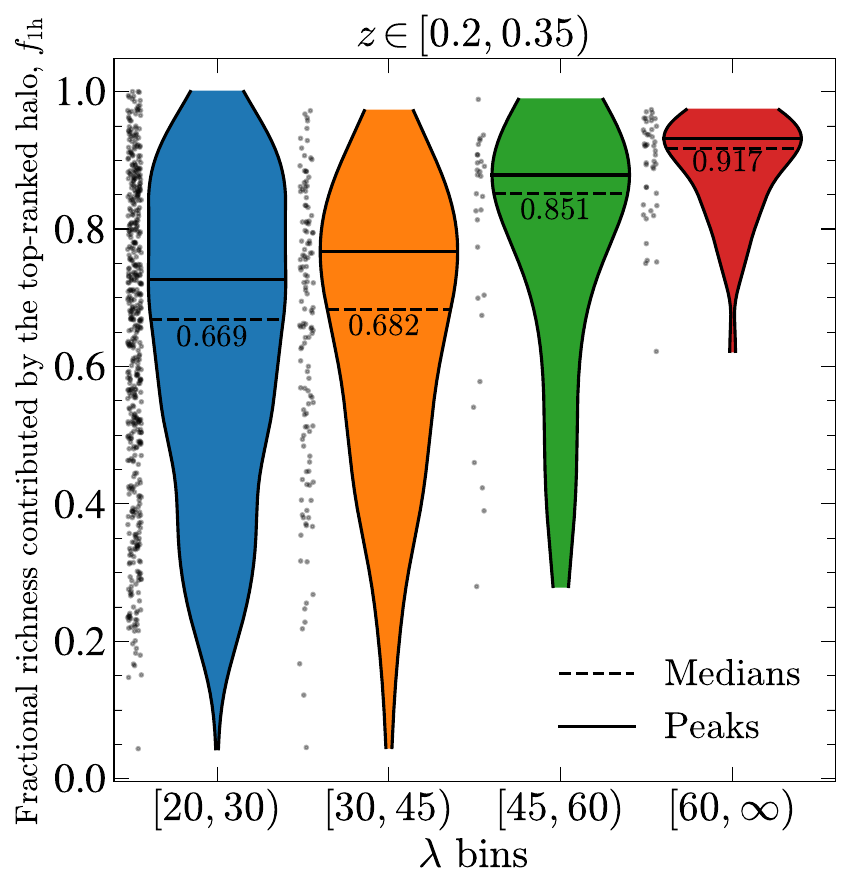}
\caption{Fractional contribution to richness by the top-ranked halo ($\fhtop$), for four richness bins at $z \in [0.2,0.35)$. Each violin shows a PDF vertically, and the scatter points represent the individual data points in each bin. The top-ranked halo contributes to 92\% and 67\% of the richness for high- and low-richness halos. Results for other redshift bins are presented in Table~\ref{tab:ranks_summary}.}
\label{fig:violin}
\end{figure}

% {\bf Contribution by the main halos.}
Figure~\ref{fig:violin} shows the fractional richness contributed by the top-ranked halo, $\fhtop$, for four richness bins at $z \in [0.2, 0.35)$. Each violin vertically displays the probability density function (PDF) of $\fhtop$ and is constructed using left-right symmetric kernel density estimation with Gaussian kernels, with a bandwidth determined by Scott's rule \citep{scott2015multivariate}. We add scatter points to show the individual data points, with horizontal random displacement to reduce overlap. For the highest-richness bin, top-ranked halos contribute 92\% of the richness with a relatively small dispersion. For the lowest-richness bin, this contribution decreases to 67\%. The tail of low $\fhtop$ clusters becomes more significant as richness decreases, indicating the larger contribution by the projected halos for low-richness clusters.

We show the results for all redshifts in Appendix~\ref{app:rank_by_mass} and Table~\ref{tab:ranks_summary}. The contribution of the top-ranked halo decreases with redshift. For example, at $z\in [0.5, 0.65)$, $\fhtop$ decreases to 74\% (richness > 60) and 26\% (richness 20 to 30). Similar deterioration with redshifts occurs for all the systematics we consider in this work and is associated with the degraded photometric redshift uncertainties at high redshifts. This trend is qualitatively consistent with the results in \cite{Costanzietal2019, Mylesetal2021, Grandisetal2021b, Grandisetal2025}. In Appendix~\ref{app:model_proj}, we present $P(\lambda|\lambTop;z)$ and fit the projection model from \citep{Costanzietal2019}.

%%%%%%%%%%%%%%%%%%%%%%%%%%%%%%

%%%%%%%%%%%%%%%%%%%%%%%%%%%%%%

%%%%%%%%%%%%%%%%%%%%%%%%%%%%%%
\begin{figure}[htbp!]
\centering
\includegraphics[width=0.49\textwidth,height=0.49\textwidth]{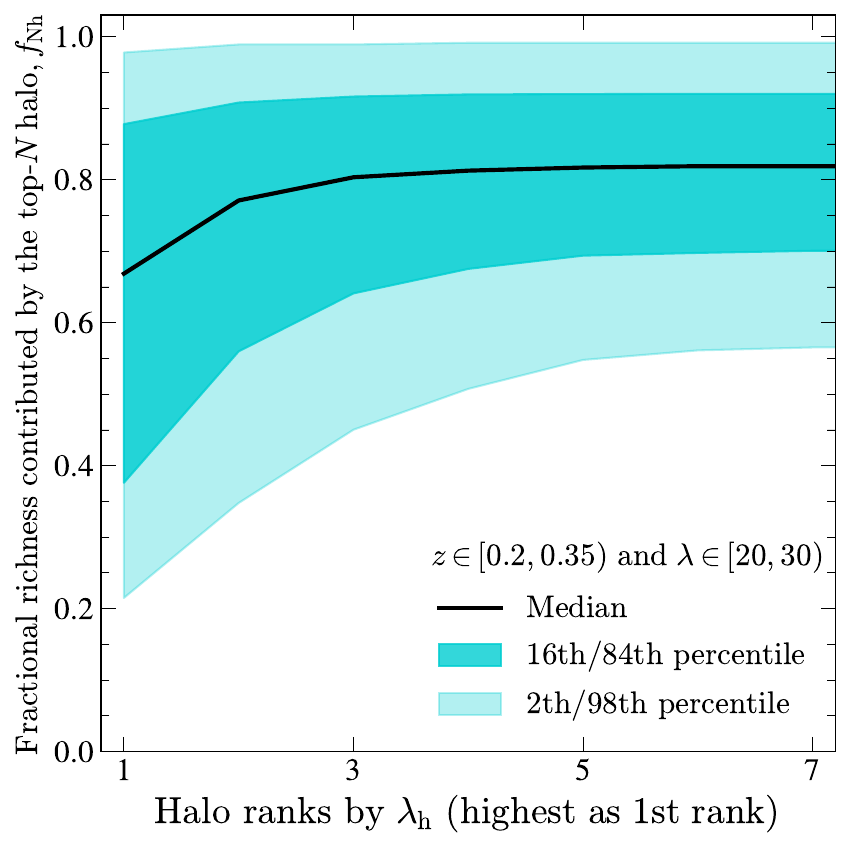}
\caption{Fractional richness contributed by the top-$N$ halo, $\fhN$. The curve and bands correspond to the median and $68\%-96\%$ intervals. The first-ranked (main) halo contributes to 67\% of the total richness, and the first- and second-ranked halos contribute to 77\%. The top three halos (main and two projected) typically account for more than 80\% of the richness. The curve does not go to 100\% because some member galaxies are not associated with resolved halos. We show the case of $z \in [0.2, 0.35)$ and $\lambda \in [20, 30)$ and present other bins in Table~\ref{tab:ranks_summary}.
}
\label{fig:fh_cum}
\end{figure}
%%%%%%%%%%%%%%%%%%%%%%%%%%%%%%

\smallskip
\subsubsection{\textbf{Contribution by the projected halos}}
% {\bf Contribution by the projected halos.}
We now consider the contribution from non-top-ranked (projected) halos. Figure~\ref{fig:fh_cum} shows the {fractional richness contributed by the top-$N$ halo,} denoted as $\fhN$. The black curve shows the median, while the band shows the 68\% and 96\% intervals. The $x$-axis corresponds to the number of top-$N$ halos, and the $y$-axis corresponds to their cumulative contribution to richness. The left edge ($N$=1, $\fhtop$) corresponds to the top-ranked halo discussed in Fig.~\ref{fig:violin}. The curve flattens at $N=3$, indicating that the top three halos (main and two projected) can account for most of the richness. As $N$ increases, we can account for a larger fraction of richness, but $\fhN$ never reaches 100\% because 19\% of the richness comes from galaxies with unresolved host halos.

%%%%%%%%%%%%%%%%%%%%%%%%%%%%%%
\begin{figure*}[htbp!]
\centering
\includegraphics[width=0.49\textwidth,height=0.49\textwidth]{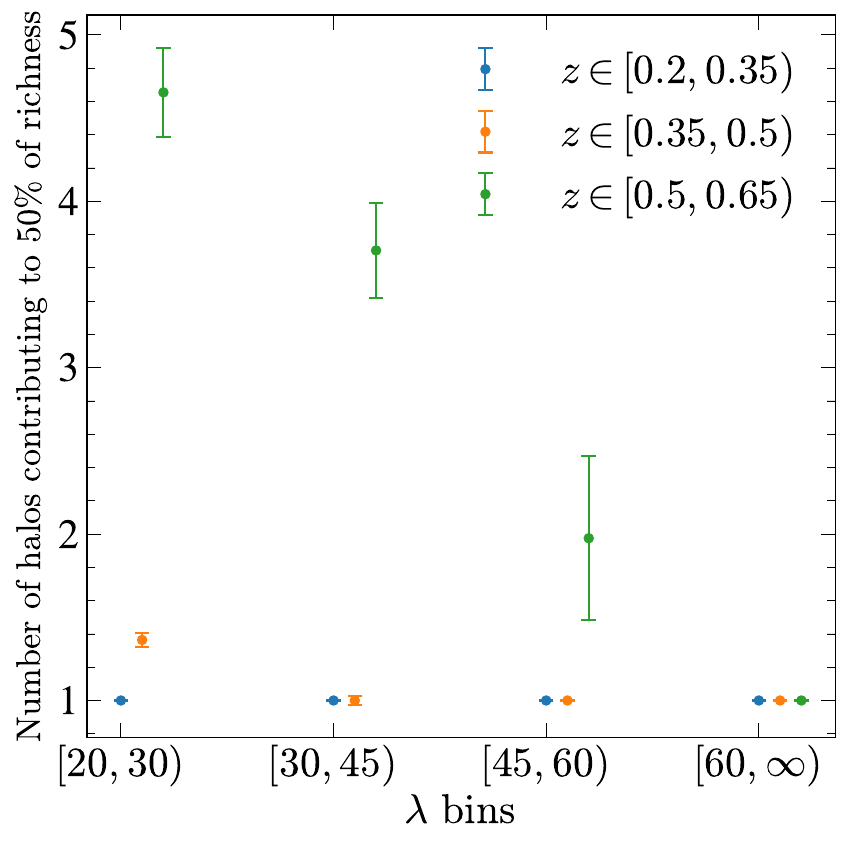}
\hfill 
\includegraphics[width=0.49\textwidth,height=0.49\textwidth]{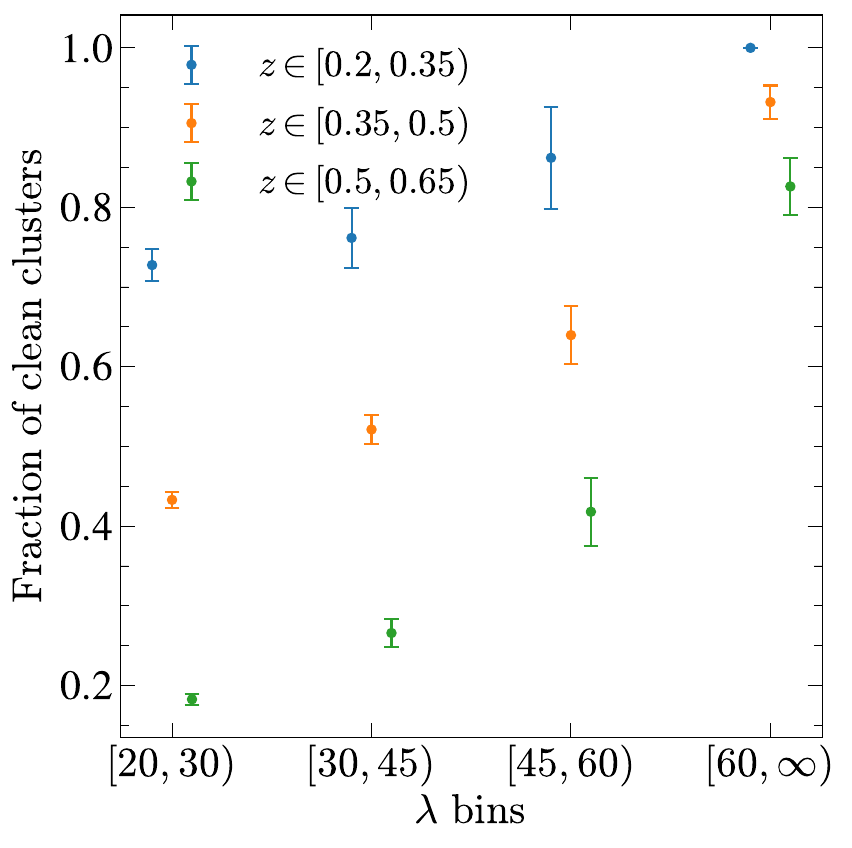}
\caption{
Left: Number of halos contributing to at least 50\% of richness, obtained by solving cumulative richness fraction $\fhN\geq 50\%$, for three cluster redshift bins and four richness bins. Except for the lowest richness and the highest redshift bin, the main halo accounts for more than 50\% of the richness.  Right: Fraction of clean clusters, defined as those with main halos contributing to more than 50\% of richness.  The clean fraction is high for low redshift (73\% to 100\%) and decreases with redshift. In both panels, the points with error bars correspond to the median and standard deviation derived from bootstrap resampling.
}
\label{fig:num_halos}
% \vspace{-50pt}
\end{figure*}
%%%%%%%%%%%%%%%%%%%%%%%%%%%%%%

%%%%%%%%%%%%%%%%%%%%%%%%%%%%%%
%%%%%%%%%%%%%%%%%%%%%%%%%%%%%%
\smallskip
\subsubsection{\textbf{How many halos contribute to a cluster substantially?}}
% {\bf How many halos contribute to a cluster substantially?}
The left-hand panel of Fig.~\ref{fig:num_halos} shows the minimum number of halos needed to account for 50\% of the cluster richness, denoted as $N_{0.5}$, calculated by linearly interpolating the median $\fhN$ as a function of $N$ and solving $\fhN \geq 0.5$. 
At the low redshift and high richness bins, $N_{0.5}=1$; that is, their main halos typically contribute to $\ge50\%$ of the richness.

To estimate the error bar of $N_{0.5}$, we use bootstrap resampling. For each redshift and richness bin, we draw with replacement the same number of clusters in that bin. We generate $10^5$ such bootstrap samples, and for each sample, we calculate the median $\fhN$ as a function of $N$ and solve for $N_{0.5}$. We use the standard deviations of these samples as our error bars.

%%%%%%%%%%%%%%%%%%%%%%%%%%%%%%
%%%%%%%%%%%%%%%%%%%%%%%%%%%%%%
\smallskip
\subsubsection{\textbf{How many clusters are clean?}}
% {\bf How many clusters are clean?}
We consider a cluster ``clean'' if its main halo contributes to more than 50\% of its richness. The right-hand panel of Fig.~\ref{fig:num_halos} shows the fraction of clean clusters in each richness and redshift bin. At low redshift, 73\% of the {$\lambda\in[20,30)$} clusters are clean, and this degrades to 18\% at high redshift. For each redshift and richness bin, we again generate $10^5$ bootstrap samples, repeat the calculations, and use their standard deviations as the error bars.

%%%%%%%%%%%%%%%%%%%%%%%%%%%%%%
%%%%%%%%%%%%%%%%%%%%%%%%%%%%%%
%\clearpage
\section{\label{sec:completeness}
Completeness of halos identified by \redmapper in Cardinal}

%%%%%%%%%%%%%%%%%%%%%%%%%%%%%%
\begin{figure}[htbp!]
\centering  \includegraphics[width=0.49\textwidth,height=0.49\textwidth]{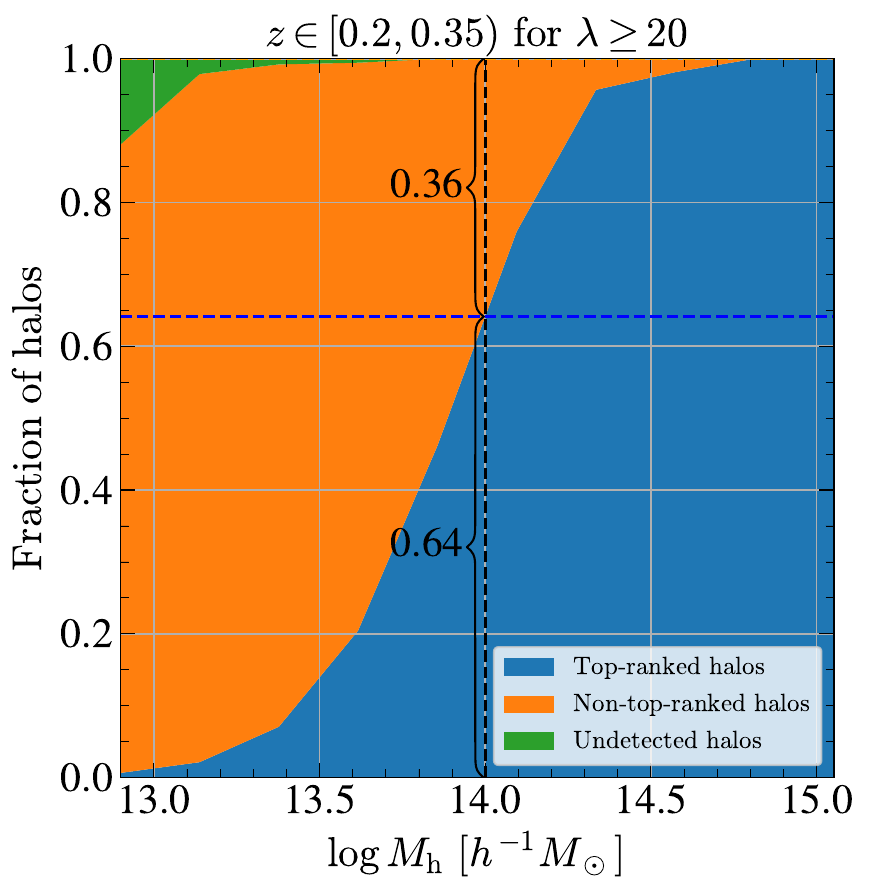}
\caption{
Completeness of halos as a function of mass, quantified by the fraction of halos that are matched as top-ranked halos of clusters (blue).  
The completeness is 100\% at the high-mass end and drops to 64\% at $10^{14}\hiMsun$. 
The orange area shows the halos that are not top-ranked of any cluster but are matched as projected halos.  The green area shows halos that do not contribute to any cluster (undetected).
}
\label{fig:completeness}
% \vspace{-50pt}
\end{figure}
%%%%%%%%%%%%%%%%%%%%%%%%%%%%%%

With our cluster-matching matching results, we would like to quantify the completeness of halos as a function of mass.  We define completeness as the fraction of halos that are identified as top-ranked halos.  
Based on our matching scheme, each halo belongs to one of the three categories:
\begin{itemize}
    \item [(1)] Top-ranked: It is the top-ranked halo of a cluster. 
    \item [(2)] Non-top-ranked: It is not the top-ranked halo of any cluster but is a non-top-ranked halo of at least one cluster.
    \item [(3)] Undetected: It does not contribute to the richness of any cluster.
\end{itemize}

We start with all isolated halos with virial mass above $6\times10^{12}~\hiMsun$ in Cardinal. To account for the survey mask, we use the HEALPix software \cite{Healpix} with $n_{\rm side}=16834$ (corresponding to a pixel size of $\sim0.04\ h^{-1}\mathrm{Mpc}$ at $z=0.3$) to exclude halos in pixels without any \redmapper members. Since halos are specified by cosmological redshifts while clusters are specified by photometric redshifts, we start with no redshift binning for halos. Using our cluster-halo matching results in Sec.~\ref{sec:matching}, we categorize each halo as top-ranked, non-top-ranked, or undetected. Because we start with no redshift binning, some of the undetected halos have significant LOS distances from clusters and should be removed. We exclude undetected halos with a redshift difference greater than $5\sigma_z$ from the nearest cluster, where $\sigma_z$ corresponds to \redmapper cluster redshift uncertainty.

For a given halo mass, we calculate the fractions of halos that are top-ranked, non-top-ranked, and undetected. Figure~\ref{fig:completeness} presents these three categories as stacked plots for $\lambda \geq 20$ \redmapper clusters in $z\in[0.2,0.35)$. At $10^{14}~\hiMsun$, approximately 64\% halos are matched as the top-ranked halos, and 36\% are matched only as non-top-ranked. The trend we observe is consistent with that presented in Fig.~8 of the DES-Y1 analysis \citep{Abbottetal2020}. At this mass, the non-top-ranked halos are usually in the vicinity of bigger halos.  Below this mass, we start to have a small fraction of undetected halos, which do not contribute galaxies to any clusters.

%%%%%%%%%%%%%%%%%%%%%%%%%%%%%%
\section{\label{sec:miscenter}Miscentering of \redmapper clusters in Cardinal}

%%%%%%%%%%%%%%%%%%%%%%%%%%%%%%
%%%%%%%%%%%%%%%%%%%%%%%%%%%%%%
\begin{figure*}[htbp!]
\centering
\includegraphics[width=0.49\textwidth,height=0.49\textwidth]{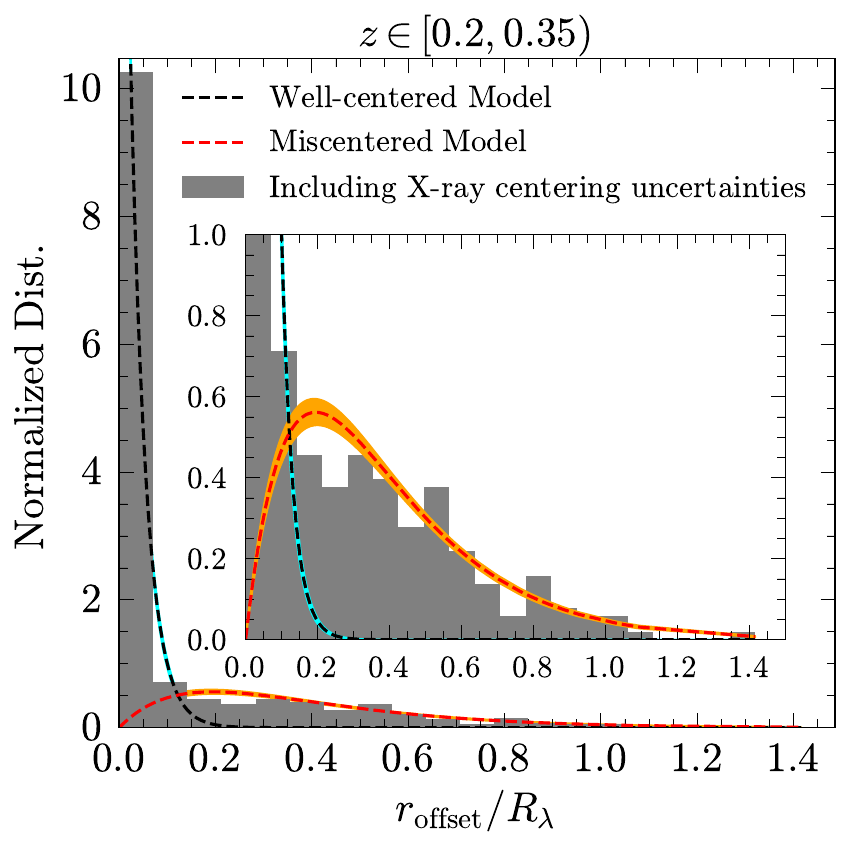}
\includegraphics[width=0.49\textwidth,height=0.49\textwidth]{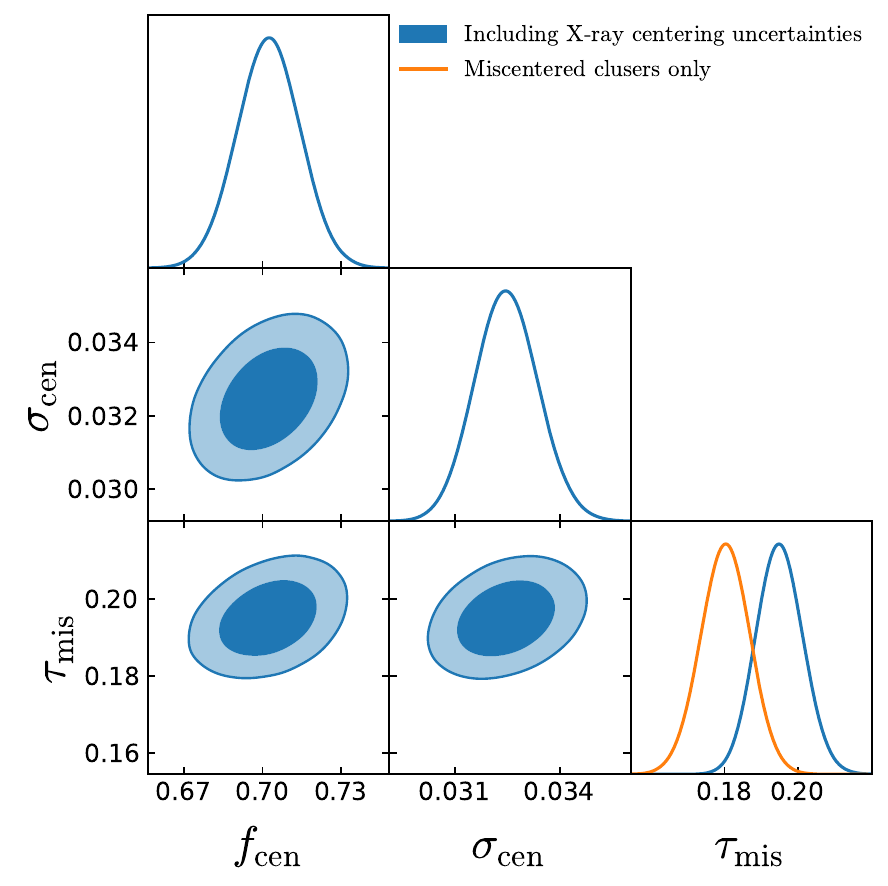}
\caption{Left: PDF of projected distance offset divided by cluster radius, for clusters in $z\in[0.2,0.35)$. The inset focuses on the miscentered population. The dashed curves show model predictions derived from posterior mean parameters in Eq.~\eqref{eq:miscent_model}, and the bands correspond to $1\sigma$ posterior predictions.  Right: Posterior distributions of the miscentering parameters for the full sample (3 parameters, blue) and the miscentered sample without X-ray offsets (1 parameter, orange).}
\label{fig:miscen}
\end{figure*}
%%%%%%%%%%%%%%%%%%%%%%%%%%%%%%
%%%%%%%%%%%%%%%%%%%%%%%%%%%%%%

Miscentering refers to the offset between the inferred cluster center and the true halo center, usually defined as the gravitational potential minimum \citep{Rozoetal2010, DES2019b, Sunayamaetal2024, DES2024, Dingetal2025}. In optical observations, this offset can occur when the color of the true central galaxy deviates from the red sequence or in merging systems where the true cluster center is ambiguous.   This offset can bias the lensing signal low at small scales and potentially bias the mass calibration \citep{Grandisetal2021b, Sommeretal2022, Bocquetetal2024, Kleinebreiletal2024, SommerSchrabbackGrandis2025}. Observationally, one often uses intracluster medium (ICM)-determined centers as proxies for the true cluster centers \cite{DES2019b, DES2024, Dingetal2025}.

In this work, we use the main halo center as the true cluster center, and we use the most probable central galaxy of \redmapper as the observed cluster center. We calculate the project distance $r_{\text{offset}}$ between these two centers. By construction, our well-centered clusters have exactly zero offsets because Cardinal puts one galaxy at each halo center. On the other hand, ICM-derived well-centered clusters can have small offsets from the halo center. Mergers, substructure, and cool cores can shift the X-ray brightness peak \citep{MartelRobichaudBarai2014}. The astrometry uncertainty in X-ray telescopes can also introduce random offsets of the center; for example, \Chandra and \XMM have astrometry uncertainties of $\approx$ 1 and 8 arcsec, respectively \citep{Saxtonetal2008, VulicGallagherBarmby2016, Medvedevetal2021}. An 8 arcsec offset at $z=0.3$ can lead to a $\approx 0.025~\hiMpc$ difference in the (physical) angular diameter distance.  To account for the small X-ray centering uncertainties, we add a small offset to our cluster centers. For each cluster, we draw a Gaussian random number with mean zero and standard deviation $8''$, take the absolute value, and convert it to the projected distance at the main halo's cosmological redshift.

Figure~\ref{fig:miscen} presents the normalized PDF of $x=r_{\text{offset}}/R_\lambda$, where $R_\lambda$ is the cluster radius defined by \redmapper. Following \cite{DES2019b}, we describe the PDF using a two-component model:
\begin{equation}
\label{eq:miscent_model}
  \begin{aligned}
     &P(x|\fcent,\sigcen,\taumis) \\
     &= \fcent P_{\text{cen}}(x|\sigcen) + 
     (1-\fcent) P_{\text{mis}}(x|\taumis),
  \end{aligned}
\end{equation}
which has a well-centered component 
\begin{equation}
     P_{\text{cen}}(x|\sigcen) = \frac{1}{\sigcen}\exp{\left(-\frac{x}{\sigcen}\right)},
     \label{eq:P_cen}
\end{equation}
and a miscentered component
\begin{equation}
     P_{\text{mis}}(x|\taumis) = \frac{x}{\taumis^2}\exp{\left(-\frac{x}{\taumis}\right)}.
     \label{eq:P_miscen}
\end{equation}
The parameter $\fcent$ is the fraction of well-centered clusters.  Here $P_{\text{cen}}$ is an exponential distribution, and $P_{\text{mis}}$ is a Gamma distribution with a shape parameter 2, a scale parameter $\taumis$, and a mean $2\taumis$.

To constrain the miscentering parameters, we fit the cumulative distribution function (CDF) instead of the PDF using Markov chain Monte Carlo (MCMC) software \textsc{emcee} \citep{emcee}. 
To estimate the covariance matrix of the CDF, we again generate bootstrap samples by resampling clusters with replacement (similar to the procedure in Sec.~\ref{sec:matching}). To stabilize the bootstrap covariance matrix, we regularize it by adding a small value---1\% of the smallest diagonal element greater than $10^{-10}$---to the diagonal. We use the same priors of the fitting parameters as used in Table 1 of \cite{DES2019b}.

The left-hand panel of Fig.~\ref{fig:miscen} presents the PDF of the centering offset for the $z\in [0.2, 0.35)$ bin.  The inset focuses on the miscentered component. 
The curves represent the model predictions derived from posterior mean parameters, while the bands are the 68\% posterior predictions derived from randomly drawing $10^5$ points from the posterior chain. 
The right-hand panel of Fig.~\ref{fig:miscen} shows the parameter constraints for the full sample with X-ray offsets (blue contours).  
For comparison, we include a one-parameter fit to only the miscentered sample, without adding X-ray offsets (orange).  The recovered $\taumis$ is consistent with that from the full sample.

Table~\ref{tab:miscent} summarizes the results in other redshift bins.  Our results are consistent with the DES-Y3 \redmapper results based on X-ray centers from \Chandra and \XMM \citep{DES2024}. Our $\taumis=0.2$ (a mean offset of 0.4$R_\lambda$) agrees well with the observational results. Our well-centered fraction $\fcent=0.7$ is slightly lower than the result $\fcent=0.8$ found in \cite{DES2024}, which is likely due to our lower mean cluster mass. Our $\sigcen$ (parameter for the well-centered clusters) is slightly lower than the observation, which is likely due to the extra X-ray centering uncertainties present in observations.

\begin{table*}[htbp!]
\centering
\setlength\tabcolsep{10pt}
\begin{threeparttable}
\caption{Miscentering parameters of \redmapper clusters from Cardinal (this work) and observations.
}
\label{tab:miscent}
\begin{tabular}{lccccc}
\toprule\toprule
Redshift & Halo Center Proxy & $\fcent$ & $\sigcen$ & $\taumis$ & Ref.\\
\midrule
$z\in[0.2,0.35)$\tnote{a} & Matched halo & $0.702\pm0.012$ & $0.0325\pm0.0009$ & $0.195\pm0.006$ & This work\\
$z\in(0.2,0.4)$ & \Chandra, \XMM (X-ray) & $0.80\pm0.06$ & $0.050\pm0.009$ & $0.21\pm0.05$ & \cite{DES2024}\\[5pt]
$z\in[0.2,0.65)$\tnote{a} & Matched halo & $0.703\pm0.004$ & $0.0370\pm0.0003$ & $0.210\pm0.002$ & This work\\
$z\in(0.2,0.65)$ & \Chandra, \XMM (X-ray) & $0.87\pm0.04$ & $0.053\pm0.006$ & $0.23\pm0.05$ & \cite{DES2024}\\
$z\in[0.2,0.7)$ & \Chandra (X-ray) & $0.835^{+0.112}_{-0.075}$ & $0.0443^{+0.0231}_{-0.0094}$ & $0.166^{+0.111}_{-0.042}$ & \cite{DES2019b}\\[5pt]
$z\in[0.35,0.5)$\tnote{a} & Matched halo & $0.748\pm0.005$ & $0.0340\pm0.0004$ & $0.214\pm0.003$ & This work\\
$z\in[0.5,0.65)$\tnote{a} & Matched halo & $0.677\pm0.005$ & $0.0401\pm0.0005$ & $0.236\pm0.003$ & This work\\[5pt]
$z\in[0.2,0.35)$\tnote{b} & Matched halo & 0.696 & -- & $0.180\pm0.007$ & This work\\
$z\in[0.35,0.5)$\tnote{b} & Matched halo & 0.710 & -- & $0.208^{+0.003}_{-0.002}$ & This work\\
$z\in[0.5,0.65)$\tnote{b} & Matched halo & 0.646 & -- & $0.215\pm0.002$ & This work\\
$z\in[0.2,0.65)$\tnote{b} & Matched halo & 0.678 & -- & $0.203^{+0.001}_{-0.002}$ & This work\\
\bottomrule\bottomrule
\end{tabular}
\begin{tablenotes}
\item [a] All clusters, with an additional offset of $8''$ mimicking X-ray centering uncertainties. 
\item [b] Miscentered clusters only, without the X-ray offset. 
\end{tablenotes}
\end{threeparttable}%
\end{table*}

\section{Comparison with previous studies}\label{sec:previous}

Ref.~\cite{Mylesetal2021} uses the spectroscopic redshifts of SDSS \redmapper cluster members to quantify projection effects. They fit a double-Gaussian model to the redshift PDF and constrain the projected fraction $f_{\rm proj}$ = 0.265 for $\lambda \in (20, 27.9]$ and 0.083 for $\lambda \in (69.3, 140]$; see their Table 2. Their $f_{\rm proj}$ is equivalent to our $1-\fhtop$, and we find 0.33 for $\lambda \in [20, 30)$ and 0.08 for $\lambda \in [60, \infty)$. Our results are consistent with theirs despite the differences in data and analysis methods.

Ref.~\citep{Farahietal2016} quantifies the top-ranked halo contribution to richness using Aardvark and Bolshoi simulations. They have used \redmapper clusters in the Aardvark simulation and find a $\fhtop$ = 0.58 for $z \in [0.1,0.3]$, $\lambda\geq20$; excluding miscentered clusters gives $\fhtop$ = 0.62. They have also used the Bolshoi galaxy mock catalogs from \cite{HearinWatson2013} and estimated richness using a $\pm 60~\hiMpc$ projection length, finding $\fhtop = 0.70$ (see their Table 2). Their Aardvark $\fhtop$ results are slightly lower than ours, while their Bolshoi result is consistent with ours. In addition, they have found a lower fraction of well-centered clusters in Aardvark (58\%) compared with our Cardinal results (70\%) and recent X-ray results (80\%). Both trends---stronger projection and miscentering---are related to the fact that Aardvark has a low number of galaxies near the centers of massive halos, an issue related to numerical disruption of subhalos \cite{Wechsleretal2022}. As a result, clusters in Aardvark have lower contributions from the top-ranked halos and thus stronger projection effects, and it is harder for \redmapper to determine their centers correctly.

\section{\label{sec:summary}Summary}

We use the Cardinal simulation to study the cluster-halo association, focusing on clusters identified by the \redmapper cluster finder under the DES-Y6 survey condition. 
For each cluster, we identify the halos hosting its member galaxies, and we rank these halos by their contribution to richness.
The halos with the largest contribution to richness are identified as the main or top-ranked halos. 
Using this cluster-halo association, we quantify three key systematics for cluster cosmology analyses.

\subsection{Contribution from main and projected halos}
% {\em Contribution from main and projected halos.}
We develop a new way to visualize the main and projected halos contributing to a cluster (Fig.~\ref{fig:halo_masses}). We find that the main halos typically contribute to 92\% to 67\% of the richness, and the top three halos (main and two projected) can account for 80\% of the total richness. At $z\approx0.3$, 73\% and 100\% are clean (with $\ge$50\% of richness from the main halo) in the lowest and highest richness bins, respectively. The impact of projection on richness becomes stronger at low richness and high redshifts. At $z\in[0.5,0.65)$, the clean clusters in the lowest richness bin is $\approx20\%$. Nevertheless, we expect that the redder bands from LSST and Roman will significantly improve cluster finding at higher redshift.

\subsection{Completeness of halos}
% {\em Completeness of halos.} 
For a given halo mass, we quantify completeness as the fraction of halos that are identified as a top-ranked halo of a cluster. Halos identified as non-top-ranked or undetected contribute to the incompleteness.  For $\lambda \geq 20$, the completeness is 64\% at $10^{14}\hiMsun$.

\subsection{Offset between cluster center and halo center}
% {\em Offset between cluster center and halo center.} 
We compare the \redmapper cluster centers with the true halo centers. On average, 70\% of our simulated clusters are well-centered, which is slightly lower than 80\% recently reported by X-ray observations from \cite{DES2024}. For the miscentered clusters, we find a mean offset of 0.4 times cluster radius, consistent with \cite{DES2024}.

Although we focus on \redmapper in this paper, our framework is applicable for benchmarking other cluster-finding algorithms, including those based on photometric redshifts \citep[][]{AMICO, WaZP, CluMPR, Doubrawaetal2024}, red-sequence galaxy properties \citep[][]{CAMIRA}, or machine learning \citep{YOLO}. 
%Our long-term goal is to use simulation-based forward modeling to capture the complex cluster-halo association \citep{Salcedoetal2024, Leeetal2024}. % Heidi: removed per referee suggestion
We provide our cluster-halo matching code online\footnote{\url{https://github.com/ShuleiCao/CHM}}.

\begin{acknowledgments}
\textit{Author Contributions:}
\textit{Data analysis and manuscript writing:} Shulei Cao and Hao-Yi Wu.
\textit{Internal reviewing:} Matteo Costanzi, Arya Farahi, and Sebastian Grandis.
\textit{Generating Cardinal simulation:} Chun-Hao To.
\textit{Substantial suggestions on the project and the draft:} David Weinberg, Gus Evrard, Eduardo Rozo, 
Andr\'es Salcedo, Lei Yang, and Conghao Zhou.

AF acknowledges support from the National Science Foundation under Cooperative Agreement 2421782 and the Simons Foundation award MPS-AI-00010515. MC is supported by the PRIN 2022 project EMC2 (code no. J53D23001620006).

The computations in this paper were performed on the high-performance computing (HPC) cluster provided by the Office of Information Technology (OIT) and the O'Donnell Data Science and Research Computing Institute (\url{https://www.smu.edu/oit/services/m3}) at Southern Methodist University.

Funding for the DES Projects has been provided by the U.S. Department of Energy, the U.S. National Science Foundation, the Ministry of Science and Education of Spain, the Science and Technology Facilities Council of the United Kingdom, the Higher Education Funding Council for England, the National Center for Supercomputing Applications at the University of Illinois at Urbana-Champaign, the Kavli Institute of Cosmological Physics at the University of Chicago, the Center for Cosmology and Astro-Particle Physics at The Ohio State University, the Mitchell Institute for Fundamental Physics and Astronomy at Texas A\&M University, Financiadora de Estudos e Projetos, Funda\c{c}\~ao Carlos Chagas Filho de Amparo \'a Pesquisa do Estado do Rio de Janeiro, Conselho Nacional de Desenvolvimento Cient\'ifico e Tecnol\'ogico and the Minist\'erio da Ci\^encia, Tecnologia e Inova\c{c}\~ao, the Deutsche Forschungsgemeinschaft and the Collaborating Institutions in the Dark Energy Survey. 

The Collaborating Institutions are Argonne National Laboratory, the University of California at Santa Cruz, the University of Cambridge, Centro de Investigaciones Energ\'eticas, Medioambientales y Tecnol\'ogicas-Madrid, the University of Chicago, University College London, the DES-Brazil Consortium, the University of Edinburgh, the Eidgenössische Technische Hochschule (ETH) Zürich, Fermi National Accelerator Laboratory, the University of Illinois at Urbana-Champaign, the Institut de Ci\`encies de l'Espai (IEEC/CSIC), the Institut de F\'isica d'Altes Energies, Lawrence Berkeley National Laboratory, the Ludwig-Maximilians Universit\"at M\"unchen and the associated Excellence Cluster Universe, the University of Michigan, the National Optical Astronomy Observatory, the University of Nottingham, The Ohio State University, the University of Pennsylvania, the University of Portsmouth, SLAC National Accelerator Laboratory, Stanford University, the University of Sussex, Texas A\&M University, and the OzDES Membership Consortium. The DES data management system is supported by the National Science Foundation under Grants No. AST-1138766 and No. AST-1536171.

\end{acknowledgments}

% \section*{DATA AVAILABILITY}
% The data that support the findings of this article are openly available \citep{Toetal2024}.

\appendix

\section{Summary of halos contributing to cluster richness}
\label{app:rank_by_mass}

Table~\ref{tab:ranks_summary} summarizes the fractional richness contributed by the top-ranked halos ($\fhtop$). We also show two versions of $\fhtop$: top-ranked halo determined by either $\lambH$-ranking or mass-ranking. For most clusters, the two ranking schemes give the same results; however, in rare cases, the most massive halo contributes to only a small fraction of members and is not a sensible choice for the main halo. Therefore, we use $\lambH$ ranking as our default method.

Table~\ref{tab:ranks_summary} also shows the number of matched halos. 
In each bin, the numbers correspond to the median, and the error bars correspond to the 68\% (16th/84th percentile) and 96\% (2nd/98th percentile) intervals.

\begin{table*}[htbp!]
\centering
\setlength\tabcolsep{18pt}
\begin{threeparttable}
\caption{Summary of the contribution of halos to richness, in three redshift bins and richness bins. We list the main halo contribution $\fhtop$, for top-ranked halos determined by $\lambH$-ranking vs.~mass-ranking. We also show the number of matched halos. The numbers and error bars correspond to the 50th, the (16th, 84th), and the (2nd, 98th) percentiles.}
\label{tab:ranks_summary}
\begin{tabular}{lcccc}
\toprule\toprule
$z$ bin & $\lambda$ bin & Ranked by & $\fhtop$ & Number of contributing halos\\
\midrule
 & $\lambda\in[20,30)$ & $\lambH$ & $66.86^{+20.91}_{-29.25}(^{+30.94}_{-45.37})$\% & $3\pm2(^{+4}_{-2})$\\[1pt]
 &  & $M_{\text{h}}$ & $66.73^{+21.04}_{-36.71}(^{+31.07}_{-63.57})$\% & \\[3pt]
 & $\lambda\in[30,45)$ & $\lambH$ & $68.24^{+17.65}_{-25.46}(^{+27.69}_{-48.57})$\% & $3^{+3}_{-1}(^{+5}_{-2})$\\[1pt]
$z\in[0.2,0.35)$ &  & $M_{\text{h}}$ & $68.24^{+17.65}_{-30.59}(^{+27.69}_{-63.35})$\% & \\[3pt]
 & $\lambda\in[45,60)$ & $\lambH$ & $85.13^{+6.46}_{-29.28}(^{+10.82}_{-51.00})$\% & $3\pm2(^{+7}_{-2})$\\[1pt]
 &  & $M_{\text{h}}$ & $85.13^{+6.46}_{-29.28}(^{+10.82}_{-83.24})$\% & \\[3pt]
 & $\lambda\in[60,\infty)$ & $\lambH$ & $91.73^{+4.09}_{-9.70}(^{+5.37}_{-18.53})$\% & $3^{+2}_{-1}(^{+4}_{-2})$\\[1pt]
 &  & $M_{\text{h}}$ & $91.08^{+4.73}_{-9.20}(^{+6.02}_{-37.55})$\% & \\
\midrule
 & $\lambda\in[20,30)$ & $\lambH$ & $44.55^{+29.46}_{-20.36}(^{+46.66}_{-30.28})$\% & $4\pm2(^{+4}_{-3})$\\[1pt]
 &  & $M_{\text{h}}$ & $41.78^{+31.92}_{-30.66}(^{+49.44}_{-40.94})$\% & \\[3pt]
 & $\lambda\in[30,45)$ & $\lambH$ & $52.09^{+25.47}_{-27.09}(^{+40.00}_{-37.60})$\% & $5\pm2(^{+5}_{-4})$\\[1pt]
$z\in[0.35,0.5)$ &  & $M_{\text{h}}$ & $51.08^{+26.37}_{-31.37}(^{+41.01}_{-49.24})$\% & \\[3pt]
 & $\lambda\in[45,60)$ & $\lambH$ & $63.05^{+22.88}_{-33.05}(^{+31.08}_{-51.17})$\% & $5\pm3(^{+8}_{-4})$\\[1pt]
 &  & $M_{\text{h}}$ & $62.63^{+23.30}_{-34.68}(^{+31.51}_{-55.25})$\% & \\[3pt]
 & $\lambda\in[60,\infty)$ & $\lambH$ & $79.23^{+10.73}_{-20.63}(^{+15.22}_{-44.89})$\% & $5\pm2(^{+6}_{-4})$\\[1pt]
 &  & $M_{\text{h}}$ & $79.23^{+10.73}_{-20.63}(^{+15.22}_{-44.89})$\% & \\
\midrule
 & $\lambda\in[20,30)$ & $\lambH$ & $26.29^{+26.57}_{-11.32}(^{+53.88}_{-17.38})$\% & $5\pm2(\pm4)$\\[1pt]
 &  & $M_{\text{h}}$ & $21.83^{+30.57}_{-17.81}(^{+58.34}_{-21.47})$\% & \\[3pt]
 & $\lambda\in[30,45)$ & $\lambH$ & $30.04^{+32.33}_{-16.09}(^{+53.96}_{-22.62})$\% & $6^{+3}_{-2}(\pm5)$\\[1pt]
$z\in[0.5,0.65)$ &  & $M_{\text{h}}$ & $28.11^{+33.97}_{-22.72}(^{+55.89}_{-27.79})$\% & \\[3pt]
 & $\lambda\in[45,60)$ & $\lambH$ & $39.08^{+35.73}_{-24.02}(^{+50.84}_{-31.67})$\% & $7\pm4(^{+8}_{-6})$\\[1pt]
 &  & $M_{\text{h}}$ & $37.03^{+36.16}_{-30.46}(^{+52.90}_{-35.79})$\% & \\[3pt]
 & $\lambda\in[60,\infty)$ & $\lambH$ & $73.67^{+12.25}_{-26.57}(^{+20.64}_{-52.85})$\% & $5^{+4}_{-2}(^{+8}_{-4})$\\[1pt]
 &  & $M_{\text{h}}$ & $73.67^{+12.25}_{-26.57}(^{+20.64}_{-72.07})$\% & \\
\bottomrule\bottomrule
\end{tabular}
\end{threeparttable}%
\end{table*}

%%%%%%%%%%%%%%%%%%%%%%%%%%%%%%
%%%%%%%%%%%%%%%%%%%%%%%%%%%%%%
\section{Modeling cluster richness with projection effects}
\label{app:model_proj}

%%%%%%%%%%%%%%%%%%%%%%%%%%%%%%
\begin{figure*}[htbp!]
\centering
    \includegraphics[width=0.32\textwidth,height=0.32\textwidth]{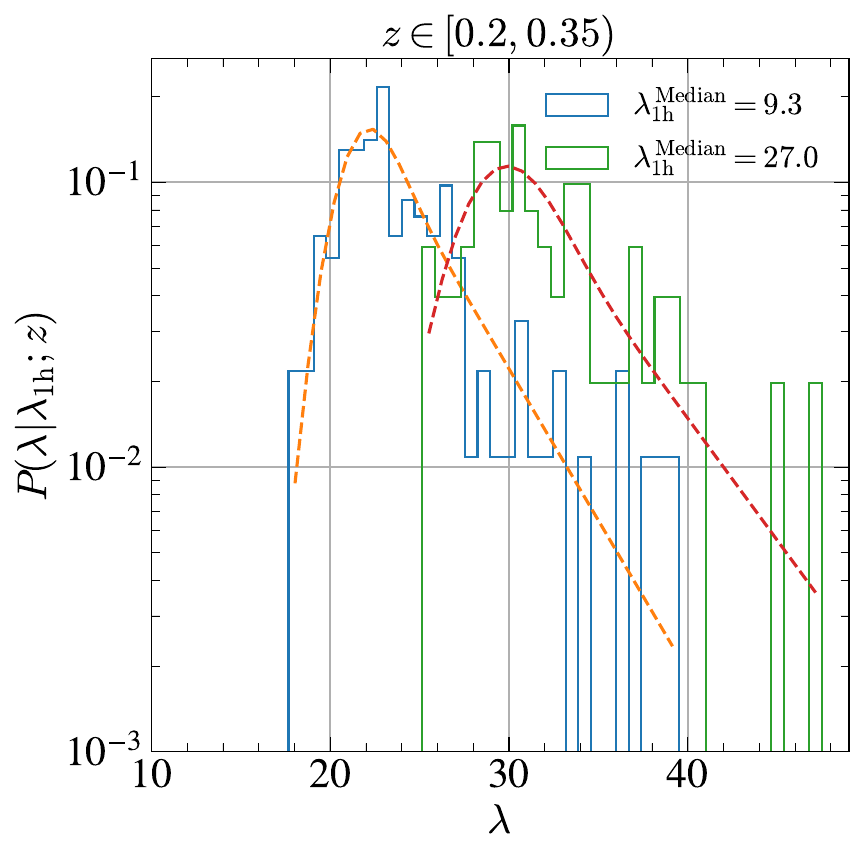}
    \hfill
    \includegraphics[width=0.32\textwidth,height=0.32\textwidth]{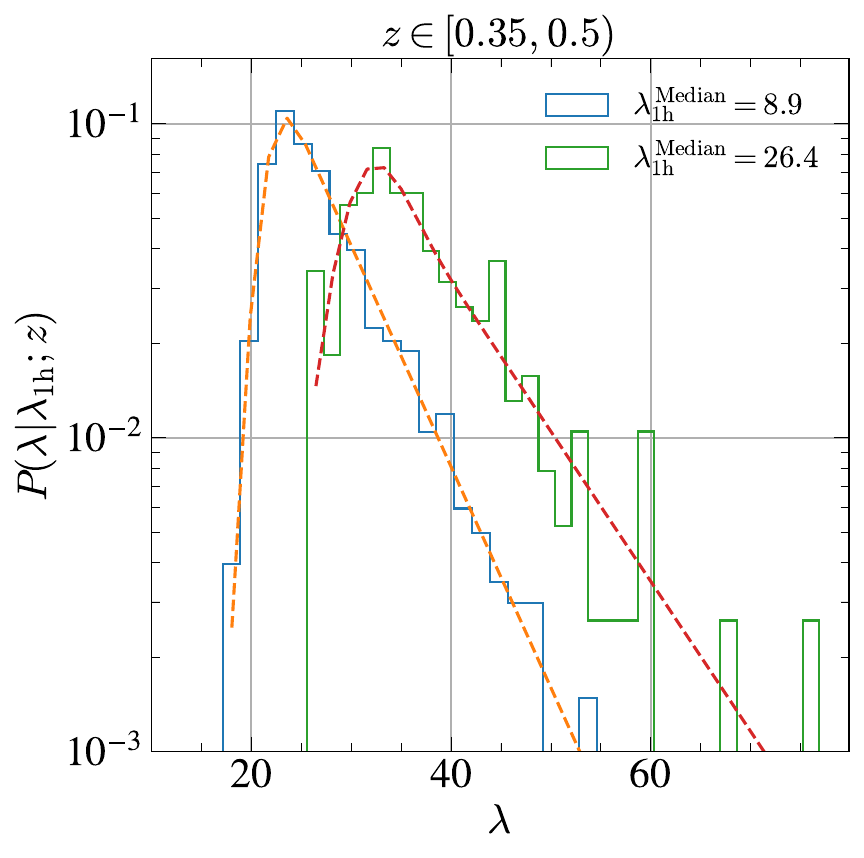}
    \hfill
    \includegraphics[width=0.32\textwidth,height=0.32\textwidth]{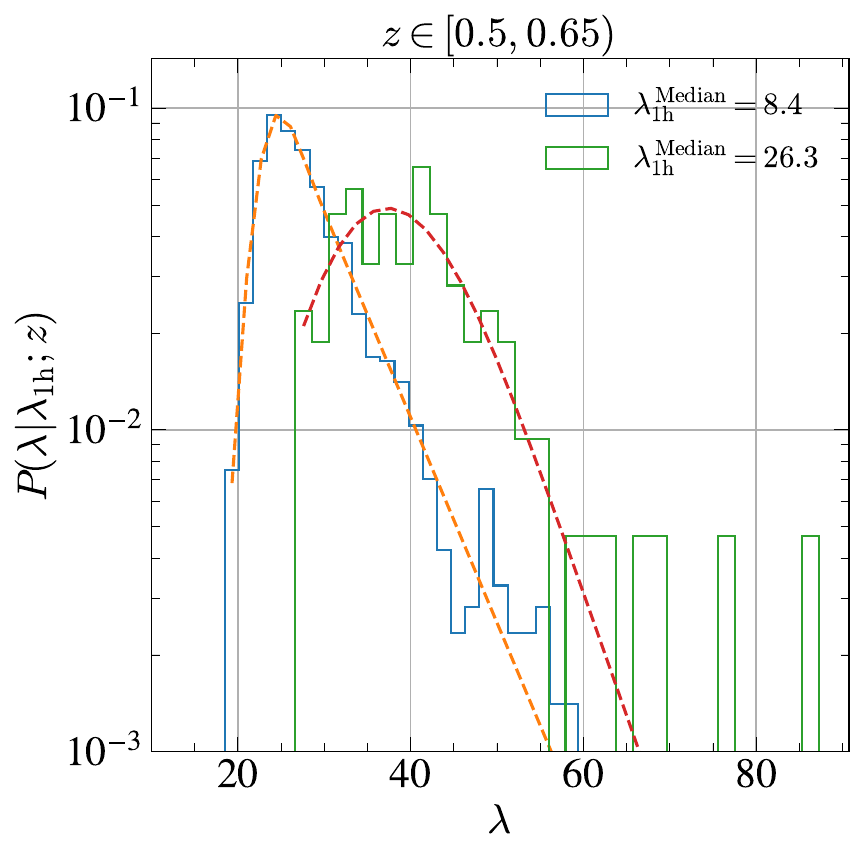}
\caption{Conditional probability distribution of observed richness given the top-ranked halo richness, $P(\lambda|\lambTop;z)$. Each histogram corresponds to clusters with a given $\lambTop$. The curves correspond to the best-fitting models to Eq.~\eqref{eq:PDF_lambda}.
}
\label{fig:PDF_lambda}
% \vspace{-50pt}
\end{figure*}
%%%%%%%%%%%%%%%%%%%%%%%%%%%%%%

Previous studies \cite{Costanzietal2019, Costanzietal2019b, Abbottetal2020} have formulated the projection effect in terms of the PDF of observed richness $\lambOb$ given true richness $\lambTr$, $P(\lambOb | \lambTr)$. In our work, the richness contributed by the top-ranked halo, $\lambTop$, can serve as a proxy for $\lambTr$. 
Figure~\ref{fig:PDF_lambda} presents $P(\lambda | \lambTop; z)$ from Cardinal. 
In each redshift range, we bin our clusters by the $\lambTop$ and calculate the PDF of $\lambda$.

Following \cite{Costanzietal2019}, we model the richness contributed by projected clusters as the convolution of a Gaussian distribution and an exponential distribution 
\begin{equation}
  \begin{aligned}
    P(\lambda|\lambTop ;z) = &(1-f_{\rm prj})\mathcal{N}(\mu,\sigma_{\rm bkg}) + f_{\rm prj}\frac{\tau_{\rm prj}}{2}\times\\
    &\exp{\left[\frac{\tau_{\rm prj}}{2}(2\mu+\tau_{\rm prj}\sigma_{\rm bkg}^2-2\lambda)\right]}\times\\
    &\mathrm{erfc}\left(\frac{\mu+\tau_{\rm prj}\sigma_{\rm bkg}^2-\lambda}{\sqrt{2}\sigma_{\rm bkg}}\right),
  \end{aligned}
  \label{eq:PDF_lambda}
\end{equation}
where $\mu=\lambTop+\Delta\mu$.
We fit the following parameters: the fraction of objects affected by projections $f_{\rm prj}$, the bias in expected richness $\Delta\mu$, the scatter for the bias $\sigma_{\rm bkg}$, and the magnitude of the projection effects $\tau_{\rm prj}$. Note that here $f_{\rm prj}$ is different from the projected fraction $f_{\rm proj}=1-\fhtop$ in the main text. This model well describes our data. We list the best-fitting parameters in Table~\ref{tab:lambda_params}.

\begin{table*}[htbp!]
\centering
\setlength\tabcolsep{15pt}
\begin{threeparttable}
\caption{Best-fitting parameters of the cluster richness projection model, $P(\lambda|\lambTop ;z)$.}
\label{tab:lambda_params}
\begin{tabular}{lccccc}
\toprule\toprule
Redshift & Halo richness & $f_{\rm prj}$ & $\Delta\mu$ & $\sigma_{\rm bkg}$ & $\tau_{\rm prj}$ \\
\midrule
$z\in[0.2,0.35)$ & $\lambTop\in[6,12)$ & 0.66 & 12.3 & 1.6 & 0.25\\
 & $\lambTop\in[24,30)$ & 0.57 & 2.1 & 2.4 & 0.20\\[5pt]
$z\in[0.35,0.5)$ & $\lambTop\in[6,12)$ & 0.90 & 13.1 & 1.6 & 0.16\\
 & $\lambTop\in[24,30)$ & 0.72 & 4.8 & 3.0 & 0.11\\[5pt]
$z\in[0.5,0.65)$ & $\lambTop\in[6,12)$ & 0.80 & 15.2 & 2.0 & 0.15\\
 & $\lambTop\in[24,30)$ & 0.999 & 6.8 & 6.7 & 0.18\\
\bottomrule\bottomrule
\end{tabular}
\end{threeparttable}%
\end{table*}

\bibliographystyle{apsrev4-2-author-truncate}
\bibliography{apssamp}% Produces the bibliography via BibTeX.

%apsrev4-2.bst 2019-01-14 (MD) hand-edited version of apsrev4-1.bst
%Control: key (0)
%Control: author (72) initials jnrlst
%Control: editor formatted (1) identically to author
%Control: production of article title (-1) disabled
%Control: page (0) single
%Control: year (1) truncated
%Control: production of eprint (0) enabled
\newcommand{\apjl}{Astrophys. J. Lett.}\newcommand{\apjs}{Astrophys. J. Suppl.}\newcommand{\mnras}{Mon. Not. R. Astron. Soc.}\newcommand{\jcap}{J. Cosmol. Astropart. Phys.}\newcommand{\aap}{Astron. Astrophys.}\newcommand{\rmxaa}{Revista Mexicana de Astronom{\'i}a y Astrof{\'i}sica}\newcommand{\phiv}{$\phi$}\newcommand{\apss}{Astrophys. Space Sci.}\newcommand{\pasp}{PASP}\newcommand{\nar}{New Astronomy Reviews}\newcommand{\pasj}{PASJ}\newcommand{\physrep}{Phys. Rep.}\newcommand{\aj}{Astron. J.}
\begin{thebibliography}{74}%
\makeatletter
\providecommand \@ifxundefined [1]{%
 \@ifx{#1\undefined}
}%
\providecommand \@ifnum [1]{%
 \ifnum #1\expandafter \@firstoftwo
 \else \expandafter \@secondoftwo
 \fi
}%
\providecommand \@ifx [1]{%
 \ifx #1\expandafter \@firstoftwo
 \else \expandafter \@secondoftwo
 \fi
}%
\providecommand \natexlab [1]{#1}%
\providecommand \enquote  [1]{``#1''}%
\providecommand \bibnamefont  [1]{#1}%
\providecommand \bibfnamefont [1]{#1}%
\providecommand \citenamefont [1]{#1}%
\providecommand \href@noop [0]{\@secondoftwo}%
\providecommand \href [0]{\begingroup \@sanitize@url \@href}%
\providecommand \@href[1]{\@@startlink{#1}\@@href}%
\providecommand \@@href[1]{\endgroup#1\@@endlink}%
\providecommand \@sanitize@url [0]{\catcode `\\12\catcode `\$12\catcode `\&12\catcode `\#12\catcode `\^12\catcode `\_12\catcode `\%12\relax}%
\providecommand \@@startlink[1]{}%
\providecommand \@@endlink[0]{}%
\providecommand \url  [0]{\begingroup\@sanitize@url \@url }%
\providecommand \@url [1]{\endgroup\@href {#1}{\urlprefix }}%
\providecommand \urlprefix  [0]{URL }%
\providecommand \Eprint [0]{\href }%
\providecommand \doibase [0]{https://doi.org/}%
\providecommand \selectlanguage [0]{\@gobble}%
\providecommand \bibinfo  [0]{\@secondoftwo}%
\providecommand \bibfield  [0]{\@secondoftwo}%
\providecommand \translation [1]{[#1]}%
\providecommand \BibitemOpen [0]{}%
\providecommand \bibitemStop [0]{}%
\providecommand \bibitemNoStop [0]{.\EOS\space}%
\providecommand \EOS [0]{\spacefactor3000\relax}%
\providecommand \BibitemShut  [1]{\csname bibitem#1\endcsname}%
\let\auto@bib@innerbib\@empty
%</preamble>
\bibitem [{\citenamefont {{Henry}}\ \emph {et~al.}(2009)\citenamefont {{Henry}}, \citenamefont {{Evrard}}, \citenamefont {{Hoekstra}}, \citenamefont {{Babul}},\ and\ \citenamefont {{Mahdavi}}}]{Henryetal2009}%
  \BibitemOpen
  \bibfield  {author} {\bibinfo {author} {\bibfnamefont {J.~P.}\ \bibnamefont {{Henry}}}, \bibinfo {author} {\bibfnamefont {A.~E.}\ \bibnamefont {{Evrard}}}, \bibinfo {author} {\bibfnamefont {H.}~\bibnamefont {{Hoekstra}}}, \bibinfo {author} {\bibfnamefont {A.}~\bibnamefont {{Babul}}},\ \bibnamefont {and}\ \bibinfo {author} {\bibfnamefont {A.}~\bibnamefont {{Mahdavi}}},\ }\href {https://doi.org/10.1088/0004-637X/691/2/1307} {\bibfield  {journal} {\bibinfo  {journal} {\apj}\ }\textbf {\bibinfo {volume} {691}},\ \bibinfo {pages} {1307} (\bibinfo {year} {2009})}\BibitemShut {NoStop}%
\bibitem [{\citenamefont {{Vikhlinin}}\ \emph {et~al.}(2009)\citenamefont {{Vikhlinin}}, \citenamefont {{Kravtsov}}, \citenamefont {{Burenin}}, \citenamefont {{Ebeling}}, \citenamefont {{Forman}}, \citenamefont {{Hornstrup}}, \citenamefont {{Jones}}, \citenamefont {{Murray}}, \citenamefont {{Nagai}}, \citenamefont {{Quintana}},\ and\ \citenamefont {{Voevodkin}}}]{Vikhlininetal2009}%
  \BibitemOpen
  \bibfield  {author} {\bibinfo {author} {\bibfnamefont {A.}~\bibnamefont {{Vikhlinin}}}, \bibnamefont {et~al.},\ }\href {https://doi.org/10.1088/0004-637X/692/2/1060} {\bibfield  {journal} {\bibinfo  {journal} {\apj}\ }\textbf {\bibinfo {volume} {692}},\ \bibinfo {pages} {1060} (\bibinfo {year} {2009})}\BibitemShut {NoStop}%
\bibitem [{\citenamefont {{Mantz}}\ \emph {et~al.}(2010)\citenamefont {{Mantz}}, \citenamefont {{Allen}}, \citenamefont {{Rapetti}},\ and\ \citenamefont {{Ebeling}}}]{MantzRapettiEbeling2010}%
  \BibitemOpen
  \bibfield  {author} {\bibinfo {author} {\bibfnamefont {A.}~\bibnamefont {{Mantz}}}, \bibinfo {author} {\bibfnamefont {S.~W.}\ \bibnamefont {{Allen}}}, \bibinfo {author} {\bibfnamefont {D.}~\bibnamefont {{Rapetti}}},\ \bibnamefont {and}\ \bibinfo {author} {\bibfnamefont {H.}~\bibnamefont {{Ebeling}}},\ }\href {https://doi.org/10.1111/j.1365-2966.2010.16992.x} {\bibfield  {journal} {\bibinfo  {journal} {\mnras}\ }\textbf {\bibinfo {volume} {406}},\ \bibinfo {pages} {1759} (\bibinfo {year} {2010})}\BibitemShut {NoStop}%
\bibitem [{\citenamefont {{Rozo}}\ \emph {et~al.}(2010)\citenamefont {{Rozo}}, \citenamefont {{Wechsler}}, \citenamefont {{Rykoff}}, \citenamefont {{Annis}}, \citenamefont {{Becker}}, \citenamefont {{Evrard}}, \citenamefont {{Frieman}}, \citenamefont {{Hansen}}, \citenamefont {{Hao}}, \citenamefont {{Johnston}}, \citenamefont {{Koester}}, \citenamefont {{McKay}}, \citenamefont {{Sheldon}},\ and\ \citenamefont {{Weinberg}}}]{Rozoetal2010}%
  \BibitemOpen
  \bibfield  {author} {\bibinfo {author} {\bibfnamefont {E.}~\bibnamefont {{Rozo}}}, \bibnamefont {et~al.},\ }\href {https://doi.org/10.1088/0004-637X/708/1/645} {\bibfield  {journal} {\bibinfo  {journal} {\apj}\ }\textbf {\bibinfo {volume} {708}},\ \bibinfo {pages} {645} (\bibinfo {year} {2010})}\BibitemShut {NoStop}%
\bibitem [{\citenamefont {{Weinberg}}\ \emph {et~al.}(2013)\citenamefont {{Weinberg}}, \citenamefont {{Mortonson}}, \citenamefont {{Eisenstein}}, \citenamefont {{Hirata}}, \citenamefont {{Riess}},\ and\ \citenamefont {{Rozo}}}]{Weinbergetal2013}%
  \BibitemOpen
  \bibfield  {author} {\bibinfo {author} {\bibfnamefont {D.~H.}\ \bibnamefont {{Weinberg}}}, \bibinfo {author} {\bibfnamefont {M.~J.}\ \bibnamefont {{Mortonson}}}, \bibinfo {author} {\bibfnamefont {D.~J.}\ \bibnamefont {{Eisenstein}}}, \bibinfo {author} {\bibfnamefont {C.}~\bibnamefont {{Hirata}}}, \bibinfo {author} {\bibfnamefont {A.~G.}\ \bibnamefont {{Riess}}},\ \bibnamefont {and}\ \bibinfo {author} {\bibfnamefont {E.}~\bibnamefont {{Rozo}}},\ }\href {https://doi.org/10.1016/j.physrep.2013.05.001} {\bibfield  {journal} {\bibinfo  {journal} {\physrep}\ }\textbf {\bibinfo {volume} {530}},\ \bibinfo {pages} {87} (\bibinfo {year} {2013})}\BibitemShut {NoStop}%
\bibitem [{\citenamefont {{To}}\ \emph {et~al.}(2021)\citenamefont {{To}}, \citenamefont {{Krause}}, \citenamefont {{Rozo}}, \citenamefont {{Wu}}, \citenamefont {{Gruen}}, \citenamefont {{Wechsler}}, \citenamefont {{Eifler}}, \citenamefont {{Rykoff}}, \citenamefont {{Costanzi}}, \citenamefont {{Becker}}, \citenamefont {{Bernstein}}, \citenamefont {{Blazek}}, \citenamefont {{Bocquet}}, \citenamefont {{Bridle}}, \citenamefont {{Cawthon}}, \citenamefont {{Choi}}, \citenamefont {{Crocce}}, \citenamefont {{Davis}}, \citenamefont {{DeRose}}, \citenamefont {{Drlica-Wagner}}, \citenamefont {{Elvin-Poole}}, \citenamefont {{Fang}}, \citenamefont {{Farahi}}, \citenamefont {{Friedrich}}, \citenamefont {{Gatti}}, \citenamefont {{Gaztanaga}}, \citenamefont {{Giannantonio}}, \citenamefont {{Hartley}}, \citenamefont {{Hoyle}}, \citenamefont {{Jarvis}}, \citenamefont {{MacCrann}}, \citenamefont {{McClintock}}, \citenamefont {{Miranda}}, \citenamefont {{Pereira}}, \citenamefont {{Park}}, \citenamefont {{Porredon}},
  \citenamefont {{Prat}}, \citenamefont {{Rau}}, \citenamefont {{Ross}}, \citenamefont {{Samuroff}}, \citenamefont {{S{\'a}nchez}}, \citenamefont {{Sevilla-Noarbe}}, \citenamefont {{Sheldon}}, \citenamefont {{Troxel}}, \citenamefont {{Varga}}, \citenamefont {{Vielzeuf}}, \citenamefont {{Zhang}}, \citenamefont {{Zuntz}}, \citenamefont {{Abbott}}, \citenamefont {{Aguena}}, \citenamefont {{Amon}}, \citenamefont {{Annis}}, \citenamefont {{Avila}}, \citenamefont {{Bertin}}, \citenamefont {{Bhargava}}, \citenamefont {{Brooks}}, \citenamefont {{Burke}}, \citenamefont {{Carnero Rosell}}, \citenamefont {{Carrasco Kind}}, \citenamefont {{Carretero}}, \citenamefont {{Chang}}, \citenamefont {{Conselice}}, \citenamefont {{da Costa}}, \citenamefont {{Davis}}, \citenamefont {{Desai}}, \citenamefont {{Diehl}}, \citenamefont {{Dietrich}}, \citenamefont {{Everett}}, \citenamefont {{Evrard}}, \citenamefont {{Ferrero}}, \citenamefont {{Flaugher}}, \citenamefont {{Fosalba}}, \citenamefont {{Frieman}}, \citenamefont
  {{Garc{\'\i}a-Bellido}}, \citenamefont {{Gruendl}}, \citenamefont {{Gutierrez}}, \citenamefont {{Hinton}}, \citenamefont {{Hollowood}}, \citenamefont {{Honscheid}}, \citenamefont {{Huterer}}, \citenamefont {{James}}, \citenamefont {{Jeltema}}, \citenamefont {{Kron}}, \citenamefont {{Kuehn}}, \citenamefont {{Kuropatkin}}, \citenamefont {{Lima}}, \citenamefont {{Maia}}, \citenamefont {{Marshall}}, \citenamefont {{Menanteau}}, \citenamefont {{Miquel}}, \citenamefont {{Morgan}}, \citenamefont {{Muir}}, \citenamefont {{Myles}}, \citenamefont {{Palmese}}, \citenamefont {{Paz-Chinch{\'o}n}}, \citenamefont {{Plazas}}, \citenamefont {{Romer}}, \citenamefont {{Roodman}}, \citenamefont {{Sanchez}}, \citenamefont {{Santiago}}, \citenamefont {{Scarpine}}, \citenamefont {{Serrano}}, \citenamefont {{Smith}}, \citenamefont {{Suchyta}}, \citenamefont {{Swanson}}, \citenamefont {{Tarle}}, \citenamefont {{Thomas}}, \citenamefont {{Tucker}}, \citenamefont {{Weller}}, \citenamefont {{Wester}}, \citenamefont {{Wilkinson}},\ and\
  \citenamefont {{DES Collaboration}}}]{DES2021}%
  \BibitemOpen
  \bibfield  {author} {\bibinfo {author} {\bibfnamefont {C.}~\bibnamefont {{To}}}, \bibnamefont {et~al.},\ }\href {https://doi.org/10.1103/PhysRevLett.126.141301} {\bibfield  {journal} {\bibinfo  {journal} {\prl}\ }\textbf {\bibinfo {volume} {126}},\ \bibinfo {eid} {141301} (\bibinfo {year} {2021})}\BibitemShut {NoStop}%
\bibitem [{\citenamefont {{Wu}}\ \emph {et~al.}(2021)\citenamefont {{Wu}}, \citenamefont {{Weinberg}}, \citenamefont {{Salcedo}},\ and\ \citenamefont {{Wibking}}}]{Wuetal2021}%
  \BibitemOpen
  \bibfield  {author} {\bibinfo {author} {\bibfnamefont {H.-Y.}\ \bibnamefont {{Wu}}}, \bibinfo {author} {\bibfnamefont {D.~H.}\ \bibnamefont {{Weinberg}}}, \bibinfo {author} {\bibfnamefont {A.~N.}\ \bibnamefont {{Salcedo}}},\ \bibnamefont {and}\ \bibinfo {author} {\bibfnamefont {B.~D.}\ \bibnamefont {{Wibking}}},\ }\href {https://doi.org/10.3847/1538-4357/abdc23} {\bibfield  {journal} {\bibinfo  {journal} {\apj}\ }\textbf {\bibinfo {volume} {910}},\ \bibinfo {eid} {28} (\bibinfo {year} {2021})}\BibitemShut {NoStop}%
\bibitem [{\citenamefont {{DES Collaboration}}\ \emph {et~al.}(2025)\citenamefont {{DES Collaboration}}, \citenamefont {{Abbott}}, \citenamefont {{Aguena}}, \citenamefont {{Alarcon}}, \citenamefont {{Anbajagane}}, \citenamefont {{Andrade-Oliveira}}, \citenamefont {{Avila}}, \citenamefont {{Bacon}}, \citenamefont {{Becker}}, \citenamefont {{Bhargava}}, \citenamefont {{Blazek}}, \citenamefont {{Bocquet}}, \citenamefont {{Brooks}}, \citenamefont {{Carnero Rosell}}, \citenamefont {{Carretero}}, \citenamefont {{Castander}}, \citenamefont {{Chang}}, \citenamefont {{Choi}}, \citenamefont {{Conselice}}, \citenamefont {{Costanzi}}, \citenamefont {{Crocce}}, \citenamefont {{da Costa}}, \citenamefont {{Pereira}}, \citenamefont {{Davis}}, \citenamefont {{Desai}}, \citenamefont {{Diehl}}, \citenamefont {{Dodelson}}, \citenamefont {{Doel}}, \citenamefont {{Elvin-Poole}}, \citenamefont {{Esteves}}, \citenamefont {{Everett}}, \citenamefont {{Farahi}}, \citenamefont {{Fert{\'e}}}, \citenamefont {{Flaugher}}, \citenamefont
  {{Garc{\'\i}a-Bellido}}, \citenamefont {{Gatti}}, \citenamefont {{Giannini}}, \citenamefont {{Giles}}, \citenamefont {{Grandis}}, \citenamefont {{Gruen}}, \citenamefont {{Gruendl}}, \citenamefont {{Gutierrez}}, \citenamefont {{Harrison}}, \citenamefont {{Hinton}}, \citenamefont {{Hollowood}}, \citenamefont {{Honscheid}}, \citenamefont {{Jeffrey}}, \citenamefont {{Jeltema}}, \citenamefont {{Krause}}, \citenamefont {{Lahav}}, \citenamefont {{Lee}}, \citenamefont {{Lidman}}, \citenamefont {{Lima}}, \citenamefont {{Lin}}, \citenamefont {{Mohr}}, \citenamefont {{Marshall}}, \citenamefont {{McCullough}}, \citenamefont {{Mena-Fern}}, \citenamefont {{Miquel}}, \citenamefont {{Muir}}, \citenamefont {{Myles}}, \citenamefont {{Ogando}}, \citenamefont {{Palmese}}, \citenamefont {{Paterno}}, \citenamefont {{Plazas Malag{\'o}n}}, \citenamefont {{Porredon}}, \citenamefont {{Prat}}, \citenamefont {{Romer}}, \citenamefont {{Roodman}}, \citenamefont {{Rozo}}, \citenamefont {{Rykoff}}, \citenamefont {{Sanchez}}, \citenamefont
  {{Sanchez Cid}}, \citenamefont {{Sevilla-Noarbe}}, \citenamefont {{Smith}}, \citenamefont {{Suchyta}}, \citenamefont {{Tarle}}, \citenamefont {{Thomas}}, \citenamefont {{To}}, \citenamefont {{Troxel}}, \citenamefont {{Vikram}}, \citenamefont {{Walker}}, \citenamefont {{Weinberg}}, \citenamefont {{Weaverdyck}}, \citenamefont {{Wechsler}}, \citenamefont {{Weller}}, \citenamefont {{Wu}}, \citenamefont {{Yamamoto}}, \citenamefont {{Yanny}}, \citenamefont {{Zhang}},\ and\ \citenamefont {{Zhou}}}]{DES2025}%
  \BibitemOpen
  \bibfield  {author} {\bibinfo {author} {\bibnamefont {{DES Collaboration}}}, \bibnamefont {et~al.},\ }\href {https://doi.org/10.48550/arXiv.2503.13632} {\bibfield  {journal} {\bibinfo  {journal} {arXiv e-prints}\ ,\ \bibinfo {eid} {arXiv:2503.13632}} (\bibinfo {year} {2025})}\BibitemShut {NoStop}%
\bibitem [{\citenamefont {{Gladders}}\ and\ \citenamefont {{Yee}}(2000)}]{GladdersMichaelYee2000}%
  \BibitemOpen
  \bibfield  {author} {\bibinfo {author} {\bibfnamefont {M.~D.}\ \bibnamefont {{Gladders}}}\ \bibnamefont {and}\ \bibinfo {author} {\bibfnamefont {H.~K.~C.}\ \bibnamefont {{Yee}}},\ }\href {https://doi.org/10.1086/301557} {\bibfield  {journal} {\bibinfo  {journal} {\aj}\ }\textbf {\bibinfo {volume} {120}},\ \bibinfo {pages} {2148} (\bibinfo {year} {2000})}\BibitemShut {NoStop}%
\bibitem [{\citenamefont {{Rykoff}}\ \emph {et~al.}(2014)\citenamefont {{Rykoff}}, \citenamefont {{Rozo}}, \citenamefont {{Busha}}, \citenamefont {{Cunha}}, \citenamefont {{Finoguenov}}, \citenamefont {{Evrard}}, \citenamefont {{Hao}}, \citenamefont {{Koester}}, \citenamefont {{Leauthaud}}, \citenamefont {{Nord}}, \citenamefont {{Pierre}}, \citenamefont {{Reddick}}, \citenamefont {{Sadibekova}}, \citenamefont {{Sheldon}},\ and\ \citenamefont {{Wechsler}}}]{Rykoffetal2014}%
  \BibitemOpen
  \bibfield  {author} {\bibinfo {author} {\bibfnamefont {E.~S.}\ \bibnamefont {{Rykoff}}}, \bibnamefont {et~al.},\ }\href {https://doi.org/10.1088/0004-637X/785/2/104} {\bibfield  {journal} {\bibinfo  {journal} {\apj}\ }\textbf {\bibinfo {volume} {785}},\ \bibinfo {eid} {104} (\bibinfo {year} {2014})}\BibitemShut {NoStop}%
\bibitem [{\citenamefont {{Gonzalez}}\ \emph {et~al.}(2019)\citenamefont {{Gonzalez}}, \citenamefont {{Gettings}}, \citenamefont {{Brodwin}}, \citenamefont {{Eisenhardt}}, \citenamefont {{Stanford}}, \citenamefont {{Wylezalek}}, \citenamefont {{Decker}}, \citenamefont {{Marrone}}, \citenamefont {{Moravec}}, \citenamefont {{O'Donnell}}, \citenamefont {{Stalder}}, \citenamefont {{Stern}}, \citenamefont {{Abdulla}}, \citenamefont {{Brown}}, \citenamefont {{Carlstrom}}, \citenamefont {{Chambers}}, \citenamefont {{Hayden}}, \citenamefont {{Lin}}, \citenamefont {{Magnier}}, \citenamefont {{Masci}}, \citenamefont {{Mantz}}, \citenamefont {{McDonald}}, \citenamefont {{Mo}}, \citenamefont {{Perlmutter}}, \citenamefont {{Wright}},\ and\ \citenamefont {{Zeimann}}}]{Gonzalezetal2019}%
  \BibitemOpen
  \bibfield  {author} {\bibinfo {author} {\bibfnamefont {A.~H.}\ \bibnamefont {{Gonzalez}}}, \bibnamefont {et~al.},\ }\href {https://doi.org/10.3847/1538-4365/aafad2} {\bibfield  {journal} {\bibinfo  {journal} {\apjs}\ }\textbf {\bibinfo {volume} {240}},\ \bibinfo {eid} {33} (\bibinfo {year} {2019})}\BibitemShut {NoStop}%
\bibitem [{\citenamefont {{Euclid Collaboration}}\ \emph {et~al.}(2019)\citenamefont {{Euclid Collaboration}}, \citenamefont {{Adam}}, \citenamefont {{Vannier}}, \citenamefont {{Maurogordato}}, \citenamefont {{Biviano}}, \citenamefont {{Adami}}, \citenamefont {{Ascaso}}, \citenamefont {{Bellagamba}}, \citenamefont {{Benoist}}, \citenamefont {{Cappi}}, \citenamefont {{D{\'\i}az-S{\'a}nchez}}, \citenamefont {{Durret}}, \citenamefont {{Farrens}}, \citenamefont {{Gonzalez}}, \citenamefont {{Iovino}}, \citenamefont {{Licitra}}, \citenamefont {{Maturi}}, \citenamefont {{Mei}}, \citenamefont {{Merson}}, \citenamefont {{Munari}}, \citenamefont {{Pell{\'o}}}, \citenamefont {{Ricci}}, \citenamefont {{Rocci}}, \citenamefont {{Roncarelli}}, \citenamefont {{Sarron}}, \citenamefont {{Amoura}}, \citenamefont {{Andreon}}, \citenamefont {{Apostolakos}}, \citenamefont {{Arnaud}}, \citenamefont {{Bardelli}}, \citenamefont {{Bartlett}}, \citenamefont {{Baugh}}, \citenamefont {{Borgani}}, \citenamefont {{Brodwin}}, \citenamefont
  {{Castander}}, \citenamefont {{Castignani}}, \citenamefont {{Cucciati}}, \citenamefont {{De Lucia}}, \citenamefont {{Dubath}}, \citenamefont {{Fosalba}}, \citenamefont {{Giocoli}}, \citenamefont {{Hoekstra}}, \citenamefont {{Mamon}}, \citenamefont {{Melin}}, \citenamefont {{Moscardini}}, \citenamefont {{Paltani}}, \citenamefont {{Radovich}}, \citenamefont {{Sartoris}}, \citenamefont {{Schultheis}}, \citenamefont {{Sereno}}, \citenamefont {{Weller}}, \citenamefont {{Burigana}}, \citenamefont {{Carvalho}}, \citenamefont {{Corcione}}, \citenamefont {{Kurki-Suonio}}, \citenamefont {{Lilje}}, \citenamefont {{Sirri}}, \citenamefont {{Toledo-Moreo}},\ and\ \citenamefont {{Zamorani}}}]{CFC}%
  \BibitemOpen
  \bibfield  {author} {\bibinfo {author} {\bibnamefont {{Euclid Collaboration}}}, \bibnamefont {et~al.},\ }\href {https://doi.org/10.1051/0004-6361/201935088} {\bibfield  {journal} {\bibinfo  {journal} {\aap}\ }\textbf {\bibinfo {volume} {627}},\ \bibinfo {eid} {A23} (\bibinfo {year} {2019})}\BibitemShut {NoStop}%
\bibitem [{\citenamefont {{Kuijken}}\ \emph {et~al.}(2019)\citenamefont {{Kuijken}}, \citenamefont {{Heymans}}, \citenamefont {{Dvornik}}, \citenamefont {{Hildebrandt}}, \citenamefont {{de Jong}}, \citenamefont {{Wright}}, \citenamefont {{Erben}}, \citenamefont {{Bilicki}}, \citenamefont {{Giblin}}, \citenamefont {{Shan}}, \citenamefont {{Getman}}, \citenamefont {{Grado}}, \citenamefont {{Hoekstra}}, \citenamefont {{Miller}}, \citenamefont {{Napolitano}}, \citenamefont {{Paolilo}}, \citenamefont {{Radovich}}, \citenamefont {{Schneider}}, \citenamefont {{Sutherland}}, \citenamefont {{Tewes}}, \citenamefont {{Tortora}}, \citenamefont {{Valentijn}},\ and\ \citenamefont {{Verdoes Kleijn}}}]{KiDS2019}%
  \BibitemOpen
  \bibfield  {author} {\bibinfo {author} {\bibfnamefont {K.}~\bibnamefont {{Kuijken}}}, \bibnamefont {et~al.},\ }\href {https://doi.org/10.1051/0004-6361/201834918} {\bibfield  {journal} {\bibinfo  {journal} {\aap}\ }\textbf {\bibinfo {volume} {625}},\ \bibinfo {eid} {A2} (\bibinfo {year} {2019})}\BibitemShut {NoStop}%
\bibitem [{\citenamefont {{B{\"o}hringer}}\ \emph {et~al.}(2001)\citenamefont {{B{\"o}hringer}}, \citenamefont {{Schuecker}}, \citenamefont {{Guzzo}}, \citenamefont {{Collins}}, \citenamefont {{Voges}}, \citenamefont {{Schindler}}, \citenamefont {{Neumann}}, \citenamefont {{Cruddace}}, \citenamefont {{De Grandi}}, \citenamefont {{Chincarini}}, \citenamefont {{Edge}}, \citenamefont {{MacGillivray}},\ and\ \citenamefont {{Shaver}}}]{Bohringeretal2001}%
  \BibitemOpen
  \bibfield  {author} {\bibinfo {author} {\bibfnamefont {H.}~\bibnamefont {{B{\"o}hringer}}}, \bibnamefont {et~al.},\ }\href {https://doi.org/10.1051/0004-6361:20010240} {\bibfield  {journal} {\bibinfo  {journal} {\aap}\ }\textbf {\bibinfo {volume} {369}},\ \bibinfo {pages} {826} (\bibinfo {year} {2001})}\BibitemShut {NoStop}%
\bibitem [{\citenamefont {{Melnyk}}\ \emph {et~al.}(2018)\citenamefont {{Melnyk}}, \citenamefont {{Elyiv}}, \citenamefont {{Smol{\v{c}}i{\'c}}}, \citenamefont {{Plionis}}, \citenamefont {{Koulouridis}}, \citenamefont {{Fotopoulou}}, \citenamefont {{Chiappetti}}, \citenamefont {{Adami}}, \citenamefont {{Baran}}, \citenamefont {{Butler}}, \citenamefont {{Delhaize}}, \citenamefont {{Delvecchio}}, \citenamefont {{Finet}}, \citenamefont {{Huynh}}, \citenamefont {{Lidman}}, \citenamefont {{Pierre}}, \citenamefont {{Pompei}}, \citenamefont {{Vignali}},\ and\ \citenamefont {{Surdej}}}]{Melnyketal2018}%
  \BibitemOpen
  \bibfield  {author} {\bibinfo {author} {\bibfnamefont {O.}~\bibnamefont {{Melnyk}}}, \bibnamefont {et~al.},\ }\href {https://doi.org/10.1051/0004-6361/201730479} {\bibfield  {journal} {\bibinfo  {journal} {\aap}\ }\textbf {\bibinfo {volume} {620}},\ \bibinfo {eid} {A6} (\bibinfo {year} {2018})}\BibitemShut {NoStop}%
\bibitem [{\citenamefont {{Kleinebreil}}\ \emph {et~al.}(2025)\citenamefont {{Kleinebreil}}, \citenamefont {{Grandis}}, \citenamefont {{Schrabback}}, \citenamefont {{Ghirardini}}, \citenamefont {{Chiu}}, \citenamefont {{Liu}}, \citenamefont {{Kluge}}, \citenamefont {{Reiprich}}, \citenamefont {{Artis}}, \citenamefont {{Bahar}}, \citenamefont {{Balzer}}, \citenamefont {{Bulbul}}, \citenamefont {{Clerc}}, \citenamefont {{Comparat}}, \citenamefont {{Garrel}}, \citenamefont {{Gruen}}, \citenamefont {{Li}}, \citenamefont {{Miyatake}}, \citenamefont {{Miyazaki}}, \citenamefont {{Ramos-Ceja}}, \citenamefont {{Sanders}}, \citenamefont {{Seppi}}, \citenamefont {{Okabe}},\ and\ \citenamefont {{Zhang}}}]{Kleinebreiletal2024}%
  \BibitemOpen
  \bibfield  {author} {\bibinfo {author} {\bibfnamefont {F.}~\bibnamefont {{Kleinebreil}}}, \bibnamefont {et~al.},\ }\href {https://doi.org/10.1051/0004-6361/202449599} {\bibfield  {journal} {\bibinfo  {journal} {\aap}\ }\textbf {\bibinfo {volume} {695}},\ \bibinfo {eid} {A216} (\bibinfo {year} {2025})}\BibitemShut {NoStop}%
\bibitem [{\citenamefont {{Bulbul}}\ \emph {et~al.}(2024)\citenamefont {{Bulbul}}, \citenamefont {{Liu}}, \citenamefont {{Kluge}}, \citenamefont {{Zhang}}, \citenamefont {{Sanders}}, \citenamefont {{Bahar}}, \citenamefont {{Ghirardini}}, \citenamefont {{Artis}}, \citenamefont {{Seppi}}, \citenamefont {{Garrel}}, \citenamefont {{Ramos-Ceja}}, \citenamefont {{Comparat}}, \citenamefont {{Balzer}}, \citenamefont {{B{\"o}ckmann}}, \citenamefont {{Br{\"u}ggen}}, \citenamefont {{Clerc}}, \citenamefont {{Dennerl}}, \citenamefont {{Dolag}}, \citenamefont {{Freyberg}}, \citenamefont {{Grandis}}, \citenamefont {{Gruen}}, \citenamefont {{Kleinebreil}}, \citenamefont {{Krippendorf}}, \citenamefont {{Lamer}}, \citenamefont {{Merloni}}, \citenamefont {{Migkas}}, \citenamefont {{Nandra}}, \citenamefont {{Pacaud}}, \citenamefont {{Predehl}}, \citenamefont {{Reiprich}}, \citenamefont {{Schrabback}}, \citenamefont {{Veronica}}, \citenamefont {{Weller}},\ and\ \citenamefont {{Zelmer}}}]{Bulbuletal2024}%
  \BibitemOpen
  \bibfield  {author} {\bibinfo {author} {\bibfnamefont {E.}~\bibnamefont {{Bulbul}}}, \bibnamefont {et~al.},\ }\href {https://doi.org/10.1051/0004-6361/202348264} {\bibfield  {journal} {\bibinfo  {journal} {\aap}\ }\textbf {\bibinfo {volume} {685}},\ \bibinfo {eid} {A106} (\bibinfo {year} {2024})}\BibitemShut {NoStop}%
\bibitem [{\citenamefont {{Vanderlinde}}\ \emph {et~al.}(2010)\citenamefont {{Vanderlinde}}, \citenamefont {{Crawford}}, \citenamefont {{de Haan}}, \citenamefont {{Dudley}}, \citenamefont {{Shaw}}, \citenamefont {{Ade}}, \citenamefont {{Aird}}, \citenamefont {{Benson}}, \citenamefont {{Bleem}}, \citenamefont {{Brodwin}}, \citenamefont {{Carlstrom}}, \citenamefont {{Chang}}, \citenamefont {{Crites}}, \citenamefont {{Desai}}, \citenamefont {{Dobbs}}, \citenamefont {{Foley}}, \citenamefont {{George}}, \citenamefont {{Gladders}}, \citenamefont {{Hall}}, \citenamefont {{Halverson}}, \citenamefont {{High}}, \citenamefont {{Holder}}, \citenamefont {{Holzapfel}}, \citenamefont {{Hrubes}}, \citenamefont {{Joy}}, \citenamefont {{Keisler}}, \citenamefont {{Knox}}, \citenamefont {{Lee}}, \citenamefont {{Leitch}}, \citenamefont {{Loehr}}, \citenamefont {{Lueker}}, \citenamefont {{Marrone}}, \citenamefont {{McMahon}}, \citenamefont {{Mehl}}, \citenamefont {{Meyer}}, \citenamefont {{Mohr}}, \citenamefont {{Montroy}},
  \citenamefont {{Ngeow}}, \citenamefont {{Padin}}, \citenamefont {{Plagge}}, \citenamefont {{Pryke}}, \citenamefont {{Reichardt}}, \citenamefont {{Rest}}, \citenamefont {{Ruel}}, \citenamefont {{Ruhl}}, \citenamefont {{Schaffer}}, \citenamefont {{Shirokoff}}, \citenamefont {{Song}}, \citenamefont {{Spieler}}, \citenamefont {{Stalder}}, \citenamefont {{Staniszewski}}, \citenamefont {{Stark}}, \citenamefont {{Stubbs}}, \citenamefont {{van Engelen}}, \citenamefont {{Vieira}}, \citenamefont {{Williamson}}, \citenamefont {{Yang}}, \citenamefont {{Zahn}},\ and\ \citenamefont {{Zenteno}}}]{Vanderlindeetal2010}%
  \BibitemOpen
  \bibfield  {author} {\bibinfo {author} {\bibfnamefont {K.}~\bibnamefont {{Vanderlinde}}}, \bibnamefont {et~al.},\ }\href {https://doi.org/10.1088/0004-637X/722/2/1180} {\bibfield  {journal} {\bibinfo  {journal} {\apj}\ }\textbf {\bibinfo {volume} {722}},\ \bibinfo {pages} {1180} (\bibinfo {year} {2010})}\BibitemShut {NoStop}%
\bibitem [{\citenamefont {{Bleem}}\ \emph {et~al.}(2015)\citenamefont {{Bleem}}, \citenamefont {{Stalder}}, \citenamefont {{de Haan}}, \citenamefont {{Aird}}, \citenamefont {{Allen}}, \citenamefont {{Applegate}}, \citenamefont {{Ashby}}, \citenamefont {{Bautz}}, \citenamefont {{Bayliss}}, \citenamefont {{Benson}}, \citenamefont {{Bocquet}}, \citenamefont {{Brodwin}}, \citenamefont {{Carlstrom}}, \citenamefont {{Chang}}, \citenamefont {{Chiu}}, \citenamefont {{Cho}}, \citenamefont {{Clocchiatti}}, \citenamefont {{Crawford}}, \citenamefont {{Crites}}, \citenamefont {{Desai}}, \citenamefont {{Dietrich}}, \citenamefont {{Dobbs}}, \citenamefont {{Foley}}, \citenamefont {{Forman}}, \citenamefont {{George}}, \citenamefont {{Gladders}}, \citenamefont {{Gonzalez}}, \citenamefont {{Halverson}}, \citenamefont {{Hennig}}, \citenamefont {{Hoekstra}}, \citenamefont {{Holder}}, \citenamefont {{Holzapfel}}, \citenamefont {{Hrubes}}, \citenamefont {{Jones}}, \citenamefont {{Keisler}}, \citenamefont {{Knox}}, \citenamefont
  {{Lee}}, \citenamefont {{Leitch}}, \citenamefont {{Liu}}, \citenamefont {{Lueker}}, \citenamefont {{Luong-Van}}, \citenamefont {{Mantz}}, \citenamefont {{Marrone}}, \citenamefont {{McDonald}}, \citenamefont {{McMahon}}, \citenamefont {{Meyer}}, \citenamefont {{Mocanu}}, \citenamefont {{Mohr}}, \citenamefont {{Murray}}, \citenamefont {{Padin}}, \citenamefont {{Pryke}}, \citenamefont {{Reichardt}}, \citenamefont {{Rest}}, \citenamefont {{Ruel}}, \citenamefont {{Ruhl}}, \citenamefont {{Saliwanchik}}, \citenamefont {{Saro}}, \citenamefont {{Sayre}}, \citenamefont {{Schaffer}}, \citenamefont {{Schrabback}}, \citenamefont {{Shirokoff}}, \citenamefont {{Song}}, \citenamefont {{Spieler}}, \citenamefont {{Stanford}}, \citenamefont {{Staniszewski}}, \citenamefont {{Stark}}, \citenamefont {{Story}}, \citenamefont {{Stubbs}}, \citenamefont {{Vanderlinde}}, \citenamefont {{Vieira}}, \citenamefont {{Vikhlinin}}, \citenamefont {{Williamson}}, \citenamefont {{Zahn}},\ and\ \citenamefont {{Zenteno}}}]{Bleemetal2015}%
  \BibitemOpen
  \bibfield  {author} {\bibinfo {author} {\bibfnamefont {L.~E.}\ \bibnamefont {{Bleem}}}, \bibnamefont {et~al.},\ }\href {https://doi.org/10.1088/0067-0049/216/2/27} {\bibfield  {journal} {\bibinfo  {journal} {\apjs}\ }\textbf {\bibinfo {volume} {216}},\ \bibinfo {eid} {27} (\bibinfo {year} {2015})}\BibitemShut {NoStop}%
\bibitem [{\citenamefont {{Planck Collaboration}}\ \emph {et~al.}(2020)\citenamefont {{Planck Collaboration}}, \citenamefont {{Aghanim}}, \citenamefont {{Akrami}}, \citenamefont {{Ashdown}}, \citenamefont {{Aumont}}, \citenamefont {{Baccigalupi}}, \citenamefont {{Ballardini}}, \citenamefont {{Banday}}, \citenamefont {{Barreiro}}, \citenamefont {{Bartolo}}, \citenamefont {{Basak}}, \citenamefont {{Battye}}, \citenamefont {{Benabed}}, \citenamefont {{Bernard}}, \citenamefont {{Bersanelli}}, \citenamefont {{Bielewicz}}, \citenamefont {{Bock}}, \citenamefont {{Bond}}, \citenamefont {{Borrill}}, \citenamefont {{Bouchet}}, \citenamefont {{Boulanger}}, \citenamefont {{Bucher}}, \citenamefont {{Burigana}}, \citenamefont {{Butler}}, \citenamefont {{Calabrese}}, \citenamefont {{Cardoso}}, \citenamefont {{Carron}}, \citenamefont {{Challinor}}, \citenamefont {{Chiang}}, \citenamefont {{Chluba}}, \citenamefont {{Colombo}}, \citenamefont {{Combet}}, \citenamefont {{Contreras}}, \citenamefont {{Crill}}, \citenamefont
  {{Cuttaia}}, \citenamefont {{de Bernardis}}, \citenamefont {{de Zotti}}, \citenamefont {{Delabrouille}}, \citenamefont {{Delouis}}, \citenamefont {{Di Valentino}}, \citenamefont {{Diego}}, \citenamefont {{Dor{\'e}}}, \citenamefont {{Douspis}}, \citenamefont {{Ducout}}, \citenamefont {{Dupac}}, \citenamefont {{Dusini}}, \citenamefont {{Efstathiou}}, \citenamefont {{Elsner}}, \citenamefont {{En{\ss}lin}}, \citenamefont {{Eriksen}}, \citenamefont {{Fantaye}}, \citenamefont {{Farhang}}, \citenamefont {{Fergusson}}, \citenamefont {{Fernandez-Cobos}}, \citenamefont {{Finelli}}, \citenamefont {{Forastieri}}, \citenamefont {{Frailis}}, \citenamefont {{Fraisse}}, \citenamefont {{Franceschi}}, \citenamefont {{Frolov}}, \citenamefont {{Galeotta}}, \citenamefont {{Galli}}, \citenamefont {{Ganga}}, \citenamefont {{G{\'e}nova-Santos}}, \citenamefont {{Gerbino}}, \citenamefont {{Ghosh}}, \citenamefont {{Gonz{\'a}lez-Nuevo}}, \citenamefont {{G{\'o}rski}}, \citenamefont {{Gratton}}, \citenamefont {{Gruppuso}}, \citenamefont
  {{Gudmundsson}}, \citenamefont {{Hamann}}, \citenamefont {{Handley}}, \citenamefont {{Hansen}}, \citenamefont {{Herranz}}, \citenamefont {{Hildebrandt}}, \citenamefont {{Hivon}}, \citenamefont {{Huang}}, \citenamefont {{Jaffe}}, \citenamefont {{Jones}}, \citenamefont {{Karakci}}, \citenamefont {{Keih{\"a}nen}}, \citenamefont {{Keskitalo}}, \citenamefont {{Kiiveri}}, \citenamefont {{Kim}}, \citenamefont {{Kisner}}, \citenamefont {{Knox}}, \citenamefont {{Krachmalnicoff}}, \citenamefont {{Kunz}}, \citenamefont {{Kurki-Suonio}}, \citenamefont {{Lagache}}, \citenamefont {{Lamarre}}, \citenamefont {{Lasenby}}, \citenamefont {{Lattanzi}}, \citenamefont {{Lawrence}}, \citenamefont {{Le Jeune}}, \citenamefont {{Lemos}}, \citenamefont {{Lesgourgues}}, \citenamefont {{Levrier}}, \citenamefont {{Lewis}}, \citenamefont {{Liguori}}, \citenamefont {{Lilje}}, \citenamefont {{Lilley}}, \citenamefont {{Lindholm}}, \citenamefont {{L{\'o}pez-Caniego}}, \citenamefont {{Lubin}}, \citenamefont {{Ma}}, \citenamefont
  {{Mac{\'\i}as-P{\'e}rez}}, \citenamefont {{Maggio}}, \citenamefont {{Maino}}, \citenamefont {{Mandolesi}}, \citenamefont {{Mangilli}}, \citenamefont {{Marcos-Caballero}}, \citenamefont {{Maris}}, \citenamefont {{Martin}}, \citenamefont {{Martinelli}}, \citenamefont {{Mart{\'\i}nez-Gonz{\'a}lez}}, \citenamefont {{Matarrese}}, \citenamefont {{Mauri}}, \citenamefont {{McEwen}}, \citenamefont {{Meinhold}}, \citenamefont {{Melchiorri}}, \citenamefont {{Mennella}}, \citenamefont {{Migliaccio}}, \citenamefont {{Millea}}, \citenamefont {{Mitra}}, \citenamefont {{Miville-Desch{\^e}nes}}, \citenamefont {{Molinari}}, \citenamefont {{Montier}}, \citenamefont {{Morgante}}, \citenamefont {{Moss}}, \citenamefont {{Natoli}}, \citenamefont {{N{\o}rgaard-Nielsen}}, \citenamefont {{Pagano}}, \citenamefont {{Paoletti}}, \citenamefont {{Partridge}}, \citenamefont {{Patanchon}}, \citenamefont {{Peiris}}, \citenamefont {{Perrotta}}, \citenamefont {{Pettorino}}, \citenamefont {{Piacentini}}, \citenamefont {{Polastri}},
  \citenamefont {{Polenta}}, \citenamefont {{Puget}}, \citenamefont {{Rachen}}, \citenamefont {{Reinecke}}, \citenamefont {{Remazeilles}}, \citenamefont {{Renzi}}, \citenamefont {{Rocha}}, \citenamefont {{Rosset}}, \citenamefont {{Roudier}}, \citenamefont {{Rubi{\~n}o-Mart{\'\i}n}}, \citenamefont {{Ruiz-Granados}}, \citenamefont {{Salvati}}, \citenamefont {{Sandri}}, \citenamefont {{Savelainen}}, \citenamefont {{Scott}}, \citenamefont {{Shellard}}, \citenamefont {{Sirignano}}, \citenamefont {{Sirri}}, \citenamefont {{Spencer}}, \citenamefont {{Sunyaev}}, \citenamefont {{Suur-Uski}}, \citenamefont {{Tauber}}, \citenamefont {{Tavagnacco}}, \citenamefont {{Tenti}}, \citenamefont {{Toffolatti}}, \citenamefont {{Tomasi}}, \citenamefont {{Trombetti}}, \citenamefont {{Valenziano}}, \citenamefont {{Valiviita}}, \citenamefont {{Van Tent}}, \citenamefont {{Vibert}}, \citenamefont {{Vielva}}, \citenamefont {{Villa}}, \citenamefont {{Vittorio}}, \citenamefont {{Wandelt}}, \citenamefont {{Wehus}}, \citenamefont {{White}},
  \citenamefont {{White}}, \citenamefont {{Zacchei}},\ and\ \citenamefont {{Zonca}}}]{Planck2020}%
  \BibitemOpen
  \bibfield  {author} {\bibinfo {author} {\bibnamefont {{Planck Collaboration}}}, \bibnamefont {et~al.},\ }\href {https://doi.org/10.1051/0004-6361/201833910} {\bibfield  {journal} {\bibinfo  {journal} {\aap}\ }\textbf {\bibinfo {volume} {641}},\ \bibinfo {eid} {A6} (\bibinfo {year} {2020})}\BibitemShut {NoStop}%
\bibitem [{\citenamefont {{Planck Collaboration}}\ \emph {et~al.}(2016)\citenamefont {{Planck Collaboration}}, \citenamefont {{Ade}}, \citenamefont {{Aghanim}}, \citenamefont {{Arnaud}}, \citenamefont {{Ashdown}}, \citenamefont {{Aumont}}, \citenamefont {{Baccigalupi}}, \citenamefont {{Banday}}, \citenamefont {{Barreiro}}, \citenamefont {{Barrena}}, \citenamefont {{Bartlett}}, \citenamefont {{Bartolo}}, \citenamefont {{Battaner}}, \citenamefont {{Battye}}, \citenamefont {{Benabed}}, \citenamefont {{Beno{\^\i}t}}, \citenamefont {{Benoit-L{\'e}vy}}, \citenamefont {{Bernard}}, \citenamefont {{Bersanelli}}, \citenamefont {{Bielewicz}}, \citenamefont {{Bikmaev}}, \citenamefont {{B{\"o}hringer}}, \citenamefont {{Bonaldi}}, \citenamefont {{Bonavera}}, \citenamefont {{Bond}}, \citenamefont {{Borrill}}, \citenamefont {{Bouchet}}, \citenamefont {{Bucher}}, \citenamefont {{Burenin}}, \citenamefont {{Burigana}}, \citenamefont {{Butler}}, \citenamefont {{Calabrese}}, \citenamefont {{Cardoso}}, \citenamefont {{Carvalho}},
  \citenamefont {{Catalano}}, \citenamefont {{Challinor}}, \citenamefont {{Chamballu}}, \citenamefont {{Chary}}, \citenamefont {{Chiang}}, \citenamefont {{Chon}}, \citenamefont {{Christensen}}, \citenamefont {{Clements}}, \citenamefont {{Colombi}}, \citenamefont {{Colombo}}, \citenamefont {{Combet}}, \citenamefont {{Comis}}, \citenamefont {{Couchot}}, \citenamefont {{Coulais}}, \citenamefont {{Crill}}, \citenamefont {{Curto}}, \citenamefont {{Cuttaia}}, \citenamefont {{Dahle}}, \citenamefont {{Danese}}, \citenamefont {{Davies}}, \citenamefont {{Davis}}, \citenamefont {{de Bernardis}}, \citenamefont {{de Rosa}}, \citenamefont {{de Zotti}}, \citenamefont {{Delabrouille}}, \citenamefont {{D{\'e}sert}}, \citenamefont {{Dickinson}}, \citenamefont {{Diego}}, \citenamefont {{Dolag}}, \citenamefont {{Dole}}, \citenamefont {{Donzelli}}, \citenamefont {{Dor{\'e}}}, \citenamefont {{Douspis}}, \citenamefont {{Ducout}}, \citenamefont {{Dupac}}, \citenamefont {{Efstathiou}}, \citenamefont {{Eisenhardt}}, \citenamefont
  {{Elsner}}, \citenamefont {{En{\ss}lin}}, \citenamefont {{Eriksen}}, \citenamefont {{Falgarone}}, \citenamefont {{Fergusson}}, \citenamefont {{Feroz}}, \citenamefont {{Ferragamo}}, \citenamefont {{Finelli}}, \citenamefont {{Forni}}, \citenamefont {{Frailis}}, \citenamefont {{Fraisse}}, \citenamefont {{Franceschi}}, \citenamefont {{Frejsel}}, \citenamefont {{Galeotta}}, \citenamefont {{Galli}}, \citenamefont {{Ganga}}, \citenamefont {{G{\'e}nova-Santos}}, \citenamefont {{Giard}}, \citenamefont {{Giraud-H{\'e}raud}}, \citenamefont {{Gjerl{\o}w}}, \citenamefont {{Gonz{\'a}lez-Nuevo}}, \citenamefont {{G{\'o}rski}}, \citenamefont {{Grainge}}, \citenamefont {{Gratton}}, \citenamefont {{Gregorio}}, \citenamefont {{Gruppuso}}, \citenamefont {{Gudmundsson}}, \citenamefont {{Hansen}}, \citenamefont {{Hanson}}, \citenamefont {{Harrison}}, \citenamefont {{Hempel}}, \citenamefont {{Henrot-Versill{\'e}}}, \citenamefont {{Hern{\'a}ndez-Monteagudo}}, \citenamefont {{Herranz}}, \citenamefont {{Hildebrandt}}, \citenamefont
  {{Hivon}}, \citenamefont {{Hobson}}, \citenamefont {{Holmes}}, \citenamefont {{Hornstrup}}, \citenamefont {{Hovest}}, \citenamefont {{Huffenberger}}, \citenamefont {{Hurier}}, \citenamefont {{Jaffe}}, \citenamefont {{Jaffe}}, \citenamefont {{Jin}}, \citenamefont {{Jones}}, \citenamefont {{Juvela}}, \citenamefont {{Keih{\"a}nen}}, \citenamefont {{Keskitalo}}, \citenamefont {{Khamitov}}, \citenamefont {{Kisner}}, \citenamefont {{Kneissl}}, \citenamefont {{Knoche}}, \citenamefont {{Kunz}}, \citenamefont {{Kurki-Suonio}}, \citenamefont {{Lagache}}, \citenamefont {{Lamarre}}, \citenamefont {{Lasenby}}, \citenamefont {{Lattanzi}}, \citenamefont {{Lawrence}}, \citenamefont {{Leonardi}}, \citenamefont {{Lesgourgues}}, \citenamefont {{Levrier}}, \citenamefont {{Liguori}}, \citenamefont {{Lilje}}, \citenamefont {{Linden-V{\o}rnle}}, \citenamefont {{L{\'o}pez-Caniego}}, \citenamefont {{Lubin}}, \citenamefont {{Mac{\'\i}as-P{\'e}rez}}, \citenamefont {{Maggio}}, \citenamefont {{Maino}}, \citenamefont {{Mak}},
  \citenamefont {{Mandolesi}}, \citenamefont {{Mangilli}}, \citenamefont {{Martin}}, \citenamefont {{Mart{\'\i}nez-Gonz{\'a}lez}}, \citenamefont {{Masi}}, \citenamefont {{Matarrese}}, \citenamefont {{Mazzotta}}, \citenamefont {{McGehee}}, \citenamefont {{Mei}}, \citenamefont {{Melchiorri}}, \citenamefont {{Melin}}, \citenamefont {{Mendes}}, \citenamefont {{Mennella}}, \citenamefont {{Migliaccio}}, \citenamefont {{Mitra}}, \citenamefont {{Miville-Desch{\^e}nes}}, \citenamefont {{Moneti}}, \citenamefont {{Montier}}, \citenamefont {{Morgante}}, \citenamefont {{Mortlock}}, \citenamefont {{Moss}}, \citenamefont {{Munshi}}, \citenamefont {{Murphy}}, \citenamefont {{Naselsky}}, \citenamefont {{Nastasi}}, \citenamefont {{Nati}}, \citenamefont {{Natoli}}, \citenamefont {{Netterfield}}, \citenamefont {{N{\o}rgaard-Nielsen}}, \citenamefont {{Noviello}}, \citenamefont {{Novikov}}, \citenamefont {{Novikov}}, \citenamefont {{Olamaie}}, \citenamefont {{Oxborrow}}, \citenamefont {{Paci}}, \citenamefont {{Pagano}},
  \citenamefont {{Pajot}}, \citenamefont {{Paoletti}}, \citenamefont {{Pasian}}, \citenamefont {{Patanchon}}, \citenamefont {{Pearson}}, \citenamefont {{Perdereau}}, \citenamefont {{Perotto}}, \citenamefont {{Perrott}}, \citenamefont {{Perrotta}}, \citenamefont {{Pettorino}}, \citenamefont {{Piacentini}}, \citenamefont {{Piat}}, \citenamefont {{Pierpaoli}}, \citenamefont {{Pietrobon}}, \citenamefont {{Plaszczynski}}, \citenamefont {{Pointecouteau}}, \citenamefont {{Polenta}}, \citenamefont {{Pratt}}, \citenamefont {{Pr{\'e}zeau}}, \citenamefont {{Prunet}},\ and\ \citenamefont {{Puget}}}]{Planck2016}%
  \BibitemOpen
  \bibfield  {author} {\bibinfo {author} {\bibnamefont {{Planck Collaboration}}}, \bibnamefont {et~al.},\ }\href {https://doi.org/10.1051/0004-6361/201525823} {\bibfield  {journal} {\bibinfo  {journal} {\aap}\ }\textbf {\bibinfo {volume} {594}},\ \bibinfo {eid} {A27} (\bibinfo {year} {2016})}\BibitemShut {NoStop}%
\bibitem [{\citenamefont {{Hilton}}\ \emph {et~al.}(2021)\citenamefont {{Hilton}}, \citenamefont {{Sif{\'o}n}}, \citenamefont {{Naess}}, \citenamefont {{Madhavacheril}}, \citenamefont {{Oguri}}, \citenamefont {{Rozo}}, \citenamefont {{Rykoff}}, \citenamefont {{Abbott}}, \citenamefont {{Adhikari}}, \citenamefont {{Aguena}}, \citenamefont {{Aiola}}, \citenamefont {{Allam}}, \citenamefont {{Amodeo}}, \citenamefont {{Amon}}, \citenamefont {{Annis}}, \citenamefont {{Ansarinejad}}, \citenamefont {{Aros-Bunster}}, \citenamefont {{Austermann}}, \citenamefont {{Avila}}, \citenamefont {{Bacon}}, \citenamefont {{Battaglia}}, \citenamefont {{Beall}}, \citenamefont {{Becker}}, \citenamefont {{Bernstein}}, \citenamefont {{Bertin}}, \citenamefont {{Bhandarkar}}, \citenamefont {{Bhargava}}, \citenamefont {{Bond}}, \citenamefont {{Brooks}}, \citenamefont {{Burke}}, \citenamefont {{Calabrese}}, \citenamefont {{Carrasco Kind}}, \citenamefont {{Carretero}}, \citenamefont {{Choi}}, \citenamefont {{Choi}}, \citenamefont
  {{Conselice}}, \citenamefont {{da Costa}}, \citenamefont {{Costanzi}}, \citenamefont {{Crichton}}, \citenamefont {{Crowley}}, \citenamefont {{D{\"u}nner}}, \citenamefont {{Denison}}, \citenamefont {{Devlin}}, \citenamefont {{Dicker}}, \citenamefont {{Diehl}}, \citenamefont {{Dietrich}}, \citenamefont {{Doel}}, \citenamefont {{Duff}}, \citenamefont {{Duivenvoorden}}, \citenamefont {{Dunkley}}, \citenamefont {{Everett}}, \citenamefont {{Ferraro}}, \citenamefont {{Ferrero}}, \citenamefont {{Fert{\'e}}}, \citenamefont {{Flaugher}}, \citenamefont {{Frieman}}, \citenamefont {{Gallardo}}, \citenamefont {{Garc{\'\i}a-Bellido}}, \citenamefont {{Gaztanaga}}, \citenamefont {{Gerdes}}, \citenamefont {{Giles}}, \citenamefont {{Golec}}, \citenamefont {{Gralla}}, \citenamefont {{Grandis}}, \citenamefont {{Gruen}}, \citenamefont {{Gruendl}}, \citenamefont {{Gschwend}}, \citenamefont {{Gutierrez}}, \citenamefont {{Han}}, \citenamefont {{Hartley}}, \citenamefont {{Hasselfield}}, \citenamefont {{Hill}}, \citenamefont
  {{Hilton}}, \citenamefont {{Hincks}}, \citenamefont {{Hinton}}, \citenamefont {{Ho}}, \citenamefont {{Honscheid}}, \citenamefont {{Hoyle}}, \citenamefont {{Hubmayr}}, \citenamefont {{Huffenberger}}, \citenamefont {{Hughes}}, \citenamefont {{Jaelani}}, \citenamefont {{Jain}}, \citenamefont {{James}}, \citenamefont {{Jeltema}}, \citenamefont {{Kent}}, \citenamefont {{Knowles}}, \citenamefont {{Koopman}}, \citenamefont {{Kuehn}}, \citenamefont {{Lahav}}, \citenamefont {{Lima}}, \citenamefont {{Lin}}, \citenamefont {{Lokken}}, \citenamefont {{Loubser}}, \citenamefont {{MacCrann}}, \citenamefont {{Maia}}, \citenamefont {{Marriage}}, \citenamefont {{Martin}}, \citenamefont {{McMahon}}, \citenamefont {{Melchior}}, \citenamefont {{Menanteau}}, \citenamefont {{Miquel}}, \citenamefont {{Miyatake}}, \citenamefont {{Moodley}}, \citenamefont {{Morgan}}, \citenamefont {{Mroczkowski}}, \citenamefont {{Nati}}, \citenamefont {{Newburgh}}, \citenamefont {{Niemack}}, \citenamefont {{Nishizawa}}, \citenamefont {{Ogando}},
  \citenamefont {{Orlowski-Scherer}}, \citenamefont {{Page}}, \citenamefont {{Palmese}}, \citenamefont {{Partridge}}, \citenamefont {{Paz-Chinch{\'o}n}}, \citenamefont {{Phakathi}}, \citenamefont {{Plazas}}, \citenamefont {{Robertson}}, \citenamefont {{Romer}}, \citenamefont {{Carnero Rosell}}, \citenamefont {{Salatino}}, \citenamefont {{Sanchez}}, \citenamefont {{Schaan}}, \citenamefont {{Schillaci}}, \citenamefont {{Sehgal}}, \citenamefont {{Serrano}}, \citenamefont {{Shin}}, \citenamefont {{Simon}}, \citenamefont {{Smith}}, \citenamefont {{Soares-Santos}}, \citenamefont {{Spergel}}, \citenamefont {{Staggs}}, \citenamefont {{Storer}}, \citenamefont {{Suchyta}}, \citenamefont {{Swanson}}, \citenamefont {{Tarle}}, \citenamefont {{Thomas}}, \citenamefont {{To}}, \citenamefont {{Trac}}, \citenamefont {{Ullom}}, \citenamefont {{Vale}}, \citenamefont {{Van Lanen}}, \citenamefont {{Vavagiakis}}, \citenamefont {{De Vicente}}, \citenamefont {{Wilkinson}}, \citenamefont {{Wollack}}, \citenamefont {{Xu}},\ and\
  \citenamefont {{Zhang}}}]{Hiltonetal2021}%
  \BibitemOpen
  \bibfield  {author} {\bibinfo {author} {\bibfnamefont {M.}~\bibnamefont {{Hilton}}}, \bibnamefont {et~al.},\ }\href {https://doi.org/10.3847/1538-4365/abd023} {\bibfield  {journal} {\bibinfo  {journal} {\apjs}\ }\textbf {\bibinfo {volume} {253}},\ \bibinfo {eid} {3} (\bibinfo {year} {2021})}\BibitemShut {NoStop}%
\bibitem [{\citenamefont {{Bleem}}\ \emph {et~al.}(2024)\citenamefont {{Bleem}}, \citenamefont {{Klein}}, \citenamefont {{Abbot}}, \citenamefont {{Ade}}, \citenamefont {{Aguena}}, \citenamefont {{Alves}}, \citenamefont {{Anderson}}, \citenamefont {{Andrade-Oliveira}}, \citenamefont {{Ansarinejad}}, \citenamefont {{Archipley}}, \citenamefont {{Ashby}}, \citenamefont {{Austermann}}, \citenamefont {{Bacon}}, \citenamefont {{Beall}}, \citenamefont {{Bender}}, \citenamefont {{Benson}}, \citenamefont {{Bianchini}}, \citenamefont {{Bocquet}}, \citenamefont {{Brooks}}, \citenamefont {{Burke}}, \citenamefont {{Calzadilla}}, \citenamefont {{Carlstrom}}, \citenamefont {{Carnero Rosell}}, \citenamefont {{Carretero}}, \citenamefont {{Chang}}, \citenamefont {{Chaubal}}, \citenamefont {{Chiang}}, \citenamefont {{Chou}}, \citenamefont {{Citron}}, \citenamefont {{Corbett Moran}}, \citenamefont {{Costanzi}}, \citenamefont {{Constanzi}}, \citenamefont {{Crawford}}, \citenamefont {{Crites}}, \citenamefont {{da Costa}},
  \citenamefont {{de Haan}}, \citenamefont {{De Vicente}}, \citenamefont {{Desai}}, \citenamefont {{Dobbs}}, \citenamefont {{Doel}}, \citenamefont {{Everett}}, \citenamefont {{Ferrero}}, \citenamefont {{Flaugher}}, \citenamefont {{Floyd}}, \citenamefont {{Friedel}}, \citenamefont {{Frieman}}, \citenamefont {{Gallicchio}}, \citenamefont {{Garc'ia-Bellido}}, \citenamefont {{Gatti}}, \citenamefont {{George}}, \citenamefont {{Giannini}}, \citenamefont {{Grandis}}, \citenamefont {{Gruen}}, \citenamefont {{Gruendl}}, \citenamefont {{Gupta}}, \citenamefont {{Gutierrez}}, \citenamefont {{Halverson}}, \citenamefont {{Hinton}}, \citenamefont {{Hinton}}, \citenamefont {{Holder}}, \citenamefont {{Hollowood}}, \citenamefont {{Holzapfel}}, \citenamefont {{Honscheid}}, \citenamefont {{Hrubes}}, \citenamefont {{Huang}}, \citenamefont {{Hubmayr}}, \citenamefont {{Irwin}}, \citenamefont {{Mena-Fern{\'a}ndez}}, \citenamefont {{James}}, \citenamefont {{K{\'e}ruzor{\'e}}}, \citenamefont {{Knox}}, \citenamefont {{Kuehn}},
  \citenamefont {{Lahav}}, \citenamefont {{Lee}}, \citenamefont {{Lee}}, \citenamefont {{Li}}, \citenamefont {{Lowitz}}, \citenamefont {{Marshal}}, \citenamefont {{McDonald}}, \citenamefont {{McMahon}}, \citenamefont {{Menanteau}}, \citenamefont {{Meyer}}, \citenamefont {{Miquel}}, \citenamefont {{Mohr}}, \citenamefont {{Montgomery}}, \citenamefont {{Myles}}, \citenamefont {{Natoli}}, \citenamefont {{Nibarger}}, \citenamefont {{Noble}}, \citenamefont {{Novosad}}, \citenamefont {{Ogando}}, \citenamefont {{Padin}}, \citenamefont {{Patil}}, \citenamefont {{Pereira}}, \citenamefont {{Pieres}}, \citenamefont {{Plazas Malag'on}}, \citenamefont {{Pryke}}, \citenamefont {{Reichardt}}, \citenamefont {{Rodr'iguez-Monroy}}, \citenamefont {{Romer}}, \citenamefont {{Ruhl}}, \citenamefont {{Saliwanchik}}, \citenamefont {{Salvati}}, \citenamefont {{Sanchez}}, \citenamefont {{Saro}}, \citenamefont {{Schaffer}}, \citenamefont {{Schrabback}}, \citenamefont {{Sevilla-Noarbe}}, \citenamefont {{Sievers}}, \citenamefont
  {{Smecher}}, \citenamefont {{Smith}}, \citenamefont {{Somboonpanyakul}}, \citenamefont {{Stalder}}, \citenamefont {{Stark}}, \citenamefont {{Suchyta}}, \citenamefont {{Swanson}}, \citenamefont {{Tarle}}, \citenamefont {{To}}, \citenamefont {{Tucker}}, \citenamefont {{Veach}}, \citenamefont {{Vieira}}, \citenamefont {{Vincenzi}}, \citenamefont {{Wang}}, \citenamefont {{Weller}}, \citenamefont {{Whitehorn}}, \citenamefont {{Wiseman}}, \citenamefont {{Wu}}, \citenamefont {{Yefremenko}}, \citenamefont {{Zebrowski}},\ and\ \citenamefont {{Zhang}}}]{SPT2024}%
  \BibitemOpen
  \bibfield  {author} {\bibinfo {author} {\bibfnamefont {L.~E.}\ \bibnamefont {{Bleem}}}, \bibnamefont {et~al.},\ }\href {https://doi.org/10.21105/astro.2311.07512} {\bibfield  {journal} {\bibinfo  {journal} {The Open Journal of Astrophysics}\ }\textbf {\bibinfo {volume} {7}},\ \bibinfo {eid} {13} (\bibinfo {year} {2024})}\BibitemShut {NoStop}%
\bibitem [{\citenamefont {{Qu}}\ \emph {et~al.}(2024)\citenamefont {{Qu}}, \citenamefont {{Sherwin}}, \citenamefont {{Madhavacheril}}, \citenamefont {{Han}}, \citenamefont {{Crowley}}, \citenamefont {{Abril-Cabezas}}, \citenamefont {{Ade}}, \citenamefont {{Aiola}}, \citenamefont {{Alford}}, \citenamefont {{Amiri}}, \citenamefont {{Amodeo}}, \citenamefont {{An}}, \citenamefont {{Atkins}}, \citenamefont {{Austermann}}, \citenamefont {{Battaglia}}, \citenamefont {{Battistelli}}, \citenamefont {{Beall}}, \citenamefont {{Bean}}, \citenamefont {{Beringue}}, \citenamefont {{Bhandarkar}}, \citenamefont {{Biermann}}, \citenamefont {{Bolliet}}, \citenamefont {{Bond}}, \citenamefont {{Cai}}, \citenamefont {{Calabrese}}, \citenamefont {{Calafut}}, \citenamefont {{Capalbo}}, \citenamefont {{Carrero}}, \citenamefont {{Carron}}, \citenamefont {{Challinor}}, \citenamefont {{Chesmore}}, \citenamefont {{Cho}}, \citenamefont {{Choi}}, \citenamefont {{Clark}}, \citenamefont {{C{\'o}rdova Rosado}}, \citenamefont {{Cothard}},
  \citenamefont {{Coughlin}}, \citenamefont {{Coulton}}, \citenamefont {{Dalal}}, \citenamefont {{Darwish}}, \citenamefont {{Devlin}}, \citenamefont {{Dicker}}, \citenamefont {{Doze}}, \citenamefont {{Duell}}, \citenamefont {{Duff}}, \citenamefont {{Duivenvoorden}}, \citenamefont {{Dunkley}}, \citenamefont {{D{\"u}nner}}, \citenamefont {{Fanfani}}, \citenamefont {{Fankhanel}}, \citenamefont {{Farren}}, \citenamefont {{Ferraro}}, \citenamefont {{Freundt}}, \citenamefont {{Fuzia}}, \citenamefont {{Gallardo}}, \citenamefont {{Garrido}}, \citenamefont {{Gluscevic}}, \citenamefont {{Golec}}, \citenamefont {{Guan}}, \citenamefont {{Halpern}}, \citenamefont {{Harrison}}, \citenamefont {{Hasselfield}}, \citenamefont {{Healy}}, \citenamefont {{Henderson}}, \citenamefont {{Hensley}}, \citenamefont {{Herv{\'\i}as-Caimapo}}, \citenamefont {{Hill}}, \citenamefont {{Hilton}}, \citenamefont {{Hilton}}, \citenamefont {{Hincks}}, \citenamefont {{Hlo{\v{z}}ek}}, \citenamefont {{Ho}}, \citenamefont {{Huber}}, \citenamefont
  {{Hubmayr}}, \citenamefont {{Huffenberger}}, \citenamefont {{Hughes}}, \citenamefont {{Irwin}}, \citenamefont {{Isopi}}, \citenamefont {{Jense}}, \citenamefont {{Keller}}, \citenamefont {{Kim}}, \citenamefont {{Knowles}}, \citenamefont {{Koopman}}, \citenamefont {{Kosowsky}}, \citenamefont {{Kramer}}, \citenamefont {{Kusiak}}, \citenamefont {{La Posta}}, \citenamefont {{Lague}}, \citenamefont {{Lakey}}, \citenamefont {{Lee}}, \citenamefont {{Li}}, \citenamefont {{Li}}, \citenamefont {{Limon}}, \citenamefont {{Lokken}}, \citenamefont {{Louis}}, \citenamefont {{Lungu}}, \citenamefont {{MacCrann}}, \citenamefont {{MacInnis}}, \citenamefont {{Maldonado}}, \citenamefont {{Maldonado}}, \citenamefont {{Mallaby-Kay}}, \citenamefont {{Marques}}, \citenamefont {{McMahon}}, \citenamefont {{Mehta}}, \citenamefont {{Menanteau}}, \citenamefont {{Moodley}}, \citenamefont {{Morris}}, \citenamefont {{Mroczkowski}}, \citenamefont {{Naess}}, \citenamefont {{Namikawa}}, \citenamefont {{Nati}}, \citenamefont {{Newburgh}},
  \citenamefont {{Nicola}}, \citenamefont {{Niemack}}, \citenamefont {{Nolta}}, \citenamefont {{Orlowski-Scherer}}, \citenamefont {{Page}}, \citenamefont {{Pandey}}, \citenamefont {{Partridge}}, \citenamefont {{Prince}}, \citenamefont {{Puddu}}, \citenamefont {{Radiconi}}, \citenamefont {{Robertson}}, \citenamefont {{Rojas}}, \citenamefont {{Sakuma}}, \citenamefont {{Salatino}}, \citenamefont {{Schaan}}, \citenamefont {{Schmitt}}, \citenamefont {{Sehgal}}, \citenamefont {{Shaikh}}, \citenamefont {{Sierra}}, \citenamefont {{Sievers}}, \citenamefont {{Sif{\'o}n}}, \citenamefont {{Simon}}, \citenamefont {{Sonka}}, \citenamefont {{Spergel}}, \citenamefont {{Staggs}}, \citenamefont {{Storer}}, \citenamefont {{Switzer}}, \citenamefont {{Tampier}}, \citenamefont {{Thornton}}, \citenamefont {{Trac}}, \citenamefont {{Treu}}, \citenamefont {{Tucker}}, \citenamefont {{Ullom}}, \citenamefont {{Vale}}, \citenamefont {{Van Engelen}}, \citenamefont {{Van Lanen}}, \citenamefont {{van Marrewijk}}, \citenamefont {{Vargas}},
  \citenamefont {{Vavagiakis}}, \citenamefont {{Wagoner}}, \citenamefont {{Wang}}, \citenamefont {{Wenzl}}, \citenamefont {{Wollack}}, \citenamefont {{Xu}}, \citenamefont {{Zago}},\ and\ \citenamefont {{Zheng}}}]{ACT2024}%
  \BibitemOpen
  \bibfield  {author} {\bibinfo {author} {\bibfnamefont {F.~J.}\ \bibnamefont {{Qu}}}, \bibnamefont {et~al.},\ }\href {https://doi.org/10.3847/1538-4357/acfe06} {\bibfield  {journal} {\bibinfo  {journal} {\apj}\ }\textbf {\bibinfo {volume} {962}},\ \bibinfo {eid} {112} (\bibinfo {year} {2024})}\BibitemShut {NoStop}%
\bibitem [{\citenamefont {{Simet}}\ \emph {et~al.}(2017)\citenamefont {{Simet}}, \citenamefont {{Battaglia}}, \citenamefont {{Mandelbaum}},\ and\ \citenamefont {{Seljak}}}]{Simet17}%
  \BibitemOpen
  \bibfield  {author} {\bibinfo {author} {\bibfnamefont {M.}~\bibnamefont {{Simet}}}, \bibinfo {author} {\bibfnamefont {N.}~\bibnamefont {{Battaglia}}}, \bibinfo {author} {\bibfnamefont {R.}~\bibnamefont {{Mandelbaum}}},\ \bibnamefont {and}\ \bibinfo {author} {\bibfnamefont {U.}~\bibnamefont {{Seljak}}},\ }\href {https://doi.org/10.1093/mnras/stw3322} {\bibfield  {journal} {\bibinfo  {journal} {\mnras}\ }\textbf {\bibinfo {volume} {466}},\ \bibinfo {pages} {3663} (\bibinfo {year} {2017})}\BibitemShut {NoStop}%
\bibitem [{\citenamefont {{Melchior}}\ \emph {et~al.}(2017)\citenamefont {{Melchior}}, \citenamefont {{Gruen}}, \citenamefont {{McClintock}}, \citenamefont {{Varga}}, \citenamefont {{Sheldon}}, \citenamefont {{Rozo}}, \citenamefont {{Amara}}, \citenamefont {{Becker}}, \citenamefont {{Benson}}, \citenamefont {{Bermeo}}, \citenamefont {{Bridle}}, \citenamefont {{Clampitt}}, \citenamefont {{Dietrich}}, \citenamefont {{Hartley}}, \citenamefont {{Hollowood}}, \citenamefont {{Jain}}, \citenamefont {{Jarvis}}, \citenamefont {{Jeltema}}, \citenamefont {{Kacprzak}}, \citenamefont {{MacCrann}}, \citenamefont {{Rykoff}}, \citenamefont {{Saro}}, \citenamefont {{Suchyta}}, \citenamefont {{Troxel}}, \citenamefont {{Zuntz}}, \citenamefont {{Bonnett}}, \citenamefont {{Plazas}}, \citenamefont {{Abbott}}, \citenamefont {{Abdalla}}, \citenamefont {{Annis}}, \citenamefont {{Benoit-L{\'e}vy}}, \citenamefont {{Bernstein}}, \citenamefont {{Bertin}}, \citenamefont {{Brooks}}, \citenamefont {{Buckley-Geer}}, \citenamefont {{Carnero
  Rosell}}, \citenamefont {{Carrasco Kind}}, \citenamefont {{Carretero}}, \citenamefont {{Cunha}}, \citenamefont {{D'Andrea}}, \citenamefont {{da Costa}}, \citenamefont {{Desai}}, \citenamefont {{Eifler}}, \citenamefont {{Flaugher}}, \citenamefont {{Fosalba}}, \citenamefont {{Garc{\'\i}a-Bellido}}, \citenamefont {{Gaztanaga}}, \citenamefont {{Gerdes}}, \citenamefont {{Gruendl}}, \citenamefont {{Gschwend}}, \citenamefont {{Gutierrez}}, \citenamefont {{Honscheid}}, \citenamefont {{James}}, \citenamefont {{Kirk}}, \citenamefont {{Krause}}, \citenamefont {{Kuehn}}, \citenamefont {{Kuropatkin}}, \citenamefont {{Lahav}}, \citenamefont {{Lima}}, \citenamefont {{Maia}}, \citenamefont {{March}}, \citenamefont {{Martini}}, \citenamefont {{Menanteau}}, \citenamefont {{Miller}}, \citenamefont {{Miquel}}, \citenamefont {{Mohr}}, \citenamefont {{Nichol}}, \citenamefont {{Ogando}}, \citenamefont {{Romer}}, \citenamefont {{Sanchez}}, \citenamefont {{Scarpine}}, \citenamefont {{Sevilla-Noarbe}}, \citenamefont {{Smith}},
  \citenamefont {{Soares-Santos}}, \citenamefont {{Sobreira}}, \citenamefont {{Swanson}}, \citenamefont {{Tarle}}, \citenamefont {{Thomas}}, \citenamefont {{Walker}}, \citenamefont {{Weller}},\ and\ \citenamefont {{Zhang}}}]{Melchior17}%
  \BibitemOpen
  \bibfield  {author} {\bibinfo {author} {\bibfnamefont {P.}~\bibnamefont {{Melchior}}}, \bibnamefont {et~al.},\ }\href {https://doi.org/10.1093/mnras/stx1053} {\bibfield  {journal} {\bibinfo  {journal} {\mnras}\ }\textbf {\bibinfo {volume} {469}},\ \bibinfo {pages} {4899} (\bibinfo {year} {2017})}\BibitemShut {NoStop}%
\bibitem [{\citenamefont {{McClintock}}\ \emph {et~al.}(2019)\citenamefont {{McClintock}}, \citenamefont {{Varga}}, \citenamefont {{Gruen}}, \citenamefont {{Rozo}}, \citenamefont {{Rykoff}}, \citenamefont {{Shin}}, \citenamefont {{Melchior}}, \citenamefont {{DeRose}}, \citenamefont {{Seitz}}, \citenamefont {{Dietrich}}, \citenamefont {{Sheldon}}, \citenamefont {{Zhang}}, \citenamefont {{von der Linden}}, \citenamefont {{Jeltema}}, \citenamefont {{Mantz}}, \citenamefont {{Romer}}, \citenamefont {{Allen}}, \citenamefont {{Becker}}, \citenamefont {{Bermeo}}, \citenamefont {{Bhargava}}, \citenamefont {{Costanzi}}, \citenamefont {{Everett}}, \citenamefont {{Farahi}}, \citenamefont {{Hamaus}}, \citenamefont {{Hartley}}, \citenamefont {{Hollowood}}, \citenamefont {{Hoyle}}, \citenamefont {{Israel}}, \citenamefont {{Li}}, \citenamefont {{MacCrann}}, \citenamefont {{Morris}}, \citenamefont {{Palmese}}, \citenamefont {{Plazas}}, \citenamefont {{Pollina}}, \citenamefont {{Rau}}, \citenamefont {{Simet}}, \citenamefont
  {{Soares-Santos}}, \citenamefont {{Troxel}}, \citenamefont {{Vergara Cervantes}}, \citenamefont {{Wechsler}}, \citenamefont {{Zuntz}}, \citenamefont {{Abbott}}, \citenamefont {{Abdalla}}, \citenamefont {{Allam}}, \citenamefont {{Annis}}, \citenamefont {{Avila}}, \citenamefont {{Bridle}}, \citenamefont {{Brooks}}, \citenamefont {{Burke}}, \citenamefont {{Carnero Rosell}}, \citenamefont {{Carrasco Kind}}, \citenamefont {{Carretero}}, \citenamefont {{Castander}}, \citenamefont {{Crocce}}, \citenamefont {{Cunha}}, \citenamefont {{D'Andrea}}, \citenamefont {{da Costa}}, \citenamefont {{Davis}}, \citenamefont {{De Vicente}}, \citenamefont {{Diehl}}, \citenamefont {{Doel}}, \citenamefont {{Drlica-Wagner}}, \citenamefont {{Evrard}}, \citenamefont {{Flaugher}}, \citenamefont {{Fosalba}}, \citenamefont {{Frieman}}, \citenamefont {{Garc{\'\i}a-Bellido}}, \citenamefont {{Gaztanaga}}, \citenamefont {{Gerdes}}, \citenamefont {{Giannantonio}}, \citenamefont {{Gruendl}}, \citenamefont {{Gutierrez}}, \citenamefont
  {{Honscheid}}, \citenamefont {{James}}, \citenamefont {{Kirk}}, \citenamefont {{Krause}}, \citenamefont {{Kuehn}}, \citenamefont {{Lahav}}, \citenamefont {{Li}}, \citenamefont {{Lima}}, \citenamefont {{March}}, \citenamefont {{Marshall}}, \citenamefont {{Menanteau}}, \citenamefont {{Miquel}}, \citenamefont {{Mohr}}, \citenamefont {{Nord}}, \citenamefont {{Ogando}}, \citenamefont {{Roodman}}, \citenamefont {{Sanchez}}, \citenamefont {{Scarpine}}, \citenamefont {{Schindler}}, \citenamefont {{Sevilla-Noarbe}}, \citenamefont {{Smith}}, \citenamefont {{Smith}}, \citenamefont {{Sobreira}}, \citenamefont {{Suchyta}}, \citenamefont {{Swanson}}, \citenamefont {{Tarle}}, \citenamefont {{Tucker}}, \citenamefont {{Vikram}}, \citenamefont {{Walker}}, \citenamefont {{Weller}},\ and\ \citenamefont {{DES Collaboration}}}]{DES2019a}%
  \BibitemOpen
  \bibfield  {author} {\bibinfo {author} {\bibfnamefont {T.}~\bibnamefont {{McClintock}}}, \bibnamefont {et~al.},\ }\href {https://doi.org/10.1093/mnras/sty2711} {\bibfield  {journal} {\bibinfo  {journal} {\mnras}\ }\textbf {\bibinfo {volume} {482}},\ \bibinfo {pages} {1352} (\bibinfo {year} {2019})}\BibitemShut {NoStop}%
\bibitem [{\citenamefont {{Cohn}}\ \emph {et~al.}(2007)\citenamefont {{Cohn}}, \citenamefont {{Evrard}}, \citenamefont {{White}}, \citenamefont {{Croton}},\ and\ \citenamefont {{Ellingson}}}]{Cohnetal2007}%
  \BibitemOpen
  \bibfield  {author} {\bibinfo {author} {\bibfnamefont {J.~D.}\ \bibnamefont {{Cohn}}}, \bibinfo {author} {\bibfnamefont {A.~E.}\ \bibnamefont {{Evrard}}}, \bibinfo {author} {\bibfnamefont {M.}~\bibnamefont {{White}}}, \bibinfo {author} {\bibfnamefont {D.}~\bibnamefont {{Croton}}},\ \bibnamefont {and}\ \bibinfo {author} {\bibfnamefont {E.}~\bibnamefont {{Ellingson}}},\ }\href {https://doi.org/10.1111/j.1365-2966.2007.12479.x} {\bibfield  {journal} {\bibinfo  {journal} {\mnras}\ }\textbf {\bibinfo {volume} {382}},\ \bibinfo {pages} {1738} (\bibinfo {year} {2007})}\BibitemShut {NoStop}%
\bibitem [{\citenamefont {{Farahi}}\ \emph {et~al.}(2016)\citenamefont {{Farahi}}, \citenamefont {{Evrard}}, \citenamefont {{Rozo}}, \citenamefont {{Rykoff}},\ and\ \citenamefont {{Wechsler}}}]{Farahietal2016}%
  \BibitemOpen
  \bibfield  {author} {\bibinfo {author} {\bibfnamefont {A.}~\bibnamefont {{Farahi}}}, \bibinfo {author} {\bibfnamefont {A.~E.}\ \bibnamefont {{Evrard}}}, \bibinfo {author} {\bibfnamefont {E.}~\bibnamefont {{Rozo}}}, \bibinfo {author} {\bibfnamefont {E.~S.}\ \bibnamefont {{Rykoff}}},\ \bibnamefont {and}\ \bibinfo {author} {\bibfnamefont {R.~H.}\ \bibnamefont {{Wechsler}}},\ }\href {https://doi.org/10.1093/mnras/stw1143} {\bibfield  {journal} {\bibinfo  {journal} {\mnras}\ }\textbf {\bibinfo {volume} {460}},\ \bibinfo {pages} {3900} (\bibinfo {year} {2016})}\BibitemShut {NoStop}%
\bibitem [{\citenamefont {{Busch}}\ and\ \citenamefont {{White}}(2017)}]{BuschWhite2017}%
  \BibitemOpen
  \bibfield  {author} {\bibinfo {author} {\bibfnamefont {P.}~\bibnamefont {{Busch}}}\ \bibnamefont {and}\ \bibinfo {author} {\bibfnamefont {S.~D.~M.}\ \bibnamefont {{White}}},\ }\href {https://doi.org/10.1093/mnras/stx1584} {\bibfield  {journal} {\bibinfo  {journal} {\mnras}\ }\textbf {\bibinfo {volume} {470}},\ \bibinfo {pages} {4767} (\bibinfo {year} {2017})}\BibitemShut {NoStop}%
\bibitem [{\citenamefont {{Zu}}\ \emph {et~al.}(2017)\citenamefont {{Zu}}, \citenamefont {{Mandelbaum}}, \citenamefont {{Simet}}, \citenamefont {{Rozo}},\ and\ \citenamefont {{Rykoff}}}]{Zuetal2017}%
  \BibitemOpen
  \bibfield  {author} {\bibinfo {author} {\bibfnamefont {Y.}~\bibnamefont {{Zu}}}, \bibinfo {author} {\bibfnamefont {R.}~\bibnamefont {{Mandelbaum}}}, \bibinfo {author} {\bibfnamefont {M.}~\bibnamefont {{Simet}}}, \bibinfo {author} {\bibfnamefont {E.}~\bibnamefont {{Rozo}}},\ \bibnamefont {and}\ \bibinfo {author} {\bibfnamefont {E.~S.}\ \bibnamefont {{Rykoff}}},\ }\href {https://doi.org/10.1093/mnras/stx1264} {\bibfield  {journal} {\bibinfo  {journal} {\mnras}\ }\textbf {\bibinfo {volume} {470}},\ \bibinfo {pages} {551} (\bibinfo {year} {2017})}\BibitemShut {NoStop}%
\bibitem [{\citenamefont {{Costanzi}}\ \emph {et~al.}(2019{\natexlab{a}})\citenamefont {{Costanzi}}, \citenamefont {{Rozo}}, \citenamefont {{Rykoff}}, \citenamefont {{Farahi}}, \citenamefont {{Jeltema}}, \citenamefont {{Evrard}}, \citenamefont {{Mantz}}, \citenamefont {{Gruen}}, \citenamefont {{Mandelbaum}}, \citenamefont {{DeRose}}, \citenamefont {{McClintock}}, \citenamefont {{Varga}}, \citenamefont {{Zhang}}, \citenamefont {{Weller}}, \citenamefont {{Wechsler}},\ and\ \citenamefont {{Aguena}}}]{Costanzietal2019}%
  \BibitemOpen
  \bibfield  {author} {\bibinfo {author} {\bibfnamefont {M.}~\bibnamefont {{Costanzi}}}, \bibnamefont {et~al.},\ }\href {https://doi.org/10.1093/mnras/sty2665} {\bibfield  {journal} {\bibinfo  {journal} {\mnras}\ }\textbf {\bibinfo {volume} {482}},\ \bibinfo {pages} {490} (\bibinfo {year} {2019}{\natexlab{a}})}\BibitemShut {NoStop}%
\bibitem [{\citenamefont {{Sunayama}}\ \emph {et~al.}(2020)\citenamefont {{Sunayama}}, \citenamefont {{Park}}, \citenamefont {{Takada}}, \citenamefont {{Kobayashi}}, \citenamefont {{Nishimichi}}, \citenamefont {{Kurita}}, \citenamefont {{More}}, \citenamefont {{Oguri}},\ and\ \citenamefont {{Osato}}}]{Sunayamaetal2020}%
  \BibitemOpen
  \bibfield  {author} {\bibinfo {author} {\bibfnamefont {T.}~\bibnamefont {{Sunayama}}}, \bibinfo {author} {\bibfnamefont {Y.}~\bibnamefont {{Park}}}, \bibinfo {author} {\bibfnamefont {M.}~\bibnamefont {{Takada}}}, \bibinfo {author} {\bibfnamefont {Y.}~\bibnamefont {{Kobayashi}}}, \bibinfo {author} {\bibfnamefont {T.}~\bibnamefont {{Nishimichi}}}, \bibinfo {author} {\bibfnamefont {T.}~\bibnamefont {{Kurita}}}, \bibinfo {author} {\bibfnamefont {S.}~\bibnamefont {{More}}}, \bibinfo {author} {\bibfnamefont {M.}~\bibnamefont {{Oguri}}},\ \bibnamefont {and}\ \bibinfo {author} {\bibfnamefont {K.}~\bibnamefont {{Osato}}},\ }\href {https://doi.org/10.1093/mnras/staa1646} {\bibfield  {journal} {\bibinfo  {journal} {\mnras}\ }\textbf {\bibinfo {volume} {496}},\ \bibinfo {pages} {4468} (\bibinfo {year} {2020})}\BibitemShut {NoStop}%
\bibitem [{\citenamefont {{Wu}}\ \emph {et~al.}(2022)\citenamefont {{Wu}}, \citenamefont {{Costanzi}}, \citenamefont {{To}}, \citenamefont {{Salcedo}}, \citenamefont {{Weinberg}}, \citenamefont {{Annis}}, \citenamefont {{Bocquet}}, \citenamefont {{da Silva Pereira}}, \citenamefont {{DeRose}}, \citenamefont {{Esteves}}, \citenamefont {{Farahi}}, \citenamefont {{Grandis}}, \citenamefont {{Rozo}}, \citenamefont {{Rykoff}}, \citenamefont {{Varga}}, \citenamefont {{Wechsler}}, \citenamefont {{Zeng}}, \citenamefont {{Zhang}}, \citenamefont {{Zhang}},\ and\ \citenamefont {{DES Collaboration}}}]{Wuetal2022}%
  \BibitemOpen
  \bibfield  {author} {\bibinfo {author} {\bibfnamefont {H.-Y.}\ \bibnamefont {{Wu}}}, \bibnamefont {et~al.},\ }\href {https://doi.org/10.1093/mnras/stac2048} {\bibfield  {journal} {\bibinfo  {journal} {\mnras}\ }\textbf {\bibinfo {volume} {515}},\ \bibinfo {pages} {4471} (\bibinfo {year} {2022})}\BibitemShut {NoStop}%
\bibitem [{\citenamefont {{Sunayama}}\ \emph {et~al.}(2024)\citenamefont {{Sunayama}}, \citenamefont {{Miyatake}}, \citenamefont {{Sugiyama}}, \citenamefont {{More}}, \citenamefont {{Li}}, \citenamefont {{Dalal}}, \citenamefont {{Rau}}, \citenamefont {{Shi}}, \citenamefont {{Chiu}}, \citenamefont {{Shirasaki}}, \citenamefont {{Zhang}},\ and\ \citenamefont {{Nishizawa}}}]{Sunayamaetal2024}%
  \BibitemOpen
  \bibfield  {author} {\bibinfo {author} {\bibfnamefont {T.}~\bibnamefont {{Sunayama}}}, \bibnamefont {et~al.},\ }\href {https://doi.org/10.1103/PhysRevD.110.083511} {\bibfield  {journal} {\bibinfo  {journal} {\prd}\ }\textbf {\bibinfo {volume} {110}},\ \bibinfo {eid} {083511} (\bibinfo {year} {2024})}\BibitemShut {NoStop}%
\bibitem [{\citenamefont {{Euclid Collaboration}}\ \emph {et~al.}(2025)\citenamefont {{Euclid Collaboration}}, \citenamefont {{Ragagnin}}, \citenamefont {{Saro}}, \citenamefont {{Andreon}}, \citenamefont {{Biviano}}, \citenamefont {{Dolag}}, \citenamefont {{Ettori}}, \citenamefont {{Giocoli}}, \citenamefont {{Le Brun}}, \citenamefont {{Mamon}}, \citenamefont {{Maughan}}, \citenamefont {{Meneghetti}}, \citenamefont {{Moscardini}}, \citenamefont {{Pacaud}}, \citenamefont {{Pratt}}, \citenamefont {{Sereno}}, \citenamefont {{Borgani}}, \citenamefont {{Calura}}, \citenamefont {{Castignani}}, \citenamefont {{De Petris}}, \citenamefont {{Eckert}}, \citenamefont {{Lesci}}, \citenamefont {{Macias-Perez}}, \citenamefont {{Maturi}}, \citenamefont {{Amara}}, \citenamefont {{Auricchio}}, \citenamefont {{Baccigalupi}}, \citenamefont {{Baldi}}, \citenamefont {{Bardelli}}, \citenamefont {{Bonino}}, \citenamefont {{Branchini}}, \citenamefont {{Brescia}}, \citenamefont {{Brinchmann}}, \citenamefont {{Camera}}, \citenamefont
  {{Capobianco}}, \citenamefont {{Carbone}}, \citenamefont {{Carretero}}, \citenamefont {{Casas}}, \citenamefont {{Castellano}}, \citenamefont {{Cavuoti}}, \citenamefont {{Cimatti}}, \citenamefont {{Colodro-Conde}}, \citenamefont {{Congedo}}, \citenamefont {{Conselice}}, \citenamefont {{Conversi}}, \citenamefont {{Copin}}, \citenamefont {{Courbin}}, \citenamefont {{Courtois}}, \citenamefont {{Da Silva}}, \citenamefont {{Degaudenzi}}, \citenamefont {{De Lucia}}, \citenamefont {{Dinis}}, \citenamefont {{Dubath}}, \citenamefont {{Dupac}}, \citenamefont {{Farina}}, \citenamefont {{Farrens}}, \citenamefont {{Ferriol}}, \citenamefont {{Frailis}}, \citenamefont {{Franceschi}}, \citenamefont {{Fumana}}, \citenamefont {{George}}, \citenamefont {{Gillis}}, \citenamefont {{Grazian}}, \citenamefont {{Grupp}}, \citenamefont {{Haugan}}, \citenamefont {{Holmes}}, \citenamefont {{Hook}}, \citenamefont {{Hormuth}}, \citenamefont {{Hornstrup}}, \citenamefont {{Jahnke}}, \citenamefont {{Keih{\"a}nen}}, \citenamefont
  {{Kermiche}}, \citenamefont {{Kiessling}}, \citenamefont {{Kilbinger}}, \citenamefont {{Kubik}}, \citenamefont {{K{\"u}mmel}}, \citenamefont {{Kunz}}, \citenamefont {{Kurki-Suonio}}, \citenamefont {{Ligori}}, \citenamefont {{Lilje}}, \citenamefont {{Lindholm}}, \citenamefont {{Lloro}}, \citenamefont {{Maino}}, \citenamefont {{Maiorano}}, \citenamefont {{Mansutti}}, \citenamefont {{Marggraf}}, \citenamefont {{Markovic}}, \citenamefont {{Martinelli}}, \citenamefont {{Martinet}}, \citenamefont {{Marulli}}, \citenamefont {{Massey}}, \citenamefont {{Maurogordato}}, \citenamefont {{Medinaceli}}, \citenamefont {{Mei}}, \citenamefont {{Mellier}}, \citenamefont {{Meylan}}, \citenamefont {{Moresco}}, \citenamefont {{Munari}}, \citenamefont {{Neissner}}, \citenamefont {{Niemi}}, \citenamefont {{Nightingale}}, \citenamefont {{Padilla}}, \citenamefont {{Paltani}}, \citenamefont {{Pasian}}, \citenamefont {{Pedersen}}, \citenamefont {{Pettorino}}, \citenamefont {{Polenta}}, \citenamefont {{Poncet}}, \citenamefont
  {{Popa}}, \citenamefont {{Pozzetti}}, \citenamefont {{Raison}}, \citenamefont {{Renzi}}, \citenamefont {{Rhodes}}, \citenamefont {{Riccio}}, \citenamefont {{Romelli}}, \citenamefont {{Roncarelli}}, \citenamefont {{Rossetti}}, \citenamefont {{Saglia}}, \citenamefont {{Sakr}}, \citenamefont {{S{\'a}nchez}}, \citenamefont {{Sapone}}, \citenamefont {{Sartoris}}, \citenamefont {{Scaramella}}, \citenamefont {{Schneider}}, \citenamefont {{Schrabback}}, \citenamefont {{Secroun}}, \citenamefont {{Sefusatti}}, \citenamefont {{Seidel}}, \citenamefont {{Serrano}}, \citenamefont {{Sirignano}}, \citenamefont {{Sirri}}, \citenamefont {{Stanco}}, \citenamefont {{Steinwagner}}, \citenamefont {{Tallada-Cresp{\'\i}}}, \citenamefont {{Tereno}}, \citenamefont {{Toledo-Moreo}}, \citenamefont {{Torradeflot}}, \citenamefont {{Tutusaus}}, \citenamefont {{Valenziano}}, \citenamefont {{Vassallo}}, \citenamefont {{Verdoes Kleijn}}, \citenamefont {{Veropalumbo}}, \citenamefont {{Wang}}, \citenamefont {{Weller}}, \citenamefont
  {{Zamorani}}, \citenamefont {{Zucca}}, \citenamefont {{Bolzonella}}, \citenamefont {{Boucaud}}, \citenamefont {{Bozzo}}, \citenamefont {{Burigana}}, \citenamefont {{Calabrese}}, \citenamefont {{Di Ferdinando}}, \citenamefont {{Escartin Vigo}}, \citenamefont {{Farinelli}}, \citenamefont {{Gracia-Carpio}}, \citenamefont {{Mauri}}, \citenamefont {{Scottez}}, \citenamefont {{Tenti}}, \citenamefont {{Viel}}, \citenamefont {{Wiesmann}}, \citenamefont {{Akrami}}, \citenamefont {{Allevato}}, \citenamefont {{Anselmi}}, \citenamefont {{Ballardini}}, \citenamefont {{Bergamini}}, \citenamefont {{Blanchard}}, \citenamefont {{Blot}}, \citenamefont {{Bruton}}, \citenamefont {{Cabanac}}, \citenamefont {{Calabro}}, \citenamefont {{Canas-Herrera}}, \citenamefont {{Cappi}}, \citenamefont {{Carvalho}}, \citenamefont {{Castro}}, \citenamefont {{Chambers}}, \citenamefont {{Contarini}}, \citenamefont {{Cooray}}, \citenamefont {{Costanzi}}, \citenamefont {{De Caro}}, \citenamefont {{de la Torre}}, \citenamefont {{Desprez}},
  \citenamefont {{D{\'\i}az-S{\'a}nchez}}, \citenamefont {{Di Domizio}}, \citenamefont {{Dole}}, \citenamefont {{Escoffier}}, \citenamefont {{Ferrari}}, \citenamefont {{Ferreira}}, \citenamefont {{Ferrero}}, \citenamefont {{Finelli}}, \citenamefont {{Fornari}}, \citenamefont {{Gabarra}}, \citenamefont {{Ganga}}, \citenamefont {{Garc{\'\i}a-Bellido}}, \citenamefont {{Gaztanaga}}, \citenamefont {{Giacomini}}, \citenamefont {{Gozaliasl}}, \citenamefont {{Hall}}, \citenamefont {{Hildebrandt}}, \citenamefont {{Hjorth}},\ and\ \citenamefont {{Jimenez Mu{\~n}oz}}}]{EuclidProj2025}%
  \BibitemOpen
  \bibfield  {author} {\bibinfo {author} {\bibnamefont {{Euclid Collaboration}}}, \bibnamefont {et~al.},\ }\href {https://doi.org/10.1051/0004-6361/202451347} {\bibfield  {journal} {\bibinfo  {journal} {\aap}\ }\textbf {\bibinfo {volume} {695}},\ \bibinfo {eid} {A282} (\bibinfo {year} {2025})}\BibitemShut {NoStop}%
\bibitem [{\citenamefont {{Rozo}}\ \emph {et~al.}(2015)\citenamefont {{Rozo}}, \citenamefont {{Rykoff}}, \citenamefont {{Becker}}, \citenamefont {{Reddick}},\ and\ \citenamefont {{Wechsler}}}]{Rozoetal2015}%
  \BibitemOpen
  \bibfield  {author} {\bibinfo {author} {\bibfnamefont {E.}~\bibnamefont {{Rozo}}}, \bibinfo {author} {\bibfnamefont {E.~S.}\ \bibnamefont {{Rykoff}}}, \bibinfo {author} {\bibfnamefont {M.}~\bibnamefont {{Becker}}}, \bibinfo {author} {\bibfnamefont {R.~M.}\ \bibnamefont {{Reddick}}},\ \bibnamefont {and}\ \bibinfo {author} {\bibfnamefont {R.~H.}\ \bibnamefont {{Wechsler}}},\ }\href {https://doi.org/10.1093/mnras/stv1560} {\bibfield  {journal} {\bibinfo  {journal} {\mnras}\ }\textbf {\bibinfo {volume} {453}},\ \bibinfo {pages} {38} (\bibinfo {year} {2015})}\BibitemShut {NoStop}%
\bibitem [{\citenamefont {{Evrard}}\ \emph {et~al.}(2014)\citenamefont {{Evrard}}, \citenamefont {{Arnault}}, \citenamefont {{Huterer}},\ and\ \citenamefont {{Farahi}}}]{Evrard14}%
  \BibitemOpen
  \bibfield  {author} {\bibinfo {author} {\bibfnamefont {A.~E.}\ \bibnamefont {{Evrard}}}, \bibinfo {author} {\bibfnamefont {P.}~\bibnamefont {{Arnault}}}, \bibinfo {author} {\bibfnamefont {D.}~\bibnamefont {{Huterer}}},\ \bibnamefont {and}\ \bibinfo {author} {\bibfnamefont {A.}~\bibnamefont {{Farahi}}},\ }\href {https://doi.org/10.1093/mnras/stu784} {\bibfield  {journal} {\bibinfo  {journal} {\mnras}\ }\textbf {\bibinfo {volume} {441}},\ \bibinfo {pages} {3562} (\bibinfo {year} {2014})}\BibitemShut {NoStop}%
\bibitem [{\citenamefont {{Becker}}\ and\ \citenamefont {{Kravtsov}}(2011)}]{BeckerKravtsov2011}%
  \BibitemOpen
  \bibfield  {author} {\bibinfo {author} {\bibfnamefont {M.~R.}\ \bibnamefont {{Becker}}}\ \bibnamefont {and}\ \bibinfo {author} {\bibfnamefont {A.~V.}\ \bibnamefont {{Kravtsov}}},\ }\href {https://doi.org/10.1088/0004-637X/740/1/25} {\bibfield  {journal} {\bibinfo  {journal} {\apj}\ }\textbf {\bibinfo {volume} {740}},\ \bibinfo {eid} {25} (\bibinfo {year} {2011})}\BibitemShut {NoStop}%
\bibitem [{\citenamefont {{Dietrich}}\ \emph {et~al.}(2019)\citenamefont {{Dietrich}}, \citenamefont {{Bocquet}}, \citenamefont {{Schrabback}}, \citenamefont {{Applegate}}, \citenamefont {{Hoekstra}}, \citenamefont {{Grandis}}, \citenamefont {{Mohr}}, \citenamefont {{Allen}}, \citenamefont {{Bayliss}}, \citenamefont {{Benson}}, \citenamefont {{Bleem}}, \citenamefont {{Brodwin}}, \citenamefont {{Bulbul}}, \citenamefont {{Capasso}}, \citenamefont {{Chiu}}, \citenamefont {{Crawford}}, \citenamefont {{Gonzalez}}, \citenamefont {{de Haan}}, \citenamefont {{Klein}}, \citenamefont {{von der Linden}}, \citenamefont {{Mantz}}, \citenamefont {{Marrone}}, \citenamefont {{McDonald}}, \citenamefont {{Raghunathan}}, \citenamefont {{Rapetti}}, \citenamefont {{Reichardt}}, \citenamefont {{Saro}}, \citenamefont {{Stalder}}, \citenamefont {{Stark}}, \citenamefont {{Stern}},\ and\ \citenamefont {{Stubbs}}}]{Dietrichetal2019}%
  \BibitemOpen
  \bibfield  {author} {\bibinfo {author} {\bibfnamefont {J.~P.}\ \bibnamefont {{Dietrich}}}, \bibnamefont {et~al.},\ }\href {https://doi.org/10.1093/mnras/sty3088} {\bibfield  {journal} {\bibinfo  {journal} {\mnras}\ }\textbf {\bibinfo {volume} {483}},\ \bibinfo {pages} {2871} (\bibinfo {year} {2019})}\BibitemShut {NoStop}%
\bibitem [{\citenamefont {{Grandis}}\ \emph {et~al.}(2021{\natexlab{a}})\citenamefont {{Grandis}}, \citenamefont {{Mohr}}, \citenamefont {{Costanzi}}, \citenamefont {{Saro}}, \citenamefont {{Bocquet}}, \citenamefont {{Klein}}, \citenamefont {{Aguena}}, \citenamefont {{Allam}}, \citenamefont {{Annis}}, \citenamefont {{Ansarinejad}}, \citenamefont {{Bacon}}, \citenamefont {{Bertin}}, \citenamefont {{Bleem}}, \citenamefont {{Brooks}}, \citenamefont {{Burke}}, \citenamefont {{Carnero Rosel}}, \citenamefont {{Carrasco Kind}}, \citenamefont {{Carretero}}, \citenamefont {{Castander}}, \citenamefont {{Choi}}, \citenamefont {{da Costa}}, \citenamefont {{De Vincente}}, \citenamefont {{Desai}}, \citenamefont {{Diehl}}, \citenamefont {{Dietrich}}, \citenamefont {{Doel}}, \citenamefont {{Eifler}}, \citenamefont {{Everett}}, \citenamefont {{Ferrero}}, \citenamefont {{Floyd}}, \citenamefont {{Fosalba}}, \citenamefont {{Frieman}}, \citenamefont {{Garc{\'\i}a-Bellido}}, \citenamefont {{Gaztanaga}}, \citenamefont {{Gruen}},
  \citenamefont {{Gruendl}}, \citenamefont {{Gschwend}}, \citenamefont {{Gupta}}, \citenamefont {{Gutierrez}}, \citenamefont {{Hinton}}, \citenamefont {{Hollowood}}, \citenamefont {{Honscheid}}, \citenamefont {{James}}, \citenamefont {{Jeltema}}, \citenamefont {{Kuehn}}, \citenamefont {{Lahav}}, \citenamefont {{Lidman}}, \citenamefont {{Lima}}, \citenamefont {{Maia}}, \citenamefont {{March}}, \citenamefont {{Marshall}}, \citenamefont {{Melchior}}, \citenamefont {{Menanteau}}, \citenamefont {{Miquel}}, \citenamefont {{Morgan}}, \citenamefont {{Myles}}, \citenamefont {{Ogando}}, \citenamefont {{Palmese}}, \citenamefont {{Paz-Chinch{\'o}n}}, \citenamefont {{Plazas}}, \citenamefont {{Reichardt}}, \citenamefont {{Romer}}, \citenamefont {{Sanchez}}, \citenamefont {{Scarpine}}, \citenamefont {{Serrano}}, \citenamefont {{Sevilla-Noarbe}}, \citenamefont {{Singh}}, \citenamefont {{Smith}}, \citenamefont {{Suchyta}}, \citenamefont {{Swanson}}, \citenamefont {{Tarle}}, \citenamefont {{Thomas}}, \citenamefont {{To}},
  \citenamefont {{Weller}}, \citenamefont {{Wilkinson}},\ and\ \citenamefont {{Wu}}}]{Grandisetal2021a}%
  \BibitemOpen
  \bibfield  {author} {\bibinfo {author} {\bibfnamefont {S.}~\bibnamefont {{Grandis}}}, \bibnamefont {et~al.},\ }\href {https://doi.org/10.1093/mnras/stab869} {\bibfield  {journal} {\bibinfo  {journal} {\mnras}\ }\textbf {\bibinfo {volume} {504}},\ \bibinfo {pages} {1253} (\bibinfo {year} {2021}{\natexlab{a}})}\BibitemShut {NoStop}%
\bibitem [{\citenamefont {{Schrabback}}\ \emph {et~al.}(2021)\citenamefont {{Schrabback}}, \citenamefont {{Bocquet}}, \citenamefont {{Sommer}}, \citenamefont {{Zohren}}, \citenamefont {{van den Busch}}, \citenamefont {{Hern{\'a}ndez-Mart{\'\i}n}}, \citenamefont {{Hoekstra}}, \citenamefont {{Raihan}}, \citenamefont {{Schirmer}}, \citenamefont {{Applegate}}, \citenamefont {{Bayliss}}, \citenamefont {{Benson}}, \citenamefont {{Bleem}}, \citenamefont {{Dietrich}}, \citenamefont {{Floyd}}, \citenamefont {{Hilbert}}, \citenamefont {{Hlavacek-Larrondo}}, \citenamefont {{McDonald}}, \citenamefont {{Saro}}, \citenamefont {{Stark}},\ and\ \citenamefont {{Weissgerber}}}]{Schrabbacketal2021}%
  \BibitemOpen
  \bibfield  {author} {\bibinfo {author} {\bibfnamefont {T.}~\bibnamefont {{Schrabback}}}, \bibnamefont {et~al.},\ }\href {https://doi.org/10.1093/mnras/stab1386} {\bibfield  {journal} {\bibinfo  {journal} {\mnras}\ }\textbf {\bibinfo {volume} {505}},\ \bibinfo {pages} {3923} (\bibinfo {year} {2021})}\BibitemShut {NoStop}%
\bibitem [{\citenamefont {{Sommer}}\ \emph {et~al.}(2022)\citenamefont {{Sommer}}, \citenamefont {{Schrabback}}, \citenamefont {{Applegate}}, \citenamefont {{Hilbert}}, \citenamefont {{Ansarinejad}}, \citenamefont {{Floyd}},\ and\ \citenamefont {{Grandis}}}]{Sommeretal2022}%
  \BibitemOpen
  \bibfield  {author} {\bibinfo {author} {\bibfnamefont {M.~W.}\ \bibnamefont {{Sommer}}}, \bibinfo {author} {\bibfnamefont {T.}~\bibnamefont {{Schrabback}}}, \bibinfo {author} {\bibfnamefont {D.~E.}\ \bibnamefont {{Applegate}}}, \bibinfo {author} {\bibfnamefont {S.}~\bibnamefont {{Hilbert}}}, \bibinfo {author} {\bibfnamefont {B.}~\bibnamefont {{Ansarinejad}}}, \bibinfo {author} {\bibfnamefont {B.}~\bibnamefont {{Floyd}}},\ \bibnamefont {and}\ \bibinfo {author} {\bibfnamefont {S.}~\bibnamefont {{Grandis}}},\ }\href {https://doi.org/10.1093/mnras/stab3052} {\bibfield  {journal} {\bibinfo  {journal} {\mnras}\ }\textbf {\bibinfo {volume} {509}},\ \bibinfo {pages} {1127} (\bibinfo {year} {2022})}\BibitemShut {NoStop}%
\bibitem [{\citenamefont {{To}}\ \emph {et~al.}(2024)\citenamefont {{To}}, \citenamefont {{DeRose}}, \citenamefont {{Wechsler}}, \citenamefont {{Rykoff}}, \citenamefont {{Wu}}, \citenamefont {{Adhikari}}, \citenamefont {{Krause}}, \citenamefont {{Rozo}},\ and\ \citenamefont {{Weinberg}}}]{Toetal2024}%
  \BibitemOpen
  \bibfield  {author} {\bibinfo {author} {\bibfnamefont {C.-H.}\ \bibnamefont {{To}}}, \bibinfo {author} {\bibfnamefont {J.}~\bibnamefont {{DeRose}}}, \bibinfo {author} {\bibfnamefont {R.~H.}\ \bibnamefont {{Wechsler}}}, \bibinfo {author} {\bibfnamefont {E.}~\bibnamefont {{Rykoff}}}, \bibinfo {author} {\bibfnamefont {H.-Y.}\ \bibnamefont {{Wu}}}, \bibinfo {author} {\bibfnamefont {S.}~\bibnamefont {{Adhikari}}}, \bibinfo {author} {\bibfnamefont {E.}~\bibnamefont {{Krause}}}, \bibinfo {author} {\bibfnamefont {E.}~\bibnamefont {{Rozo}}},\ \bibnamefont {and}\ \bibinfo {author} {\bibfnamefont {D.~H.}\ \bibnamefont {{Weinberg}}},\ }\href {https://doi.org/10.3847/1538-4357/ad0e61} {\bibfield  {journal} {\bibinfo  {journal} {\apj}\ }\textbf {\bibinfo {volume} {961}},\ \bibinfo {eid} {59} (\bibinfo {year} {2024})}\BibitemShut {NoStop}%
\bibitem [{\citenamefont {{Rykoff}}\ \emph {et~al.}(2016)\citenamefont {{Rykoff}}, \citenamefont {{Rozo}}, \citenamefont {{Hollowood}}, \citenamefont {{Bermeo-Hernandez}}, \citenamefont {{Jeltema}}, \citenamefont {{Mayers}}, \citenamefont {{Romer}}, \citenamefont {{Rooney}}, \citenamefont {{Saro}}, \citenamefont {{Vergara Cervantes}}, \citenamefont {{Wechsler}}, \citenamefont {{Wilcox}}, \citenamefont {{Abbott}}, \citenamefont {{Abdalla}}, \citenamefont {{Allam}}, \citenamefont {{Annis}}, \citenamefont {{Benoit-L{\'e}vy}}, \citenamefont {{Bernstein}}, \citenamefont {{Bertin}}, \citenamefont {{Brooks}}, \citenamefont {{Burke}}, \citenamefont {{Capozzi}}, \citenamefont {{Carnero Rosell}}, \citenamefont {{Carrasco Kind}}, \citenamefont {{Castander}}, \citenamefont {{Childress}}, \citenamefont {{Collins}}, \citenamefont {{Cunha}}, \citenamefont {{D'Andrea}}, \citenamefont {{da Costa}}, \citenamefont {{Davis}}, \citenamefont {{Desai}}, \citenamefont {{Diehl}}, \citenamefont {{Dietrich}}, \citenamefont {{Doel}},
  \citenamefont {{Evrard}}, \citenamefont {{Finley}}, \citenamefont {{Flaugher}}, \citenamefont {{Fosalba}}, \citenamefont {{Frieman}}, \citenamefont {{Glazebrook}}, \citenamefont {{Goldstein}}, \citenamefont {{Gruen}}, \citenamefont {{Gruendl}}, \citenamefont {{Gutierrez}}, \citenamefont {{Hilton}}, \citenamefont {{Honscheid}}, \citenamefont {{Hoyle}}, \citenamefont {{James}}, \citenamefont {{Kay}}, \citenamefont {{Kuehn}}, \citenamefont {{Kuropatkin}}, \citenamefont {{Lahav}}, \citenamefont {{Lewis}}, \citenamefont {{Lidman}}, \citenamefont {{Lima}}, \citenamefont {{Maia}}, \citenamefont {{Mann}}, \citenamefont {{Marshall}}, \citenamefont {{Martini}}, \citenamefont {{Melchior}}, \citenamefont {{Miller}}, \citenamefont {{Miquel}}, \citenamefont {{Mohr}}, \citenamefont {{Nichol}}, \citenamefont {{Nord}}, \citenamefont {{Ogando}}, \citenamefont {{Plazas}}, \citenamefont {{Reil}}, \citenamefont {{Sahl{\'e}n}}, \citenamefont {{Sanchez}}, \citenamefont {{Santiago}}, \citenamefont {{Scarpine}}, \citenamefont
  {{Schubnell}}, \citenamefont {{Sevilla-Noarbe}}, \citenamefont {{Smith}}, \citenamefont {{Soares-Santos}}, \citenamefont {{Sobreira}}, \citenamefont {{Stott}}, \citenamefont {{Suchyta}}, \citenamefont {{Swanson}}, \citenamefont {{Tarle}}, \citenamefont {{Thomas}}, \citenamefont {{Tucker}}, \citenamefont {{Uddin}}, \citenamefont {{Viana}}, \citenamefont {{Vikram}}, \citenamefont {{Walker}}, \citenamefont {{Zhang}},\ and\ \citenamefont {{DES Collaboration}}}]{DES2016}%
  \BibitemOpen
  \bibfield  {author} {\bibinfo {author} {\bibfnamefont {E.~S.}\ \bibnamefont {{Rykoff}}}, \bibnamefont {et~al.},\ }\href {https://doi.org/10.3847/0067-0049/224/1/1} {\bibfield  {journal} {\bibinfo  {journal} {\apjs}\ }\textbf {\bibinfo {volume} {224}},\ \bibinfo {eid} {1} (\bibinfo {year} {2016})}\BibitemShut {NoStop}%
\bibitem [{\citenamefont {{Croton}}\ \emph {et~al.}(2006)\citenamefont {{Croton}}, \citenamefont {{Springel}}, \citenamefont {{White}}, \citenamefont {{De Lucia}}, \citenamefont {{Frenk}}, \citenamefont {{Gao}}, \citenamefont {{Jenkins}}, \citenamefont {{Kauffmann}}, \citenamefont {{Navarro}},\ and\ \citenamefont {{Yoshida}}}]{Crotonetal2006}%
  \BibitemOpen
  \bibfield  {author} {\bibinfo {author} {\bibfnamefont {D.~J.}\ \bibnamefont {{Croton}}}, \bibinfo {author} {\bibfnamefont {V.}~\bibnamefont {{Springel}}}, \bibinfo {author} {\bibfnamefont {S.~D.~M.}\ \bibnamefont {{White}}}, \bibinfo {author} {\bibfnamefont {G.}~\bibnamefont {{De Lucia}}}, \bibinfo {author} {\bibfnamefont {C.~S.}\ \bibnamefont {{Frenk}}}, \bibinfo {author} {\bibfnamefont {L.}~\bibnamefont {{Gao}}}, \bibinfo {author} {\bibfnamefont {A.}~\bibnamefont {{Jenkins}}}, \bibinfo {author} {\bibfnamefont {G.}~\bibnamefont {{Kauffmann}}}, \bibinfo {author} {\bibfnamefont {J.~F.}\ \bibnamefont {{Navarro}}},\ \bibnamefont {and}\ \bibinfo {author} {\bibfnamefont {N.}~\bibnamefont {{Yoshida}}},\ }\href {https://doi.org/10.1111/j.1365-2966.2005.09675.x} {\bibfield  {journal} {\bibinfo  {journal} {\mnras}\ }\textbf {\bibinfo {volume} {365}},\ \bibinfo {pages} {11} (\bibinfo {year} {2006})}\BibitemShut {NoStop}%
\bibitem [{\citenamefont {{DeRose}}\ \emph {et~al.}(2019)\citenamefont {{DeRose}}, \citenamefont {{Wechsler}}, \citenamefont {{Becker}}, \citenamefont {{Busha}}, \citenamefont {{Rykoff}}, \citenamefont {{MacCrann}}, \citenamefont {{Erickson}}, \citenamefont {{Evrard}}, \citenamefont {{Kravtsov}}, \citenamefont {{Gruen}}, \citenamefont {{Allam}}, \citenamefont {{Avila}}, \citenamefont {{Bridle}}, \citenamefont {{Brooks}}, \citenamefont {{Buckley-Geer}}, \citenamefont {{Carnero Rosell}}, \citenamefont {{Carrasco Kind}}, \citenamefont {{Carretero}}, \citenamefont {{Castander}}, \citenamefont {{Cawthon}}, \citenamefont {{Crocce}}, \citenamefont {{da Costa}}, \citenamefont {{Davis}}, \citenamefont {{De Vicente}}, \citenamefont {{Dietrich}}, \citenamefont {{Doel}}, \citenamefont {{Drlica-Wagner}}, \citenamefont {{Fosalba}}, \citenamefont {{Frieman}}, \citenamefont {{Garcia-Bellido}}, \citenamefont {{Gutierrez}}, \citenamefont {{Hartley}}, \citenamefont {{Hollowood}}, \citenamefont {{Hoyle}}, \citenamefont
  {{James}}, \citenamefont {{Krause}}, \citenamefont {{Kuehn}}, \citenamefont {{Kuropatkin}}, \citenamefont {{Lima}}, \citenamefont {{Maia}}, \citenamefont {{Menanteau}}, \citenamefont {{Miller}}, \citenamefont {{Miquel}}, \citenamefont {{Ogando}}, \citenamefont {{Plazas Malag{\'o}n}}, \citenamefont {{Romer}}, \citenamefont {{Sanchez}}, \citenamefont {{Schindler}}, \citenamefont {{Serrano}}, \citenamefont {{Sevilla-Noarbe}}, \citenamefont {{Smith}}, \citenamefont {{Suchyta}}, \citenamefont {{Swanson}}, \citenamefont {{Tarle}},\ and\ \citenamefont {{Vikram}}}]{DeRoseetal2019}%
  \BibitemOpen
  \bibfield  {author} {\bibinfo {author} {\bibfnamefont {J.}~\bibnamefont {{DeRose}}}, \bibnamefont {et~al.},\ }\href {https://doi.org/10.48550/arXiv.1901.02401} {\bibfield  {journal} {\bibinfo  {journal} {arXiv e-prints}\ ,\ \bibinfo {eid} {arXiv:1901.02401}} (\bibinfo {year} {2019})}\BibitemShut {NoStop}%
\bibitem [{\citenamefont {{Wechsler}}\ \emph {et~al.}(2022)\citenamefont {{Wechsler}}, \citenamefont {{DeRose}}, \citenamefont {{Busha}}, \citenamefont {{Becker}}, \citenamefont {{Rykoff}},\ and\ \citenamefont {{Evrard}}}]{Wechsleretal2022}%
  \BibitemOpen
  \bibfield  {author} {\bibinfo {author} {\bibfnamefont {R.~H.}\ \bibnamefont {{Wechsler}}}, \bibinfo {author} {\bibfnamefont {J.}~\bibnamefont {{DeRose}}}, \bibinfo {author} {\bibfnamefont {M.~T.}\ \bibnamefont {{Busha}}}, \bibinfo {author} {\bibfnamefont {M.~R.}\ \bibnamefont {{Becker}}}, \bibinfo {author} {\bibfnamefont {E.}~\bibnamefont {{Rykoff}}},\ \bibnamefont {and}\ \bibinfo {author} {\bibfnamefont {A.}~\bibnamefont {{Evrard}}},\ }\href {https://doi.org/10.3847/1538-4357/ac5b0a} {\bibfield  {journal} {\bibinfo  {journal} {\apj}\ }\textbf {\bibinfo {volume} {931}},\ \bibinfo {eid} {145} (\bibinfo {year} {2022})}\BibitemShut {NoStop}%
\bibitem [{\citenamefont {{Behroozi}}\ \emph {et~al.}(2013)\citenamefont {{Behroozi}}, \citenamefont {{Wechsler}},\ and\ \citenamefont {{Wu}}}]{BehrooziWechslerWu2013}%
  \BibitemOpen
  \bibfield  {author} {\bibinfo {author} {\bibfnamefont {P.~S.}\ \bibnamefont {{Behroozi}}}, \bibinfo {author} {\bibfnamefont {R.~H.}\ \bibnamefont {{Wechsler}}},\ \bibnamefont {and}\ \bibinfo {author} {\bibfnamefont {H.-Y.}\ \bibnamefont {{Wu}}},\ }\href {https://doi.org/10.1088/0004-637X/762/2/109} {\bibfield  {journal} {\bibinfo  {journal} {\apj}\ }\textbf {\bibinfo {volume} {762}},\ \bibinfo {eid} {109} (\bibinfo {year} {2013})}\BibitemShut {NoStop}%
\bibitem [{\citenamefont {{Bryan}}\ and\ \citenamefont {{Norman}}(1998)}]{BryanNorman1998}%
  \BibitemOpen
  \bibfield  {author} {\bibinfo {author} {\bibfnamefont {G.~L.}\ \bibnamefont {{Bryan}}}\ \bibnamefont {and}\ \bibinfo {author} {\bibfnamefont {M.~L.}\ \bibnamefont {{Norman}}},\ }\href {https://doi.org/10.1086/305262} {\bibfield  {journal} {\bibinfo  {journal} {\apj}\ }\textbf {\bibinfo {volume} {495}},\ \bibinfo {pages} {80} (\bibinfo {year} {1998})}\BibitemShut {NoStop}%
\bibitem [{\citenamefont {{Abbott}}\ \emph {et~al.}(2020)\citenamefont {{Abbott}}, \citenamefont {{Aguena}}, \citenamefont {{Alarcon}}, \citenamefont {{Allam}}, \citenamefont {{Allen}}, \citenamefont {{Annis}}, \citenamefont {{Avila}}, \citenamefont {{Bacon}}, \citenamefont {{Bechtol}}, \citenamefont {{Bermeo}}, \citenamefont {{Bernstein}}, \citenamefont {{Bertin}}, \citenamefont {{Bhargava}}, \citenamefont {{Bocquet}}, \citenamefont {{Brooks}}, \citenamefont {{Brout}}, \citenamefont {{Buckley-Geer}}, \citenamefont {{Burke}}, \citenamefont {{Carnero Rosell}}, \citenamefont {{Carrasco Kind}}, \citenamefont {{Carretero}}, \citenamefont {{Castander}}, \citenamefont {{Cawthon}}, \citenamefont {{Chang}}, \citenamefont {{Chen}}, \citenamefont {{Choi}}, \citenamefont {{Costanzi}}, \citenamefont {{Crocce}}, \citenamefont {{da Costa}}, \citenamefont {{Davis}}, \citenamefont {{De Vicente}}, \citenamefont {{DeRose}}, \citenamefont {{Desai}}, \citenamefont {{Diehl}}, \citenamefont {{Dietrich}}, \citenamefont
  {{Dodelson}}, \citenamefont {{Doel}}, \citenamefont {{Drlica-Wagner}}, \citenamefont {{Eckert}}, \citenamefont {{Eifler}}, \citenamefont {{Elvin-Poole}}, \citenamefont {{Estrada}}, \citenamefont {{Everett}}, \citenamefont {{Evrard}}, \citenamefont {{Farahi}}, \citenamefont {{Ferrero}}, \citenamefont {{Flaugher}}, \citenamefont {{Fosalba}}, \citenamefont {{Frieman}}, \citenamefont {{Garc{\'\i}a-Bellido}}, \citenamefont {{Gatti}}, \citenamefont {{Gaztanaga}}, \citenamefont {{Gerdes}}, \citenamefont {{Giannantonio}}, \citenamefont {{Giles}}, \citenamefont {{Grandis}}, \citenamefont {{Gruen}}, \citenamefont {{Gruendl}}, \citenamefont {{Gschwend}}, \citenamefont {{Gutierrez}}, \citenamefont {{Hartley}}, \citenamefont {{Hinton}}, \citenamefont {{Hollowood}}, \citenamefont {{Honscheid}}, \citenamefont {{Hoyle}}, \citenamefont {{Huterer}}, \citenamefont {{James}}, \citenamefont {{Jarvis}}, \citenamefont {{Jeltema}}, \citenamefont {{Johnson}}, \citenamefont {{Johnson}}, \citenamefont {{Kent}}, \citenamefont
  {{Krause}}, \citenamefont {{Kron}}, \citenamefont {{Kuehn}}, \citenamefont {{Kuropatkin}}, \citenamefont {{Lahav}}, \citenamefont {{Li}}, \citenamefont {{Lidman}}, \citenamefont {{Lima}}, \citenamefont {{Lin}}, \citenamefont {{MacCrann}}, \citenamefont {{Maia}}, \citenamefont {{Mantz}}, \citenamefont {{Marshall}}, \citenamefont {{Martini}}, \citenamefont {{Mayers}}, \citenamefont {{Melchior}}, \citenamefont {{Mena-Fern{\'a}ndez}}, \citenamefont {{Menanteau}}, \citenamefont {{Miquel}}, \citenamefont {{Mohr}}, \citenamefont {{Nichol}}, \citenamefont {{Nord}}, \citenamefont {{Ogando}}, \citenamefont {{Palmese}}, \citenamefont {{Paz-Chinch{\'o}n}}, \citenamefont {{Plazas}}, \citenamefont {{Prat}}, \citenamefont {{Rau}}, \citenamefont {{Romer}}, \citenamefont {{Roodman}}, \citenamefont {{Rooney}}, \citenamefont {{Rozo}}, \citenamefont {{Rykoff}}, \citenamefont {{Sako}}, \citenamefont {{Samuroff}}, \citenamefont {{S{\'a}nchez}}, \citenamefont {{Sanchez}}, \citenamefont {{Saro}}, \citenamefont {{Scarpine}},
  \citenamefont {{Schubnell}}, \citenamefont {{Scolnic}}, \citenamefont {{Serrano}}, \citenamefont {{Sevilla-Noarbe}}, \citenamefont {{Sheldon}}, \citenamefont {{Smith}}, \citenamefont {{Smith}}, \citenamefont {{Suchyta}}, \citenamefont {{Swanson}}, \citenamefont {{Tarle}}, \citenamefont {{Thomas}}, \citenamefont {{To}}, \citenamefont {{Troxel}}, \citenamefont {{Tucker}}, \citenamefont {{Varga}}, \citenamefont {{von der Linden}}, \citenamefont {{Walker}}, \citenamefont {{Wechsler}}, \citenamefont {{Weller}}, \citenamefont {{Wilkinson}}, \citenamefont {{Wu}}, \citenamefont {{Yanny}}, \citenamefont {{Zhang}}, \citenamefont {{Zhang}}, \citenamefont {{Zuntz}},\ and\ \citenamefont {{DES Collaboration}}}]{Abbottetal2020}%
  \BibitemOpen
  \bibfield  {author} {\bibinfo {author} {\bibfnamefont {T.~M.~C.}\ \bibnamefont {{Abbott}}}, \bibnamefont {et~al.},\ }\href {https://doi.org/10.1103/PhysRevD.102.023509} {\bibfield  {journal} {\bibinfo  {journal} {\prd}\ }\textbf {\bibinfo {volume} {102}},\ \bibinfo {eid} {023509} (\bibinfo {year} {2020})}\BibitemShut {NoStop}%
\bibitem [{\citenamefont {Scott}(2015)}]{scott2015multivariate}%
  \BibitemOpen
  \bibfield  {author} {\bibinfo {author} {\bibfnamefont {D.~W.}\ \bibnamefont {Scott}},\ }\href@noop {} {\emph {\bibinfo {title} {Multivariate density estimation: theory, practice, and visualization}}}\ (\bibinfo  {publisher} {John Wiley \& Sons},\ \bibinfo {year} {2015})\BibitemShut {NoStop}%
\bibitem [{\citenamefont {{Myles}}\ \emph {et~al.}(2021)\citenamefont {{Myles}}, \citenamefont {{Gruen}}, \citenamefont {{Mantz}}, \citenamefont {{Allen}}, \citenamefont {{Morris}}, \citenamefont {{Rykoff}}, \citenamefont {{Costanzi}}, \citenamefont {{To}}, \citenamefont {{DeRose}}, \citenamefont {{Wechsler}}, \citenamefont {{Rozo}}, \citenamefont {{Jeltema}}, \citenamefont {{Carrasco}}, \citenamefont {{Kremin}},\ and\ \citenamefont {{Kron}}}]{Mylesetal2021}%
  \BibitemOpen
  \bibfield  {author} {\bibinfo {author} {\bibfnamefont {J.}~\bibnamefont {{Myles}}}, \bibnamefont {et~al.},\ }\href {https://doi.org/10.1093/mnras/stab1243} {\bibfield  {journal} {\bibinfo  {journal} {\mnras}\ }\textbf {\bibinfo {volume} {505}},\ \bibinfo {pages} {33} (\bibinfo {year} {2021})}\BibitemShut {NoStop}%
\bibitem [{\citenamefont {{Grandis}}\ \emph {et~al.}(2021{\natexlab{b}})\citenamefont {{Grandis}}, \citenamefont {{Bocquet}}, \citenamefont {{Mohr}}, \citenamefont {{Klein}},\ and\ \citenamefont {{Dolag}}}]{Grandisetal2021b}%
  \BibitemOpen
  \bibfield  {author} {\bibinfo {author} {\bibfnamefont {S.}~\bibnamefont {{Grandis}}}, \bibinfo {author} {\bibfnamefont {S.}~\bibnamefont {{Bocquet}}}, \bibinfo {author} {\bibfnamefont {J.~J.}\ \bibnamefont {{Mohr}}}, \bibinfo {author} {\bibfnamefont {M.}~\bibnamefont {{Klein}}},\ \bibnamefont {and}\ \bibinfo {author} {\bibfnamefont {K.}~\bibnamefont {{Dolag}}},\ }\href {https://doi.org/10.1093/mnras/stab2414} {\bibfield  {journal} {\bibinfo  {journal} {\mnras}\ }\textbf {\bibinfo {volume} {507}},\ \bibinfo {pages} {5671} (\bibinfo {year} {2021}{\natexlab{b}})}\BibitemShut {NoStop}%
\bibitem [{\citenamefont {{Grandis}}\ \emph {et~al.}(2025)\citenamefont {{Grandis}}, \citenamefont {{Costanzi}}, \citenamefont {{Mohr}}, \citenamefont {{Bleem}}, \citenamefont {{Wu}}, \citenamefont {{Aguena}}, \citenamefont {{Allam}}, \citenamefont {{Andrade-Oliveira}}, \citenamefont {{Bocquet}}, \citenamefont {{Brooks}}, \citenamefont {{Carnero Rosell}}, \citenamefont {{Carretero}}, \citenamefont {{da Costa}}, \citenamefont {{Pereira}}, \citenamefont {{Davis}}, \citenamefont {{Desai}}, \citenamefont {{Diehl}}, \citenamefont {{Doel}}, \citenamefont {{Everett}}, \citenamefont {{Flaugher}}, \citenamefont {{Frieman}}, \citenamefont {{Garc{\'\i}a-Bellido}}, \citenamefont {{Gaztanaga}}, \citenamefont {{Gruen}}, \citenamefont {{Gruendl}}, \citenamefont {{Gutierrez}}, \citenamefont {{Hinton}}, \citenamefont {{Hlacacek-Larrondo}}, \citenamefont {{Hollowood}}, \citenamefont {{Honscheid}}, \citenamefont {{James}}, \citenamefont {{Klein}}, \citenamefont {{Marshall}}, \citenamefont {{Mena-Fern{\'a}ndez}}, \citenamefont
  {{Miquel}}, \citenamefont {{Palmese}}, \citenamefont {{Plazas Malag{\'o}n}}, \citenamefont {{Reichardt}}, \citenamefont {{Romer}}, \citenamefont {{Samuroff}}, \citenamefont {{Sanchez Cid}}, \citenamefont {{Sanchez}}, \citenamefont {{Santiago}}, \citenamefont {{Saro}}, \citenamefont {{Sevilla-Noarbe}}, \citenamefont {{Smith}}, \citenamefont {{Soares-Santos}}, \citenamefont {{Sommer}}, \citenamefont {{Suchyta}}, \citenamefont {{Tarle}}, \citenamefont {{To}}, \citenamefont {{Tucker}}, \citenamefont {{Weaverdyck}}, \citenamefont {{Weller}},\ and\ \citenamefont {{Wiseman}}}]{Grandisetal2025}%
  \BibitemOpen
  \bibfield  {author} {\bibinfo {author} {\bibfnamefont {S.}~\bibnamefont {{Grandis}}}, \bibnamefont {et~al.},\ }\href {https://doi.org/10.1051/0004-6361/202554177} {\bibfield  {journal} {\bibinfo  {journal} {\aap}\ }\textbf {\bibinfo {volume} {700}},\ \bibinfo {eid} {A15} (\bibinfo {year} {2025})}\BibitemShut {NoStop}%
\bibitem [{\citenamefont {{G{\'o}rski}}\ \emph {et~al.}(2005)\citenamefont {{G{\'o}rski}}, \citenamefont {{Hivon}}, \citenamefont {{Banday}}, \citenamefont {{Wandelt}}, \citenamefont {{Hansen}}, \citenamefont {{Reinecke}},\ and\ \citenamefont {{Bartelmann}}}]{Healpix}%
  \BibitemOpen
  \bibfield  {author} {\bibinfo {author} {\bibfnamefont {K.~M.}\ \bibnamefont {{G{\'o}rski}}}, \bibinfo {author} {\bibfnamefont {E.}~\bibnamefont {{Hivon}}}, \bibinfo {author} {\bibfnamefont {A.~J.}\ \bibnamefont {{Banday}}}, \bibinfo {author} {\bibfnamefont {B.~D.}\ \bibnamefont {{Wandelt}}}, \bibinfo {author} {\bibfnamefont {F.~K.}\ \bibnamefont {{Hansen}}}, \bibinfo {author} {\bibfnamefont {M.}~\bibnamefont {{Reinecke}}},\ \bibnamefont {and}\ \bibinfo {author} {\bibfnamefont {M.}~\bibnamefont {{Bartelmann}}},\ }\href {https://doi.org/10.1086/427976} {\bibfield  {journal} {\bibinfo  {journal} {\apj}\ }\textbf {\bibinfo {volume} {622}},\ \bibinfo {pages} {759} (\bibinfo {year} {2005})}\BibitemShut {NoStop}%
\bibitem [{\citenamefont {{Zhang}}\ \emph {et~al.}(2019)\citenamefont {{Zhang}}, \citenamefont {{Jeltema}}, \citenamefont {{Hollowood}}, \citenamefont {{Everett}}, \citenamefont {{Rozo}}, \citenamefont {{Farahi}}, \citenamefont {{Bermeo}}, \citenamefont {{Bhargava}}, \citenamefont {{Giles}}, \citenamefont {{Romer}}, \citenamefont {{Wilkinson}}, \citenamefont {{Rykoff}}, \citenamefont {{Mantz}}, \citenamefont {{Diehl}}, \citenamefont {{Evrard}}, \citenamefont {{Stern}}, \citenamefont {{Gruen}}, \citenamefont {{von der Linden}}, \citenamefont {{Splettstoesser}}, \citenamefont {{Chen}}, \citenamefont {{Costanzi}}, \citenamefont {{Allen}}, \citenamefont {{Collins}}, \citenamefont {{Hilton}}, \citenamefont {{Klein}}, \citenamefont {{Mann}}, \citenamefont {{Manolopoulou}}, \citenamefont {{Morris}}, \citenamefont {{Mayers}}, \citenamefont {{Sahlen}}, \citenamefont {{Stott}}, \citenamefont {{Vergara Cervantes}}, \citenamefont {{Viana}}, \citenamefont {{Wechsler}}, \citenamefont {{Allam}}, \citenamefont {{Avila}},
  \citenamefont {{Bechtol}}, \citenamefont {{Bertin}}, \citenamefont {{Brooks}}, \citenamefont {{Burke}}, \citenamefont {{Carnero Rosell}}, \citenamefont {{Carrasco Kind}}, \citenamefont {{Carretero}}, \citenamefont {{Castander}}, \citenamefont {{da Costa}}, \citenamefont {{De Vicente}}, \citenamefont {{Desai}}, \citenamefont {{Dietrich}}, \citenamefont {{Doel}}, \citenamefont {{Flaugher}}, \citenamefont {{Fosalba}}, \citenamefont {{Frieman}}, \citenamefont {{Garc{\'\i}a-Bellido}}, \citenamefont {{Gaztanaga}}, \citenamefont {{Gruendl}}, \citenamefont {{Gschwend}}, \citenamefont {{Gutierrez}}, \citenamefont {{Hartley}}, \citenamefont {{Honscheid}}, \citenamefont {{Hoyle}}, \citenamefont {{Krause}}, \citenamefont {{Kuehn}}, \citenamefont {{Kuropatkin}}, \citenamefont {{Lima}}, \citenamefont {{Maia}}, \citenamefont {{Marshall}}, \citenamefont {{Melchior}}, \citenamefont {{Menanteau}}, \citenamefont {{Miller}}, \citenamefont {{Miquel}}, \citenamefont {{Ogando}}, \citenamefont {{Plazas}}, \citenamefont
  {{Sanchez}}, \citenamefont {{Scarpine}}, \citenamefont {{Schindler}}, \citenamefont {{Serrano}}, \citenamefont {{Sevilla-Noarbe}}, \citenamefont {{Smith}}, \citenamefont {{Soares-Santos}}, \citenamefont {{Suchyta}}, \citenamefont {{Swanson}}, \citenamefont {{Tarle}}, \citenamefont {{Thomas}}, \citenamefont {{Tucker}}, \citenamefont {{Vikram}}, \citenamefont {{Wester}},\ and\ \citenamefont {{DES Collaboration}}}]{DES2019b}%
  \BibitemOpen
  \bibfield  {author} {\bibinfo {author} {\bibfnamefont {Y.}~\bibnamefont {{Zhang}}}, \bibnamefont {et~al.},\ }\href {https://doi.org/10.1093/mnras/stz1361} {\bibfield  {journal} {\bibinfo  {journal} {\mnras}\ }\textbf {\bibinfo {volume} {487}},\ \bibinfo {pages} {2578} (\bibinfo {year} {2019})}\BibitemShut {NoStop}%
\bibitem [{\citenamefont {{Kelly}}\ \emph {et~al.}(2024)\citenamefont {{Kelly}}, \citenamefont {{Jobel}}, \citenamefont {{Eiger}}, \citenamefont {{Abd}}, \citenamefont {{Jeltema}}, \citenamefont {{Giles}}, \citenamefont {{Hollowood}}, \citenamefont {{Wilkinson}}, \citenamefont {{Turner}}, \citenamefont {{Bhargava}}, \citenamefont {{Everett}}, \citenamefont {{Farahi}}, \citenamefont {{Romer}}, \citenamefont {{Rykoff}}, \citenamefont {{Wang}}, \citenamefont {{Bocquet}}, \citenamefont {{Cross}}, \citenamefont {{Faridjoo}}, \citenamefont {{Franco}}, \citenamefont {{Gardner}}, \citenamefont {{Kwiecien}}, \citenamefont {{Laubner}}, \citenamefont {{McDaniel}}, \citenamefont {{O'Donnell}}, \citenamefont {{Sanchez}}, \citenamefont {{Schmidt}}, \citenamefont {{Sripada}}, \citenamefont {{Swart}}, \citenamefont {{Upsdell}}, \citenamefont {{Webber}}, \citenamefont {{Aguena}}, \citenamefont {{Allam}}, \citenamefont {{Alves}}, \citenamefont {{Bacon}}, \citenamefont {{Brooks}}, \citenamefont {{Burke}}, \citenamefont
  {{Carnero Rosell}}, \citenamefont {{Carretero}}, \citenamefont {{Collins}}, \citenamefont {{Costanzi}}, \citenamefont {{da Costa}}, \citenamefont {{Pereira}}, \citenamefont {{Davis}}, \citenamefont {{Doel}}, \citenamefont {{Ferrero}}, \citenamefont {{Frieman}}, \citenamefont {{Garc{\'\i}a-Bellido}}, \citenamefont {{Giannini}}, \citenamefont {{Gruen}}, \citenamefont {{Gruendl}}, \citenamefont {{Hilton}}, \citenamefont {{Hinton}}, \citenamefont {{Honscheid}}, \citenamefont {{James}}, \citenamefont {{Kuehn}}, \citenamefont {{Mann}}, \citenamefont {{Marshall}}, \citenamefont {{Mena-Fern{\'a}ndez}}, \citenamefont {{Miller}}, \citenamefont {{Miquel}}, \citenamefont {{Myles}}, \citenamefont {{Palmese}}, \citenamefont {{Pieres}}, \citenamefont {{Plazas Malag{\'o}n}}, \citenamefont {{Rooney}}, \citenamefont {{Sahlen}}, \citenamefont {{Sanchez}}, \citenamefont {{Sanchez Cid}}, \citenamefont {{Schubnell}}, \citenamefont {{Sevilla-Noarbe}}, \citenamefont {{Smith}}, \citenamefont {{Stott}}, \citenamefont {{Suchyta}},
  \citenamefont {{Swanson}}, \citenamefont {{Tarle}}, \citenamefont {{To}}, \citenamefont {{Viana}}, \citenamefont {{Weaverdyck}}, \citenamefont {{Wiseman}},\ and\ \citenamefont {{DES Collaboration}}}]{DES2024}%
  \BibitemOpen
  \bibfield  {author} {\bibinfo {author} {\bibfnamefont {P.~M.}\ \bibnamefont {{Kelly}}}, \bibnamefont {et~al.},\ }\href {https://doi.org/10.1093/mnras/stae1786} {\bibfield  {journal} {\bibinfo  {journal} {\mnras}\ }\textbf {\bibinfo {volume} {533}},\ \bibinfo {pages} {572} (\bibinfo {year} {2024})}\BibitemShut {NoStop}%
\bibitem [{\citenamefont {{Ding}}\ \emph {et~al.}(2025)\citenamefont {{Ding}}, \citenamefont {{Dalal}}, \citenamefont {{Sunayama}}, \citenamefont {{Strauss}}, \citenamefont {{Oguri}}, \citenamefont {{Okabe}}, \citenamefont {{Hilton}}, \citenamefont {{Monteiro-Oliveira}}, \citenamefont {{Sif{\'o}n}},\ and\ \citenamefont {{Staggs}}}]{Dingetal2025}%
  \BibitemOpen
  \bibfield  {author} {\bibinfo {author} {\bibfnamefont {J.}~\bibnamefont {{Ding}}}, \bibinfo {author} {\bibfnamefont {R.}~\bibnamefont {{Dalal}}}, \bibinfo {author} {\bibfnamefont {T.}~\bibnamefont {{Sunayama}}}, \bibinfo {author} {\bibfnamefont {M.~A.}\ \bibnamefont {{Strauss}}}, \bibinfo {author} {\bibfnamefont {M.}~\bibnamefont {{Oguri}}}, \bibinfo {author} {\bibfnamefont {N.}~\bibnamefont {{Okabe}}}, \bibinfo {author} {\bibfnamefont {M.}~\bibnamefont {{Hilton}}}, \bibinfo {author} {\bibfnamefont {R.}~\bibnamefont {{Monteiro-Oliveira}}}, \bibinfo {author} {\bibfnamefont {C.}~\bibnamefont {{Sif{\'o}n}}},\ \bibnamefont {and}\ \bibinfo {author} {\bibfnamefont {S.~T.}\ \bibnamefont {{Staggs}}},\ }\href {https://doi.org/10.1093/mnras/stae2601} {\bibfield  {journal} {\bibinfo  {journal} {\mnras}\ }\textbf {\bibinfo {volume} {536}},\ \bibinfo {pages} {572} (\bibinfo {year} {2025})}\BibitemShut {NoStop}%
\bibitem [{\citenamefont {{Bocquet}}\ \emph {et~al.}(2024)\citenamefont {{Bocquet}}, \citenamefont {{Grandis}}, \citenamefont {{Bleem}}, \citenamefont {{Klein}}, \citenamefont {{Mohr}}, \citenamefont {{Aguena}}, \citenamefont {{Alarcon}}, \citenamefont {{Allam}}, \citenamefont {{Allen}}, \citenamefont {{Alves}}, \citenamefont {{Amon}}, \citenamefont {{Ansarinejad}}, \citenamefont {{Bacon}}, \citenamefont {{Bayliss}}, \citenamefont {{Bechtol}}, \citenamefont {{Becker}}, \citenamefont {{Benson}}, \citenamefont {{Bernstein}}, \citenamefont {{Brodwin}}, \citenamefont {{Brooks}}, \citenamefont {{Campos}}, \citenamefont {{Canning}}, \citenamefont {{Carlstrom}}, \citenamefont {{Carnero Rosell}}, \citenamefont {{Carrasco Kind}}, \citenamefont {{Carretero}}, \citenamefont {{Cawthon}}, \citenamefont {{Chang}}, \citenamefont {{Chen}}, \citenamefont {{Choi}}, \citenamefont {{Cordero}}, \citenamefont {{Costanzi}}, \citenamefont {{da Costa}}, \citenamefont {{Pereira}}, \citenamefont {{Davis}}, \citenamefont {{DeRose}},
  \citenamefont {{Desai}}, \citenamefont {{de Haan}}, \citenamefont {{De Vicente}}, \citenamefont {{Diehl}}, \citenamefont {{Dodelson}}, \citenamefont {{Doel}}, \citenamefont {{Doux}}, \citenamefont {{Drlica-Wagner}}, \citenamefont {{Eckert}}, \citenamefont {{Elvin-Poole}}, \citenamefont {{Everett}}, \citenamefont {{Ferrero}}, \citenamefont {{Fert{\'e}}}, \citenamefont {{Flores}}, \citenamefont {{Frieman}}, \citenamefont {{Garc{\'\i}a-Bellido}}, \citenamefont {{Gatti}}, \citenamefont {{Giannini}}, \citenamefont {{Gladders}}, \citenamefont {{Gruen}}, \citenamefont {{Gruendl}}, \citenamefont {{Harrison}}, \citenamefont {{Hartley}}, \citenamefont {{Herner}}, \citenamefont {{Hinton}}, \citenamefont {{Hollowood}}, \citenamefont {{Holzapfel}}, \citenamefont {{Honscheid}}, \citenamefont {{Huang}}, \citenamefont {{Huff}}, \citenamefont {{James}}, \citenamefont {{Jarvis}}, \citenamefont {{Khullar}}, \citenamefont {{Kim}}, \citenamefont {{Kraft}}, \citenamefont {{Kuehn}}, \citenamefont {{Kuropatkin}}, \citenamefont
  {{K{\'e}ruzor{\'e}}}, \citenamefont {{Lee}}, \citenamefont {{Leget}}, \citenamefont {{MacCrann}}, \citenamefont {{Mahler}}, \citenamefont {{Mantz}}, \citenamefont {{Marshall}}, \citenamefont {{McCullough}}, \citenamefont {{McDonald}}, \citenamefont {{Mena-Fern{\'a}ndez}}, \citenamefont {{Miquel}}, \citenamefont {{Myles}}, \citenamefont {{Navarro-Alsina}}, \citenamefont {{Ogando}}, \citenamefont {{Palmese}}, \citenamefont {{Pandey}}, \citenamefont {{Pieres}}, \citenamefont {{Plazas Malag{\'o}n}}, \citenamefont {{Prat}}, \citenamefont {{Raveri}}, \citenamefont {{Reichardt}}, \citenamefont {{Roberson}}, \citenamefont {{Rollins}}, \citenamefont {{Romer}}, \citenamefont {{Romero}}, \citenamefont {{Roodman}}, \citenamefont {{Ross}}, \citenamefont {{Rykoff}}, \citenamefont {{Salvati}}, \citenamefont {{S{\'a}nchez}}, \citenamefont {{Sanchez}}, \citenamefont {{Sanchez Cid}}, \citenamefont {{Saro}}, \citenamefont {{Schrabback}}, \citenamefont {{Schubnell}}, \citenamefont {{Secco}}, \citenamefont {{Sevilla-Noarbe}},
  \citenamefont {{Sharon}}, \citenamefont {{Sheldon}}, \citenamefont {{Shin}}, \citenamefont {{Smith}}, \citenamefont {{Somboonpanyakul}}, \citenamefont {{Stalder}}, \citenamefont {{Stark}}, \citenamefont {{Strazzullo}}, \citenamefont {{Suchyta}}, \citenamefont {{Swanson}}, \citenamefont {{Tarle}}, \citenamefont {{To}}, \citenamefont {{Troxel}}, \citenamefont {{Tutusaus}}, \citenamefont {{Varga}}, \citenamefont {{von der Linden}}, \citenamefont {{Weaverdyck}}, \citenamefont {{Weller}}, \citenamefont {{Wiseman}}, \citenamefont {{Yanny}}, \citenamefont {{Yin}}, \citenamefont {{Young}}, \citenamefont {{Zhang}}, \citenamefont {{Zuntz}}, \citenamefont {{(The DES}},\ and\ \citenamefont {{SPT Collaborations)}}}]{Bocquetetal2024}%
  \BibitemOpen
  \bibfield  {author} {\bibinfo {author} {\bibfnamefont {S.}~\bibnamefont {{Bocquet}}}, \bibnamefont {et~al.},\ }\href {https://doi.org/10.1103/PhysRevD.110.083509} {\bibfield  {journal} {\bibinfo  {journal} {\prd}\ }\textbf {\bibinfo {volume} {110}},\ \bibinfo {eid} {083509} (\bibinfo {year} {2024})}\BibitemShut {NoStop}%
\bibitem [{\citenamefont {{Sommer}}\ \emph {et~al.}(2025)\citenamefont {{Sommer}}, \citenamefont {{Schrabback}},\ and\ \citenamefont {{Grandis}}}]{SommerSchrabbackGrandis2025}%
  \BibitemOpen
  \bibfield  {author} {\bibinfo {author} {\bibfnamefont {M.~W.}\ \bibnamefont {{Sommer}}}, \bibinfo {author} {\bibfnamefont {T.}~\bibnamefont {{Schrabback}}},\ \bibnamefont {and}\ \bibinfo {author} {\bibfnamefont {S.}~\bibnamefont {{Grandis}}},\ }\href {https://doi.org/10.1093/mnrasl/slaf007} {\bibfield  {journal} {\bibinfo  {journal} {\mnras}\ }\textbf {\bibinfo {volume} {538}},\ \bibinfo {pages} {L50} (\bibinfo {year} {2025})}\BibitemShut {NoStop}%
\bibitem [{\citenamefont {{Martel}}\ \emph {et~al.}(2014)\citenamefont {{Martel}}, \citenamefont {{Robichaud}},\ and\ \citenamefont {{Barai}}}]{MartelRobichaudBarai2014}%
  \BibitemOpen
  \bibfield  {author} {\bibinfo {author} {\bibfnamefont {H.}~\bibnamefont {{Martel}}}, \bibinfo {author} {\bibfnamefont {F.}~\bibnamefont {{Robichaud}}},\ \bibnamefont {and}\ \bibinfo {author} {\bibfnamefont {P.}~\bibnamefont {{Barai}}},\ }\href {https://doi.org/10.1088/0004-637X/786/2/79} {\bibfield  {journal} {\bibinfo  {journal} {\apj}\ }\textbf {\bibinfo {volume} {786}},\ \bibinfo {eid} {79} (\bibinfo {year} {2014})}\BibitemShut {NoStop}%
\bibitem [{\citenamefont {{Saxton}}\ \emph {et~al.}(2008)\citenamefont {{Saxton}}, \citenamefont {{Read}}, \citenamefont {{Esquej}}, \citenamefont {{Freyberg}}, \citenamefont {{Altieri}},\ and\ \citenamefont {{Bermejo}}}]{Saxtonetal2008}%
  \BibitemOpen
  \bibfield  {author} {\bibinfo {author} {\bibfnamefont {R.~D.}\ \bibnamefont {{Saxton}}}, \bibinfo {author} {\bibfnamefont {A.~M.}\ \bibnamefont {{Read}}}, \bibinfo {author} {\bibfnamefont {P.}~\bibnamefont {{Esquej}}}, \bibinfo {author} {\bibfnamefont {M.~J.}\ \bibnamefont {{Freyberg}}}, \bibinfo {author} {\bibfnamefont {B.}~\bibnamefont {{Altieri}}},\ \bibnamefont {and}\ \bibinfo {author} {\bibfnamefont {D.}~\bibnamefont {{Bermejo}}},\ }\href {https://doi.org/10.1051/0004-6361:20079193} {\bibfield  {journal} {\bibinfo  {journal} {\aap}\ }\textbf {\bibinfo {volume} {480}},\ \bibinfo {pages} {611} (\bibinfo {year} {2008})}\BibitemShut {NoStop}%
\bibitem [{\citenamefont {{Vulic}}\ \emph {et~al.}(2016)\citenamefont {{Vulic}}, \citenamefont {{Gallagher}},\ and\ \citenamefont {{Barmby}}}]{VulicGallagherBarmby2016}%
  \BibitemOpen
  \bibfield  {author} {\bibinfo {author} {\bibfnamefont {N.}~\bibnamefont {{Vulic}}}, \bibinfo {author} {\bibfnamefont {S.~C.}\ \bibnamefont {{Gallagher}}},\ \bibnamefont {and}\ \bibinfo {author} {\bibfnamefont {P.}~\bibnamefont {{Barmby}}},\ }\href {https://doi.org/10.1093/mnras/stw1523} {\bibfield  {journal} {\bibinfo  {journal} {\mnras}\ }\textbf {\bibinfo {volume} {461}},\ \bibinfo {pages} {3443} (\bibinfo {year} {2016})}\BibitemShut {NoStop}%
\bibitem [{\citenamefont {{Medvedev}}\ \emph {et~al.}(2021)\citenamefont {{Medvedev}}, \citenamefont {{Gilfanov}}, \citenamefont {{Sazonov}}, \citenamefont {{Schartel}},\ and\ \citenamefont {{Sunyaev}}}]{Medvedevetal2021}%
  \BibitemOpen
  \bibfield  {author} {\bibinfo {author} {\bibfnamefont {P.}~\bibnamefont {{Medvedev}}}, \bibinfo {author} {\bibfnamefont {M.}~\bibnamefont {{Gilfanov}}}, \bibinfo {author} {\bibfnamefont {S.}~\bibnamefont {{Sazonov}}}, \bibinfo {author} {\bibfnamefont {N.}~\bibnamefont {{Schartel}}},\ \bibnamefont {and}\ \bibinfo {author} {\bibfnamefont {R.}~\bibnamefont {{Sunyaev}}},\ }\href {https://doi.org/10.1093/mnras/stab773} {\bibfield  {journal} {\bibinfo  {journal} {\mnras}\ }\textbf {\bibinfo {volume} {504}},\ \bibinfo {pages} {576} (\bibinfo {year} {2021})}\BibitemShut {NoStop}%
\bibitem [{\citenamefont {{Foreman-Mackey}}\ \emph {et~al.}(2013)\citenamefont {{Foreman-Mackey}}, \citenamefont {{Hogg}}, \citenamefont {{Lang}},\ and\ \citenamefont {{Goodman}}}]{emcee}%
  \BibitemOpen
  \bibfield  {author} {\bibinfo {author} {\bibfnamefont {D.}~\bibnamefont {{Foreman-Mackey}}}, \bibinfo {author} {\bibfnamefont {D.~W.}\ \bibnamefont {{Hogg}}}, \bibinfo {author} {\bibfnamefont {D.}~\bibnamefont {{Lang}}},\ \bibnamefont {and}\ \bibinfo {author} {\bibfnamefont {J.}~\bibnamefont {{Goodman}}},\ }\href {https://doi.org/10.1086/670067} {\bibfield  {journal} {\bibinfo  {journal} {\pasp}\ }\textbf {\bibinfo {volume} {125}},\ \bibinfo {pages} {306} (\bibinfo {year} {2013})}\BibitemShut {NoStop}%
\bibitem [{\citenamefont {{Hearin}}\ and\ \citenamefont {{Watson}}(2013)}]{HearinWatson2013}%
  \BibitemOpen
  \bibfield  {author} {\bibinfo {author} {\bibfnamefont {A.~P.}\ \bibnamefont {{Hearin}}}\ \bibnamefont {and}\ \bibinfo {author} {\bibfnamefont {D.~F.}\ \bibnamefont {{Watson}}},\ }\href {https://doi.org/10.1093/mnras/stt1374} {\bibfield  {journal} {\bibinfo  {journal} {\mnras}\ }\textbf {\bibinfo {volume} {435}},\ \bibinfo {pages} {1313} (\bibinfo {year} {2013})}\BibitemShut {NoStop}%
\bibitem [{\citenamefont {{Bellagamba}}\ \emph {et~al.}(2018)\citenamefont {{Bellagamba}}, \citenamefont {{Roncarelli}}, \citenamefont {{Maturi}},\ and\ \citenamefont {{Moscardini}}}]{AMICO}%
  \BibitemOpen
  \bibfield  {author} {\bibinfo {author} {\bibfnamefont {F.}~\bibnamefont {{Bellagamba}}}, \bibinfo {author} {\bibfnamefont {M.}~\bibnamefont {{Roncarelli}}}, \bibinfo {author} {\bibfnamefont {M.}~\bibnamefont {{Maturi}}},\ \bibnamefont {and}\ \bibinfo {author} {\bibfnamefont {L.}~\bibnamefont {{Moscardini}}},\ }\href {https://doi.org/10.1093/mnras/stx2701} {\bibfield  {journal} {\bibinfo  {journal} {\mnras}\ }\textbf {\bibinfo {volume} {473}},\ \bibinfo {pages} {5221} (\bibinfo {year} {2018})}\BibitemShut {NoStop}%
\bibitem [{\citenamefont {{Aguena}}\ \emph {et~al.}(2021)\citenamefont {{Aguena}}, \citenamefont {{Benoist}}, \citenamefont {{da Costa}}, \citenamefont {{Ogando}}, \citenamefont {{Gschwend}}, \citenamefont {{Sampaio-Santos}}, \citenamefont {{Lima}}, \citenamefont {{Maia}}, \citenamefont {{Allam}}, \citenamefont {{Avila}}, \citenamefont {{Bacon}}, \citenamefont {{Bertin}}, \citenamefont {{Bhargava}}, \citenamefont {{Brooks}}, \citenamefont {{Carnero Rosell}}, \citenamefont {{Carrasco Kind}}, \citenamefont {{Carretero}}, \citenamefont {{Costanzi}}, \citenamefont {{De Vicente}}, \citenamefont {{Desai}}, \citenamefont {{Diehl}}, \citenamefont {{Doel}}, \citenamefont {{Everett}}, \citenamefont {{Evrard}}, \citenamefont {{Ferrero}}, \citenamefont {{Fert{\'e}}}, \citenamefont {{Flaugher}}, \citenamefont {{Fosalba}}, \citenamefont {{Frieman}}, \citenamefont {{Garc{\'\i}a-Bellido}}, \citenamefont {{Giles}}, \citenamefont {{Gruendl}}, \citenamefont {{Gutierrez}}, \citenamefont {{Hinton}}, \citenamefont {{Hollowood}},
  \citenamefont {{Honscheid}}, \citenamefont {{James}}, \citenamefont {{Jeltema}}, \citenamefont {{Kuehn}}, \citenamefont {{Kuropatkin}}, \citenamefont {{Lahav}}, \citenamefont {{Melchior}}, \citenamefont {{Miquel}}, \citenamefont {{Morgan}}, \citenamefont {{Palmese}}, \citenamefont {{Paz-Chinch{\'o}n}}, \citenamefont {{Plazas}}, \citenamefont {{Romer}}, \citenamefont {{Sanchez}}, \citenamefont {{Santiago}}, \citenamefont {{Schubnell}}, \citenamefont {{Serrano}}, \citenamefont {{Sevilla-Noarbe}}, \citenamefont {{Smith}}, \citenamefont {{Soares-Santos}}, \citenamefont {{Suchyta}}, \citenamefont {{Tarle}}, \citenamefont {{To}}, \citenamefont {{Tucker}},\ and\ \citenamefont {{Wilkinson}}}]{WaZP}%
  \BibitemOpen
  \bibfield  {author} {\bibinfo {author} {\bibfnamefont {M.}~\bibnamefont {{Aguena}}}, \bibnamefont {et~al.},\ }\href {https://doi.org/10.1093/mnras/stab264} {\bibfield  {journal} {\bibinfo  {journal} {\mnras}\ }\textbf {\bibinfo {volume} {502}},\ \bibinfo {pages} {4435} (\bibinfo {year} {2021})}\BibitemShut {NoStop}%
\bibitem [{\citenamefont {{Yantovski-Barth}}\ \emph {et~al.}(2024)\citenamefont {{Yantovski-Barth}}, \citenamefont {{Newman}}, \citenamefont {{Dey}}, \citenamefont {{Andrews}}, \citenamefont {{Eracleous}}, \citenamefont {{Golden-Marx}},\ and\ \citenamefont {{Zhou}}}]{CluMPR}%
  \BibitemOpen
  \bibfield  {author} {\bibinfo {author} {\bibfnamefont {M.~J.}\ \bibnamefont {{Yantovski-Barth}}}, \bibinfo {author} {\bibfnamefont {J.~A.}\ \bibnamefont {{Newman}}}, \bibinfo {author} {\bibfnamefont {B.}~\bibnamefont {{Dey}}}, \bibinfo {author} {\bibfnamefont {B.~H.}\ \bibnamefont {{Andrews}}}, \bibinfo {author} {\bibfnamefont {M.}~\bibnamefont {{Eracleous}}}, \bibinfo {author} {\bibfnamefont {J.}~\bibnamefont {{Golden-Marx}}},\ \bibnamefont {and}\ \bibinfo {author} {\bibfnamefont {R.}~\bibnamefont {{Zhou}}},\ }\href {https://doi.org/10.1093/mnras/stae956} {\bibfield  {journal} {\bibinfo  {journal} {\mnras}\ }\textbf {\bibinfo {volume} {531}},\ \bibinfo {pages} {2285} (\bibinfo {year} {2024})}\BibitemShut {NoStop}%
\bibitem [{\citenamefont {{Doubrawa}}\ \emph {et~al.}(2024)\citenamefont {{Doubrawa}}, \citenamefont {{Cypriano}}, \citenamefont {{Finoguenov}}, \citenamefont {{Lopes}}, \citenamefont {{Gonzalez}}, \citenamefont {{Maturi}}, \citenamefont {{Dupke}}, \citenamefont {{Gonz{\'a}lez Delgado}}, \citenamefont {{Abramo}}, \citenamefont {{Benitez}}, \citenamefont {{Bonoli}}, \citenamefont {{Carneiro}}, \citenamefont {{Cenarro}}, \citenamefont {{Crist{\'o}bal-Hornillos}}, \citenamefont {{Ederoclite}}, \citenamefont {{Hern{\'a}n-Caballero}}, \citenamefont {{L{\'o}pez-Sanjuan}}, \citenamefont {{Mar{\'\i}n-Franch}}, \citenamefont {{Mendes de Oliveira}}, \citenamefont {{Moles}}, \citenamefont {{Sodr{\'e}}}, \citenamefont {{Taylor}}, \citenamefont {{Varela}},\ and\ \citenamefont {{V{\'a}zquez Rami{\'o}}}}]{Doubrawaetal2024}%
  \BibitemOpen
  \bibfield  {author} {\bibinfo {author} {\bibfnamefont {L.}~\bibnamefont {{Doubrawa}}}, \bibnamefont {et~al.},\ }\href {https://doi.org/10.1051/0004-6361/202349019} {\bibfield  {journal} {\bibinfo  {journal} {\aap}\ }\textbf {\bibinfo {volume} {685}},\ \bibinfo {eid} {A98} (\bibinfo {year} {2024})}\BibitemShut {NoStop}%
\bibitem [{\citenamefont {{Oguri}}(2014)}]{CAMIRA}%
  \BibitemOpen
  \bibfield  {author} {\bibinfo {author} {\bibfnamefont {M.}~\bibnamefont {{Oguri}}},\ }\href {https://doi.org/10.1093/mnras/stu1446} {\bibfield  {journal} {\bibinfo  {journal} {\mnras}\ }\textbf {\bibinfo {volume} {444}},\ \bibinfo {pages} {147} (\bibinfo {year} {2014})}\BibitemShut {NoStop}%
\bibitem [{\citenamefont {{Grishin}}\ \emph {et~al.}(2025)\citenamefont {{Grishin}}, \citenamefont {{Mei}}, \citenamefont {{Ilic}}, \citenamefont {{Aguena}}, \citenamefont {{Boutigny}}, \citenamefont {{Paturel}},\ and\ \citenamefont {{LSST Dark Energy Science Collaboration}}}]{YOLO}%
  \BibitemOpen
  \bibfield  {author} {\bibinfo {author} {\bibfnamefont {K.}~\bibnamefont {{Grishin}}}, \bibinfo {author} {\bibfnamefont {S.}~\bibnamefont {{Mei}}}, \bibinfo {author} {\bibfnamefont {S.}~\bibnamefont {{Ilic}}}, \bibinfo {author} {\bibfnamefont {M.}~\bibnamefont {{Aguena}}}, \bibinfo {author} {\bibfnamefont {D.}~\bibnamefont {{Boutigny}}}, \bibinfo {author} {\bibfnamefont {M.}~\bibnamefont {{Paturel}}},\ \bibnamefont {and}\ \bibinfo {author} {\bibnamefont {{LSST Dark Energy Science Collaboration}}},\ }\href {https://doi.org/10.1051/0004-6361/202452119} {\bibfield  {journal} {\bibinfo  {journal} {\aap}\ }\textbf {\bibinfo {volume} {695}},\ \bibinfo {eid} {A246} (\bibinfo {year} {2025})}\BibitemShut {NoStop}%
\bibitem [{\citenamefont {{Costanzi}}\ \emph {et~al.}(2019{\natexlab{b}})\citenamefont {{Costanzi}}, \citenamefont {{Rozo}}, \citenamefont {{Simet}}, \citenamefont {{Zhang}}, \citenamefont {{Evrard}}, \citenamefont {{Mantz}}, \citenamefont {{Rykoff}}, \citenamefont {{Jeltema}}, \citenamefont {{Gruen}}, \citenamefont {{Allen}}, \citenamefont {{McClintock}}, \citenamefont {{Romer}}, \citenamefont {{von der Linden}}, \citenamefont {{Farahi}}, \citenamefont {{DeRose}}, \citenamefont {{Varga}}, \citenamefont {{Weller}}, \citenamefont {{Giles}}, \citenamefont {{Hollowood}}, \citenamefont {{Bhargava}}, \citenamefont {{Bermeo-Hernandez}}, \citenamefont {{Chen}}, \citenamefont {{Abbott}}, \citenamefont {{Abdalla}}, \citenamefont {{Avila}}, \citenamefont {{Bechtol}}, \citenamefont {{Brooks}}, \citenamefont {{Buckley-Geer}}, \citenamefont {{Burke}}, \citenamefont {{Rosell}}, \citenamefont {{Kind}}, \citenamefont {{Carretero}}, \citenamefont {{Crocce}}, \citenamefont {{Cunha}}, \citenamefont {{da Costa}}, \citenamefont
  {{Davis}}, \citenamefont {{De Vicente}}, \citenamefont {{Diehl}}, \citenamefont {{Dietrich}}, \citenamefont {{Doel}}, \citenamefont {{Eifler}}, \citenamefont {{Estrada}}, \citenamefont {{Flaugher}}, \citenamefont {{Fosalba}}, \citenamefont {{Frieman}}, \citenamefont {{Garc{\'\i}a-Bellido}}, \citenamefont {{Gaztanaga}}, \citenamefont {{Gerdes}}, \citenamefont {{Giannantonio}}, \citenamefont {{Gruendl}}, \citenamefont {{Gschwend}}, \citenamefont {{Gutierrez}}, \citenamefont {{Hartley}}, \citenamefont {{Honscheid}}, \citenamefont {{Hoyle}}, \citenamefont {{James}}, \citenamefont {{Krause}}, \citenamefont {{Kuehn}}, \citenamefont {{Kuropatkin}}, \citenamefont {{Lima}}, \citenamefont {{Lin}}, \citenamefont {{Maia}}, \citenamefont {{March}}, \citenamefont {{Marshall}}, \citenamefont {{Martini}}, \citenamefont {{Menanteau}}, \citenamefont {{Miller}}, \citenamefont {{Miquel}}, \citenamefont {{Mohr}}, \citenamefont {{Ogando}}, \citenamefont {{Plazas}}, \citenamefont {{Roodman}}, \citenamefont {{Sanchez}},
  \citenamefont {{Scarpine}}, \citenamefont {{Schindler}}, \citenamefont {{Schubnell}}, \citenamefont {{Serrano}}, \citenamefont {{Sevilla-Noarbe}}, \citenamefont {{Sheldon}}, \citenamefont {{Smith}}, \citenamefont {{Soares-Santos}}, \citenamefont {{Sobreira}}, \citenamefont {{Suchyta}}, \citenamefont {{Swanson}}, \citenamefont {{Tarle}}, \citenamefont {{Thomas}},\ and\ \citenamefont {{Wechsler}}}]{Costanzietal2019b}%
  \BibitemOpen
  \bibfield  {author} {\bibinfo {author} {\bibfnamefont {M.}~\bibnamefont {{Costanzi}}}, \bibnamefont {et~al.},\ }\href {https://doi.org/10.1093/mnras/stz1949} {\bibfield  {journal} {\bibinfo  {journal} {\mnras}\ }\textbf {\bibinfo {volume} {488}},\ \bibinfo {pages} {4779} (\bibinfo {year} {2019}{\natexlab{b}})}\BibitemShut {NoStop}%
\end{thebibliography}%

\end{document}